\documentclass[final,3p,times,compress]{elsarticle}

\usepackage{amssymb}
\usepackage{lipsum}
\usepackage{color}
\usepackage{amsthm}
\usepackage{braket}
\usepackage{mathrsfs}
\usepackage{nccmath}
\usepackage{bm}
\usepackage{mleftright}
\usepackage[colorlinks=true]{hyperref}

\journal{Annals of Physics}

\begin{document}

\begin{frontmatter}



\title{Path-Integral Formulation of Truncated Wigner Approximation \\
for Bosonic Markovian Open Quantum Systems}


\author[first]{Toma~Yoneya}
\author[second]{Kazuya~Fujimoto}
\author[first,third]{Yuki~Kawaguchi}
\affiliation[first]{organization={Depertment of Applied Physics, Nagoya University},
            city={Nagoya},
            state={464-8603},
            country={Japan}}

\affiliation[second]{organization={Department of Physics, Institute of Science Tokyo},
            city={Tokyo},
            state={152-8551},
            country={Japan}}

\affiliation[third]{organization={Research Center for Crystalline Materials Engineering, Nagoya University},
            city={Nagoya},
            state={464-8603},
            country={Japan}}
\begin{abstract}
The truncated Wigner approximation (TWA) enables us to investigate bosonic quantum many-body dynamics, including open quantum systems described by the Gorini-Kossakowski-Sudarshan-Lindblad (GKSL) equation.
In the TWA, the Weyl-Wigner transformation, a way of mapping from quantum-mechanical operators to $c$-numbers, of the GKSL equation leads to the Fokker-Planck equation, which we calculate by reducing it to the corresponding stochastic differential equations.
However, the Fokker-Planck equation is not always reduced to the stochastic differential equations depending on details of jump operators.
In this work, we clarify the condition for obtaining the stochastic differential equations from the Fokker-Planck equation and derive analytical expressions of these equations for a system with an arbitrary Hamiltonian with jump operators that do not couple different states.
This result enables us to shortcut the conventional complicated calculations in applying the TWA.
In the course of the derivation, we formulate the GKSL equation by using the path-integral representation based on the Weyl-Wigner transformation, which gives us a clear interpretation of the relation between the TWA and quantum fluctuations and allows us to calculate the non-equal time correlation functions in the TWA.
In the benchmark calculations, we numerically confirm that the relaxation dynamics of physical quantities including the non-equal time correlation functions obtained in our formulation agrees well with the exact ones in the numerically solvable models.
\end{abstract}




\end{frontmatter}

\tableofcontents



\section{Introduction}
\label{introduction}
Considering that physical systems generally involve many-body interactions of particles and couplings with environments, understanding open quantum many-body systems \cite{Breuer} is one of the fundamental issues in condensed matter physics.
However, because the size of the Hilbert space exponentially grows with system size in many-body systems, the numerically exact calculation is limited to small systems. So far, various numerical techniques, such as the quantum Monte Carlo wave function method \cite{Dalibard,Dum,Klaus,Howard,Plenio,DaleyMonte} and the matrix-product-state approaches \cite{Vidal2003,Vidal2004,White,Daley,Verstraete,Zwolak}, have been developed to investigate open quantum many-body systems, in particular, to simulate the Gorini-Kossakowski-Sudarshan-Lindblad (GKSL) equation \cite{Gorini,Lindblad} describing the Markovian dynamics of open quantum systems. While these approaches can predict quantum dynamics with high accuracy, their application is still limited to low-dimensional or finite-size systems, and capturing the underlying physics from these simulations is not easy.
Although the mean-field approximation \cite{Iacopo,DiehlMF,Tomadin} is useful to discuss the physical picture of the system, it completely ignores the effects of quantum fluctuations and may miss the interesting quantum effects.
To take into account the effects of quantum fluctuations, the cluster mean-field \cite{Jin} and dynamical mean-field approximations \cite{Scarlatella} have been extended to the open quantum systems.

The truncated Wigner approximation (TWA) is another a method that gives physical picture of quantum dynamics while accounting for quantum fluctuations \cite{Hillery,Steel,Alice,Blakie,Polkovnikov2010}.
It can describe quantum dynamics of weakly interacting bosonic many-body systems of large size.
According to the Weyl-Wigner transformation, operators are mapped into $c$-number functions and the density operator is represented as the Wigner function in the phase space.
In the TWA for isolated systems, we take into account fluctuations of the initial state by solving the Hamilton equation with various initial conditions which are sampled from the Wigner function for the initial state.
Owing to that the TWA incorporates the effects of quantum fluctuations based on the classical motion of the system, it provides us a physically understandable interpretation of the many-body phenomena.
Recently, the TWA has been generalized to describe the many-body dynamics of spins \cite{Schachenmayer,Davidson2015,Wurtz,Zhu} and fermions \cite{Davidson2017}, and the performance of these TWAs are being investigated by comparing with experiments of large systems \cite{Orioli,Fersterer,Takasu,Nagao2021,Christopher2022,Christopher2023,Nagao2024}.

Such TWA methods can be applied to the open quantum many-body dynamics described by the GKSL equation \cite{Iacopo,Gardiner,Milburn}.
In the TWA, the GKSL equation is approximated into the Fokker-Planck equation for the Wigner function, which we calculate by solving the corresponding stochastic differential equations with various initial conditions which are sampled from the initial Wigner function.
Here, the Fokker-Planck equation is not always reduced into the stochastic differential equations depending on details of jump operators that characterize couplings with environments \cite{Plimak,Huber}.
However, to the best of our knowledge, the condition against the jump operators for obtaining the stochastic differential equations from the Fokker-Planck equation and the analytical expressions of these equations have not been derived for arbitrary Hamiltonian and jump operators.
Consequently, we need to derive the Fokker-Planck equation and stochastic differential equations model by model before simulating the GKSL equation using the TWA, and complicated calculations are required in the course of the derivations.

In this work, we clarify the condition against the jump operators for obtaining the stochastic differential equations from the Fokker-Planck equation and derive the analytical expressions of these equations for arbitrary Hamiltonian and the jump operators that do not induce transitions between different degrees of freedom.
In the course of the derivations, we use the path-integral representation based on the Weyl-Wigner transformation \cite{Polkovnikov2010,Polkovnikov2009}.
In isolated systems, this path-integral representation gives us a clear interpretation of the relation between the TWA and quantum fluctuations and allows us to calculate the non-equal time correlation functions in the TWA.
As a first step of this work, we formulate the path-integral representation of the GKSL equation based on the Weyl-Wigner transformation for systems with a single degree of freedom and derive the Fokker-Planck equation and stochastic differential equations.
The extension to systems with multiple degrees of freedom is straightforward when the jump operators are assumed not to mix different degrees of freedom.
As benchmark calculations, we investigate the relaxation dynamics of the scalar and spin-1 Bose-Einstein condensates without spatial degrees of freedom.
Comparing the relaxation dynamics of some observables including non-equal time correlation functions using the TWA with the numerically exact ones, we confirm that our results well reproduce the exact dynamics.
In the end of this work, we consider the systems with Hermitian jump operators and show that quantum fluctuations work strongly to the relaxation dynamics resulting in failing of the mean-field approximation to reproduce the exact dynamics of the GKSL equation.

This paper is organized as follows.
In Sec.~\ref{sec:Target of this paper}, we introduce the GKSL equation and the systems what we consider.
In Sec.~\ref{sec:Reviews of Weyl-Wigner transformation}, we briefly review the Weyl-Wigner transformation.
In Sec.~\ref{sec:Wigner functional representation of Markovian open quantum systems}, we formulate the path-integral representation of the GKSL equation based on the Weyl-Wigner transformation and derive the Fokker-Planck equation and stochastic differential equations.
In the course of the derivations, we provide the path-integral interpretation of the relation between the TWA and quantum fluctuations in detail.
The formula for calculating the non-equal time correlation functions is also in this section.
We show some benchmark calculations in Sec.~\ref{sec:Benchmark calculations}.
In Sec.~\ref{sec:Discussions}, we discuss the relaxation dynamics of the GKSL equation with Hermitian jump operators in the TWA.
Summary and conclusions are in Sec.~\ref{sec:Summary and conclusions}.


\section{\label{sec:Target of this paper}Target of this paper}
This paper aims to formulate the TWA for a bosonic open quantum system described by the GKSL equation. In this section, we introduce the GKSL equation and summarize the conditions under which our results are applicable.


\subsection{\label{subsec:Gorini-Kossakowski-Sudarshan-Lindblad equation}Gorini-Kossakowski-Sudarshan-Lindblad equation}
We first provide the general description of open quantum systems.
Following the conventional literature \cite{Breuer}, we divide the total system into a system we focus on and an environment that couples with the system. The total system is assumed to be isolated from other systems such that its density operator $\hat{\rho}_{\rm tot}$ obeys the von Neumann equation.
The dynamical map $\hat{\mathcal{V}}(t,t_0)$ that propagates the system's density operator as $\hat{\rho}(t_0)\to\hat{\rho}(t)$ is defined by
\begin{align}
    \label{eq:def_of_dynamical_map}
    \hat{\rho}(t) = \hat{\mathcal{V}}(t,t_0)\mleft[\hat{\rho}(t_0)\mright] = {\rm Tr}_{{\rm B}}\mleft[\hat{U}(t,t_0)\hat{\rho}_{{\rm tot}}(t_0)\hat{U}(t,t_0)\mright],
\end{align}
where ${\rm Tr}_{{\rm B}}[\dots]$ means the partial trace with respect to the degrees of freedom of the environment, and $\hat{U}(t,t_0)$ is an unitary time-evolution operator generated by the total system's Hamiltonian.
Suppose there is no entanglement between the system and the environment in the initial state, i.e., $\hat{\rho}_{\rm tot}(t_0) = \hat{\rho}(t_0)\otimes\hat{\rho}_{\rm B}(t_0)$ with $\rho_{\rm B}(t_0)$ being the initial density operator of the environment, the dynamical map becomes a completely positive and trace-preserving (CPTP) map and is written in the Kraus representation
\begin{align}
    \label{eq:Kraus_representation}
    \hat{\rho}(t) = \hat{\mathcal{V}}(t,t_0)\mleft[\hat{\rho}(t_0)\mright] = \sum_k\hat{M}_k(t,t_0)\hat{\rho}(t_0)\hat{M}^{\dagger}_k(t,t_0),
\end{align}
where $\hat{M}_k(t,t_0)$ is the Kraus operator satisfying
\begin{align}
    \sum_k\hat{M}_k^{\dagger}(t,t_0)\hat{M}_k(t,t_0) = \hat{1}
\end{align}
with $\hat{1}$ being the identity operator.
This completeness condition ensures the trace-preserving property of the density operator, and the Kraus representation also guarantees the positive-semidefiniteness of $\hat{\rho}(t)$ and $\hat{\rho}_{\rm tot}(t)$. 
It is shown that the dynamical map is written in the Kraus representation if it is CPTP \cite{Choi}.

A system described by the Kraus representation generally exhibits non-Markovian dynamics, where the quantum state depends on its history.
In this paper, however, we restrict ourselves to considering only Markovian open quantum systems, i.e., we suppose the dynamics of quantum states depend only on their instantaneous state.
For a Markovian open quantum system, the equation of motion for the density operator reduces to the Gorini-Kossakowski-Sudarshan-Lindblad (GKSL) equation \cite{Gorini,Lindblad}:
\begin{align}
    \label{eq:action_of_Lindbladia}
    \frac{d\hat{\rho}(t)}{dt} = -\frac{i}{\hbar}\mleft[\hat{H},\hat{\rho}(t)\mright]_- + \sum_k\gamma_k\mleft(\hat{L}_k\hat{\rho}(t)\hat{L}^{\dagger}_k - \frac{1}{2}\mleft[\hat{L}^{\dagger}_k\hat{L}_k,\hat{\rho}(t)\mright]_+\mright),
\end{align}
where $[\cdots]_{\mp}$ denote the commutator $(-)$ and anti-commutator $(+)$.
The first term of the right-hand side describes unitary dynamics generated by the system's Hamiltonian $\hat{H}$, and the second term describes non-unitary dynamics, where the jump operator $\hat{L}_k$ characterizes the interaction between the system and the environment, $\gamma_k$ represents the strength, and the subscript $k$ distinguishes a variety of couplings with the environment.
Whereas we can include the term $\gamma_k/2[L_k^\dagger L_k,\rho(t)]_+$ to the first term as a non-Hermitian part of the Hamiltonian, the term $\gamma_k L_k \rho L_k^\dagger$ describes the dynamics never expressed in the Schrodinger-type equation.
Below, we refer to the former as the quantum diffusion term and the latter as the quantum jump term.


\subsection{\label{sec:setup}Setup}
In this paper, we consider a system satisfying following conditions:
\begin{enumerate}
    \item \textbf{The system involves only bosonic degrees of freedom}
    \begin{itemize}
        \item[] Hamiltonian $\hat{H}$ and jump operators $\hat{L}_k$ for $\forall k$ are composed of bosonic creation and annihilation operators $\hat{a}^{\dagger}_m$ and $\hat{a}_m$ satisfying the commutation relation $[\hat{a}_{m},\hat{a}^{\dagger}_{m'}]_- = \delta_{mm'}$, where the subscripts $m$ and $m'$ denote internal or external degrees of freedom of bosons. In particular, $\hat{H}$ can include higher-body interaction terms within this condition.
    \end{itemize}
    \item \textbf{The jump operators do not couple different degrees of freedom}
    \begin{itemize}
        \item[] Each jump operator $\hat{L}_k$ involves only one bosonic state.
        To specify the involved state $m$, we relabel the subscript $k$ with $k,m$ and expand the jump operator as
        $\hat{L}_{k,m} = \sum_{p,q}l_{kmpq}\hat{a}^{\dagger p}_m\hat{a}^q_m$, where the new subscript $k$ runs $1,2,\cdots$ for each $m$.
    \end{itemize}
\end{enumerate}



\section{\label{sec:Reviews of Weyl-Wigner transformation} Review of Weyl-Wigner transformation}
Before moving on to the path-integral formulation of the TWA, we briefly summarize the basic properties of the Weyl-Wigner transformation, which is a mapping from an operator to a $c$-number function in the phase space \cite{Hillery}. 
This is the starting point of the TWA.
In this section, we consider a system with a single degree of freedom (see Sec.~\ref{subsec:Multiple degrees of freedom} for the Weyl-Wigner transformation of a system with multiple degrees of freedom).


\subsection{Weyl-Wigner transformation and Wigner function}
The Weyl-Wigner transformation maps an arbitrary operator $\hat{A}$ consisting of bosonic creation and annihilation operators $\hat{a}^{\dagger}$ and $\hat{a}$ satisfying the commutation relation $[\hat{a},\hat{a}^{\dagger}]_- = 1$ into a $c$-number function defined by
\begin{align}
    \label{eq:def_of_Weyl_Wigner_ transformation}
    \hat{A}\mapsto A_W(\alpha) = \int\frac{d^2\eta}{\pi}\chi_{A}(\eta)e^{\alpha^*\eta - \alpha\eta^*},
\end{align}
where $\alpha = \alpha^{\rm re} + i\alpha^{\rm im}\in\mathbb{C}$ ($\alpha^{\rm re},\alpha^{\rm im}\in\mathbb{R}$),
$\int d^2\eta = \int_{-\infty}^\infty d\eta^{\rm re} \int_{-\infty}^\infty d\eta^{\rm im}$ with $\eta=\eta^{\rm re} + i\eta^{\rm im}$ ($\eta^{\rm re}, \eta^{\rm im}\in\mathbb{R}$), and $\chi_{A}(\eta)$ is a characteristic function given by
\begin{align}
    \label{eq:charactratic_function_appendix}
    \chi_{A}(\eta) = {\rm Tr}\mleft[\hat{A}\hat{D}^{\dagger}(\eta)\mright]
\end{align}
with the displacement operator $\hat{D}(\eta) = e^{\eta\hat{a}^{\dagger} - \eta^*\hat{a}}$.
We note that $A_W(\alpha)$ is a function of two variables of $\alpha$ and $\alpha^*$, but we omit $\alpha^*$ in this paper for brevity.
The $c$-number function $A_W(\alpha)$ is referred to as the Weyl-Wigner representation of $\hat{A}$.

We specifically refer to the Weyl-Wigner representation of the system's density operator $\hat{\rho}(t)$ as the Wigner function, which is given by
\begin{align}
    \label{eq:Weyl_Wigner_transformation_rho}
    W(\alpha,t) &= \int\frac{d^2\eta}{\pi}\chi_{\rho}(\eta)e^{\alpha^*\eta - \alpha\eta^*}, \\
    \chi_{\rho}(\eta) &= {\rm Tr}\mleft[\hat{\rho}(t)\hat{D}^{\dagger}(\eta)\mright].
\end{align}
By using the relation
\begin{align}
    \label{eq:Tr_AB_Weyl_Wigner}
    {\rm Tr}\mleft[\hat{A}\hat{B}\mright] = \int\frac{d^2\alpha}{\pi}A_W(\alpha)B_W(\alpha)
\end{align}
with $\hat{B} = \hat{\rho}(t)$, we can calculate a physical quantity $\braket{\hat{A}(t)} = {\rm Tr}[\hat{A}\hat{\rho}(t)]$ in the Weyl-Wigner representation as
\begin{align}
    \label{eq:expectation_value_phase_space}
    \Braket{\hat{A}(t)} = \int\frac{d^2\alpha}{\pi}A_W(\alpha)W(\alpha,t).
\end{align}
In addition, by replacing $\hat{A}$ with an identity operator $\hat{1}$ in Eq.~\eqref{eq:expectation_value_phase_space} and using the fact ${\rm Tr}[\hat{\rho}(t)] = 1$, we obtain the normalization condition for the Wigner function:
\begin{align}
    \label{eq:normalizatino_conditoin_of_Wigner_function}
    \int\frac{d^2\alpha}{\pi}W(\alpha,t) = 1.
\end{align}
According to Eqs.~\eqref{eq:expectation_value_phase_space} and \eqref{eq:normalizatino_conditoin_of_Wigner_function}, the Wigner function behaves like a probability distribution function.
However, since the Wigner function in general takes negative values and does not preserve positivity, the Wigner function is often refereed to as a quasi probability distribution function. 


\subsection{Operator ordering}
The Weyl-Wigner transformation maps a set of certain ordered products of non-commutable operators, so-called Weyl-ordered product,  into a product of $c$-numbers. Here, the Weyl ordering for the bosonic operators is defined by
\begin{align}
    \label{eq:Weyl_ordering}
    \mleft\{\hat{a}^{\dagger p}\hat{a}^q\mright\}_{\rm Weyl} = \mleft.(-1)^q\frac{\partial^{p+q}}{\partial\alpha^p\partial\alpha^{*q}}\hat{D}(\alpha)\mright|_{\alpha = 0}.
\end{align}
The Weyl-orderd product consists of all possible ordering products of $\hat{a}$ and $\hat{a}^{\dagger}$, e.g., $\{\hat{a}^{\dagger}\hat{a}\}_{\rm Weyl} = (\hat{a}^{\dagger}\hat{a} + \hat{a}\hat{a}^{\dagger})/2$ and $\{\hat{a}^{\dagger 2}\hat{a}^{2}\}_{\rm Weyl} = (\hat{a}^{\dagger 2}\hat{a}^{2} + \hat{a}^{\dagger}\hat{a}\hat{a}^{\dagger}\hat{a} + \hat{a}^{\dagger}\hat{a}^2\hat{a}^{\dagger} + \hat{a}\hat{a}^{\dagger 2}\hat{a} + \hat{a}\hat{a}^{\dagger}\hat{a}\hat{a}^{\dagger} + \hat{a}^{2}\hat{a}^{\dagger 2})/6$. 
The Weyl-Wigner transformation maps a Weyl-ordered product to a complex $c$-number as
\begin{align}
    \label{eq:Weyl_Wigner_transformation_of_Weyl_ordering}
    \mleft\{\hat{a}^{\dagger p}\hat{a}^q\mright\}_{\rm Weyl}\longmapsto \alpha^{*p}\alpha^q.
\end{align}
In order to obtain the analytical expression of $A_W(\alpha)$, we first expand $\hat{A}$ in the Weyl ordering as $\hat{A} = \sum_{p,q}A_{pq}\{\hat{a}^{\dagger p}\hat{a}^q\}_{\rm Weyl}$ and replace the operators to the $c$-numbers according to Eq.~\eqref{eq:Weyl_Wigner_transformation_of_Weyl_ordering}, obtaining 
$A_W(\alpha) = \sum_{p,q}A_{pq}\alpha^{*p}\alpha^q$.

\subsection{Weyl-Wigner transformation of a product of two operators}
The Weyl-Wigner transformation of a product of two operators $[\hat{A}\hat{B}]_W(\alpha)$ is given by the well-known formula \cite{Hillery}
\begin{gather}
    \label{eq:Weyl-Wigner_transformation_of_a_product_of_two_operator}
    \mleft[\hat{A}\hat{B}\mright]_W(\alpha) = A_W(\alpha)e^{\hat{\phi}/2}B_W(\alpha),
\end{gather}
where $\hat{\phi}$ is the differential operator defined by
\begin{align}
    \label{eq:def_of_hat_phi}
    \hat{\phi} = \frac{\overleftarrow{\partial}}{\partial \alpha}\frac{\overrightarrow{\partial}}{\partial \alpha^*} - \frac{\overleftarrow{\partial}}{\partial \alpha^*}\frac{\overrightarrow{\partial}}{\partial \alpha}
\end{align}
with the arrows above the derivative symbols indicating which function, left or right, is to be differentiated.
We also use the following abbreviation
\begin{gather}
    \label{eq:Moyal_product}
    A_W(\alpha)e^{\hat{\phi}/2}B_W(\alpha) = A_W(\alpha)\star B_W(\alpha),
\end{gather}
where $\star = e^{\hat{\phi}/2}$ is referred to as the Moyal product.
In particular, taking one of $\hat{A}$ and $\hat{B}$ as $\hat{a}^{\dagger}$ or $\hat{a}$, we obtain the following formulas:
\begin{align}
    \label{eq:Bopp_operator}
\begin{aligned}
    \mleft[\hat{a}\hat{A}\mright]_W(\alpha) = \mleft(\alpha + \frac{1}{2}\frac{\partial}{\partial \alpha^*}\mright)A_W(\alpha),\quad \mleft[\hat{a}^{\dagger}\hat{A}\mright]_W(\alpha) = \mleft(\alpha^* - \frac{1}{2}\frac{\partial}{\partial \alpha}\mright)A_W(\alpha), \\
    \mleft[\hat{A}\hat{a}\mright]_W(\alpha) = \mleft(\alpha - \frac{1}{2}\frac{\partial}{\partial \alpha^*}\mright)A_W(\alpha),\quad \mleft[\hat{A}\hat{a}^{\dagger}\mright]_W(\alpha) = \mleft(\alpha^* + \frac{1}{2}\frac{\partial}{\partial \alpha}\mright)A_W(\alpha),
\end{aligned}
\end{align}
where the operators acting on $A_W(\alpha)$ on the right-hand side of Eq.~\eqref{eq:Bopp_operator} are referred to as the Bopp operators.
The Bopp operators and the Moyal product are widely used to systematically calculate the Weyl-Wigner representations of the von Neumann and GKSL equations \cite{Hillery,Steel,Alice,Blakie,Polkovnikov2010,Gardiner,Milburn}.

An another representation of $[\hat{A}\hat{B}]_W(\alpha)$ is the integral representation: 
\begin{align}
    \label{eq:AB_Weyl_Wigner_representation_1}
    \mleft[\hat{A}\hat{B}\mright]_W(\alpha) &= \int\frac{d^2\alpha_0d^2\eta}{\pi^2}A_W(\alpha_0)B_W\mleft(\alpha_0 + \frac{\eta}{2}\mright)e^{\eta^*(\alpha - \alpha_0) - \eta(\alpha^* - \alpha^*_0)} \\
    \label{eq:AB_Weyl_Wigner_representation_2}
    &= \int\frac{d^2\alpha_0d^2\eta}{\pi^2}A_W\mleft(\alpha_0 - \frac{\eta}{2}\mright)B_W\mleft(\alpha_0\mright)e^{\eta^*(\alpha - \alpha_0) - \eta(\alpha^* - \alpha^*_0)},
\end{align}
where the first and second lines are transformed to each other by changing the integration variables.
The detailed derivations of above-discussed properties of the Weyl-Wigner representation are summarized in Refs.~\cite{Hillery,Nagao}.

\section{\label{sec:Wigner functional representation of Markovian open quantum systems} Wigner functional representation of Markovian open quantum systems}

In the following, we formulate the TWA for the GKSL equation from the path-integral approach.
We first derive the path-integral representation of the GKSL equation in Secs.~\ref{subsec:Markov condition in the phase space} and \ref{subsec:Path-integral representation} using the Wigner function and then formulate the TWA in Sec.~\ref{subsec:Equation of motion in the phase space}.
The results in Secs.~\ref{subsec:Markov condition in the phase space}-\ref{subsec:Equation of motion in the phase space} are for a system with a single-degree of freedom,
which we extend to a system with multiple degrees of freedom in Sec.~\ref{subsec:Extension to a multi-mode system}.
The path-integral formulation enables us to calculate non-equal time correlation functions, which will be explained in Sec.~\ref{subsec:Non-equal time correlation functions}.


\subsection{\label{subsec:Markov condition in the phase space} Markov condition in the phase space}
The Weyl-Wigner transformation of the Kraus representation of Eq.~\eqref{eq:Kraus_representation} is
\begin{align}
    \label{eq:Wigner_function_and_propagator}
    W\mleft(\alpha_{\rm f},t\mright) = \int\frac{d^2\alpha_0}{\pi}\varUpsilon\mleft(\alpha_{\rm f},t~;\alpha_0,t_0\mright)W\mleft(\alpha_0,t_0\mright),
\end{align}
where the propagator $\varUpsilon\mleft(\alpha_{\rm f},t~;\alpha_0,t_0\mright)$ is given by
\begin{align}
    \label{propagator_kraus_representation}
    \varUpsilon\mleft(\alpha_{\rm f},t~;\alpha_0,t_0\mright) = \int\frac{d^2\xi d^2\eta }{\pi^2}\sum_k{\rm Tr}\mleft[\hat{D}^{\dagger}(\xi )\hat{M}_k\mleft(t,t_0\mright)\hat{D}^{\dagger}\mleft(\eta \mright)\hat{M}^{\dagger}_k\mleft(t,t_0\mright)\mright]e^{-\alpha_{\rm f}\xi^* + \alpha_{\rm f}^*\xi }e^{-\alpha_0\eta^* + \alpha^*_0\eta }.
\end{align}
We provide the derivations of Eq.~\eqref{propagator_kraus_representation} in \ref{The propagator of the Wigner function}.
We can show that the propagator satisfies the Markov condition
\begin{align}
    \label{eq:Markov_condition_phase_space}
    \varUpsilon\mleft(\alpha_{\rm f},t~;\alpha_0,t_0\mright) = \int\frac{d^2\alpha_j}{\pi}\varUpsilon(\alpha_{\rm f},t~;\alpha_j,t_j)\varUpsilon(\alpha_j,t_j~;\alpha_0,t_0)
\end{align}
when the dynamical map \eqref{eq:def_of_dynamical_map} satisfies the Markov condition. The detailed derivation is given in \ref{Markov condition for the propagator}.


\subsection{\label{subsec:Path-integral representation} Path-integral representation}
Using Eq.~\eqref{eq:Markov_condition_phase_space}, we can write the time-evolved Wigner function $W(\alpha_{\rm f},t)$ as an infinite product of the infinitesimal-time propagators:
\begin{align}
    \label{eq:Wigner_functional_Markovian_open_system_using_propagator}
    W\mleft(\alpha_{\rm f},t\mright) = \lim_{\Delta t\to 0}\prod_{j=0}^{N_t - 1}\int\frac{d^2\alpha_j}{\pi}\varUpsilon(\alpha_{j+1},t_j + \Delta t~;\alpha_j,t_j)W\mleft(\alpha_0,t_0\mright),
\end{align}
where we discretize the time interval $[t_0,t]$ into $N_t$ intervals with width $\Delta t$:
\begin{align}
    N_t=\frac{t-t_0}{\Delta t},\quad
    t_j = t_0 + j\Delta t,\quad
    t_{N_t} = t,\quad
    \alpha_{N_t} = \alpha_{\rm f}.
\end{align}
Equation~\eqref{eq:Wigner_functional_Markovian_open_system_using_propagator} is the path integral representation of the Wigner function.
Below, we derive the detailed expression of the propagator $\varUpsilon(\alpha_{j+1},t_j + \Delta t~;\alpha_j,t_j)$ for the GKSL equation.
By substituting the propagator into Eq.~\eqref{eq:Wigner_functional_Markovian_open_system_using_propagator}, we obtain the path-integral representation of the GKSL equation.

According to the GKSL equation \eqref{eq:action_of_Lindbladia}, we obtain
\begin{align}
    \label{eq:GKSL_infinitesimal_time_evolution}
    \hat{\rho}(t_j + \Delta t) = \hat{\rho}(t_j) - \frac{i\Delta t}{\hbar}\mleft(\hat{H}\hat{\rho}(t_j) - \hat{\rho}(t_j)\hat{H}\mright) + \sum_k\gamma_k\Delta t\mleft(\hat{L}_k\hat{\rho}(t_j)\hat{L}^{\dagger}_k - \frac{1}{2}\hat{L}^{\dagger}_k\hat{L}_k\hat{\rho}(t_j) - \frac{1}{2}\hat{\rho}(t_j)\hat{L}^{\dagger}_k\hat{L}_k\mright),
\end{align}
where we consider the infinitesimal time interval $\Delta t \ll 1$ and neglect the higher-order term of $\Delta t$.
The Weyl-Wigner representation of Eq.~\eqref{eq:GKSL_infinitesimal_time_evolution} reads
\begin{align}
    \label{eq:GKSL_Wely_Wigner_representation}
    W(\alpha_{j+1},t_j+\Delta t) =& W(\alpha_{j+1},t_j) - \frac{i\Delta t}{\hbar}\mleft(\mleft[\hat{H}\hat{\rho}(t_j)\mright]_W(\alpha_{j+1}) - \mleft[\hat{\rho}(t_j)\hat{H}\mright]_W(\alpha_{j+1})\mright) \nonumber \\
    &+ \sum_k\gamma_k\Delta t\mleft(\mleft[\hat{L}_k\hat{\rho}(t_j)\hat{L}^{\dagger}_k\mright]_W(\alpha_{j+1}) - \frac{1}{2}\mleft[\hat{L}^{\dagger}_k\hat{L}_k\hat{\rho}(t_j)\mright]_W(\alpha_{j+1}) - \frac{1}{2}\mleft[\hat{\rho}(t_j)\hat{L}^{\dagger}_k\hat{L}_k\mright]_W(\alpha_{j+1})\mright).
\end{align}
The integral expressions of the Weyl-Wigner representation of the unitary and quantum diffusion terms are straightforwardly obtained by using the formulas \eqref{eq:AB_Weyl_Wigner_representation_1} and \eqref{eq:AB_Weyl_Wigner_representation_2} as
\begin{align}
    \label{eq:Weyl_Wigner_representation_of_unitary_1}
    \mleft[\hat{H}\hat{\rho}(t_j)\mright]_W(\alpha_{j+1}) &= \int\frac{d^2 \alpha_j d^2\eta_{j+1}}{\pi^2}e^{\eta_{j+1}^*(\alpha_{j+1} - \alpha_j) - \eta_{j+1}(\alpha_{j+1}^* - \alpha_j^*)}H_W\mleft(\alpha_j - \frac{\eta_{j+1}}{2}\mright)W(\alpha_j,t_j), \\
    \label{eq:Weyl_Wigner_representation_of_unitary_2}
    \mleft[\hat{\rho}(t_j)\hat{H}\mright]_W(\alpha_{j+1}) &= \int\frac{d^2 \alpha_j d^2\eta_{j+1}}{\pi^2}e^{\eta_{j+1}^*(\alpha_{j+1} - \alpha_j) - \eta_{j+1}(\alpha_{j+1}^* - \alpha_j^*)}H_W\mleft(\alpha_j + \frac{\eta_{j+1}}{2}\mright)W(\alpha_j,t_j), \\
    \label{eq:Weyl_Wigner_representation_of_nonunitary_2}
    \mleft[\hat{L}^{\dagger}_k\hat{L}_k\hat{\rho}(t_j)\mright]_W(\alpha_{j+1}) &= \int\frac{d^2\alpha_jd^2\eta_{j+1}}{\pi^2}e^{\eta^*_{j+1}(\alpha_{j+1} - \alpha_j) - \eta_{j+1}(\alpha^*_{j+1} - \alpha^*_j)}\mleft[L^{\ast}_{kW}\mleft(\alpha_j - \frac{\eta_{j+1}}{2}\mright)\star L_{kW}\mleft(\alpha_j - \frac{\eta_{j+1}}{2}\mright)\mright]W(\alpha_j,t_j), \\
    \label{eq:Weyl_Wigner_representation_of_nonunitary_3}
    \mleft[\hat{\rho}(t_j)\hat{L}^{\dagger}_k\hat{L}_k\mright]_W(\alpha_{j+1}) &= \int\frac{d^2\alpha_jd^2\eta_{j+1}}{\pi^2}e^{\eta^*_{j+1}(\alpha_{j+1} - \alpha_j) - \eta_{j+1}(\alpha^*_{j+1} - \alpha^*_j)}\mleft[L^{\ast}_{kW}\mleft(\alpha_j + \frac{\eta_{j+1}}{2}\mright)\star L_{kW}\mleft(\alpha_j + \frac{\eta_{j+1}}{2}\mright)\mright]W(\alpha_j,t_j),
\end{align}
where we decompose the Weyl-Wigner representation of $\hat{L}_k^{\dagger}\hat{L}_k$ using Eqs.~\eqref{eq:Weyl-Wigner_transformation_of_a_product_of_two_operator} and \eqref{eq:Moyal_product}.
On the other hand, since the jump operator is applied from both sides of the density operator in the quantum jump term $\gamma_k\hat{L}_k \hat{\rho} \hat{L}_k ^{\dagger} $, it is non-trivial whether its Wey-Wigner representation is written in the integral form of $W(\alpha_j, t_j)$.
We, however, find that the Weyl-Wigner representation of a product of three operators can be cast into
\begin{align}
    \label{eq:ABC_Weyl_Wigner_representation}
    \mleft[\hat{A}\hat{B}\hat{C}\mright]_W(\alpha) = \int\frac{d^2\alpha_0 d^2\eta}{\pi^2}e^{\eta^*(\alpha - \alpha_0) - \eta(\alpha^* - \alpha_0^*)}\mleft[C_W\mleft(\alpha_0 + \frac{\eta}{2}\mright)\star^{e}A_W\mleft(\alpha_0 - \frac{\eta}{2}\mright)\mright]B_W(\alpha_0),
\end{align}
where we define the extended Moyal product\footnote{$\star^{e}$ is referred to as the Moyal product on the double phase-space in Ref.~\cite{Graefe}.} $\star^{e}$ as
\begin{align}
    \label{eq:extended_Moyal_product}
    A_W(\alpha)\star^{e} B_W(\beta) = A_W(\alpha){\rm exp}\mleft\{\frac{1}{2}\Biggl(\frac{\overleftarrow{\partial}}{\partial \alpha}\frac{\overrightarrow{\partial}}{\partial \beta^*} - \frac{\overleftarrow{\partial}}{\partial \alpha^*}\frac{\overrightarrow{\partial}}{\partial \beta}\Biggr)\mright\}B_W(\beta).
\end{align}
When we choose $\alpha = \beta$, Eq.~\eqref{eq:extended_Moyal_product} reduces to the Moyal product in Eq.~\eqref{eq:Moyal_product}.
We provide the derivation of Eq.~\eqref{eq:ABC_Weyl_Wigner_representation} in \ref{Weyl-Wigner representation of a product of three operators}.
Using Eq.~\eqref{eq:ABC_Weyl_Wigner_representation}, we obtain the Weyl-Wigner representation of the quantum jump term as
\begin{align}
    \label{eq:Weyl_Wigner_representation_of_nonunitary_1}
    \mleft[\hat{L}_k\hat{\rho}(t_j)\hat{L}^{\dagger}_k\mright]_W(\alpha_{j+1}) = \int\frac{d^2\alpha_jd^2\eta_{j+1}}{\pi^2}e^{\eta^*_{j+1}(\alpha_{j+1} - \alpha_j) - \eta_{j+1}(\alpha^*_{j+1} - \alpha^*_j)}L^{\ast}_{kW}\mleft(\alpha_j + \frac{\eta_{j+1}}{2}\mright)\star^{e} L_{kW}\mleft(\alpha_j - \frac{\eta_{j+1}}{2}\mright)W(\alpha_j,t_j).
\end{align}

Substituting Eqs.~\eqref{eq:Weyl_Wigner_representation_of_unitary_1}-\eqref{eq:Weyl_Wigner_representation_of_nonunitary_3} and \eqref{eq:Weyl_Wigner_representation_of_nonunitary_1} into the right-hand side of Eq.~\eqref{eq:GKSL_Wely_Wigner_representation}, we obtain
\begin{align}
    \label{eq:GKSL_Wigner_Weyl_representation_infinitesimal_time_interval_tochu}
    &W(\alpha_{j+1},t_j + \Delta t) = W(\alpha_{j+1},t_j) + \int\frac{d^2\alpha_jd^2\eta_{j+1}}{\pi^2}e^{\eta^*_{j+1}(\alpha_{j+1} - \alpha_j) - \eta_{j+1}(\alpha^*_{j+1} - \alpha^*_j)} \nonumber\\
    &\hphantom{W(\alpha_{j+1},t_j + \Delta t) =} \times\frac{i\Delta t}{\hbar}\mleft\{\sum_{n=0,1}(-1)^nH_W\mleft(\alpha_j + (-1)^n\frac{\eta_{j+1}}{2}\mright) - i\hbar\mathfrak{L}\mleft(\alpha_j + \frac{\eta_{j+1}}{2},\alpha_j - \frac{\eta_{j+1}}{2}\mright)\mright\}W(\alpha_j,t_j),
\end{align}
where $\mathfrak{L}$ is given by
\begin{align}
    \mathfrak{L}(\alpha,\beta) = \sum_k\gamma_k\mleft\{L^{\ast}_{kW}(\alpha)\star^{e} L_{kW}(\beta) - \frac{1}{2}L^{\ast}_{kW}(\alpha)\star L_{kW}(\alpha) - \frac{1}{2}L^{\ast}_{kW}(\beta)\star L_{kW}(\beta)\mright\}.
\end{align}
Using the Dirac delta function in the phase space
\begin{align}
    \label{eq:Dirac_delta_function_phase_space}
    \int \frac{d^2\eta}{\pi^2}e^{\eta^*\alpha-\eta\alpha^*}=\delta^{(2)}(\alpha)=\delta(\alpha^{\rm re})\delta(\alpha^{\rm im}),
\end{align}
we can rewrite the first term of Eq.~\eqref{eq:GKSL_Wigner_Weyl_representation_infinitesimal_time_interval_tochu} as
\begin{align}
    W(\alpha_{j+1},t_j) = 
    \int d^2\alpha_j\delta^{(2)}(\alpha_{j+1} - \alpha_j)W(\alpha_j,t_j)
    = \int\frac{d^2\alpha_jd^2\eta_{j+1}}{\pi^2}e^{\eta^*_{j+1}(\alpha_{j+1} - \alpha_j) - \eta_{j+1}(\alpha^*_{j+1} - \alpha^*_j)}W(\alpha_j,t_j),
\end{align}
obtaining
\begin{align}
    &W(\alpha_{j+1},t_j + \Delta t) = \int\frac{d^2\alpha_jd^2\eta_{j+1}}{\pi^2}e^{\eta^*_{j+1}(\alpha_{j+1} - \alpha_j) - \eta_{j+1}(\alpha^*_{j+1} - \alpha^*_j)} \nonumber\\
    &\hphantom{W(\alpha_{j+1},t_j + \Delta t) = \int} \times\mleft[1 + \frac{i\Delta t}{\hbar}\mleft\{\sum_{n=0,1}(-1)^nH_W\mleft(\alpha_j + (-1)^n\frac{\eta_{j+1}}{2}\mright) - i\hbar\mathfrak{L}\mleft(\alpha_j + \frac{\eta_{j+1}}{2},\alpha_j - \frac{\eta_{j+1}}{2}\mright)\mright\}\mright]W(\alpha_j,t_j) \\
    \label{eq:GKSL_Wigner_Weyl_representation_infinitesimal_time_interval}
    &\hphantom{W(\alpha_{j+1},t_j + \Delta t)}=\int\frac{d^2\alpha_jd^2\eta_{j+1}}{\pi^2}{\rm exp}\Biggl[\eta_{j+1}^*(\alpha_{j+1} - \alpha_j) - \eta_{j+1}(\alpha_{j+1}^* - \alpha_j^*)\Biggr.\nonumber \\
    &\hphantom{W(\alpha_{j+1},t_j + \Delta t) = \int}\mleft.+ \frac{i\Delta t}{\hbar}\mleft\{\sum_{n=0,1}(-1)^nH_W\mleft(\alpha_j + (-1)^n\frac{\eta_{j+1}}{2}\mright) - i\hbar\mathfrak{L}\mleft(\alpha_j + \frac{\eta_{j+1}}{2},\alpha_j - \frac{\eta_{j+1}}{2}\mright)\mright\} + o(\Delta t)\mright]W(\alpha_j,t_j).
\end{align}
Comparing the right-hand side of Eqs.~\eqref{eq:GKSL_Wigner_Weyl_representation_infinitesimal_time_interval} and \eqref{eq:Wigner_function_and_propagator}, we obtain the propagator $\varUpsilon(\alpha_{j+1},t_j + \Delta t~;\alpha_j,t_j)$ as
\begin{align}
    \label{eq:propagator_infinitesimal_time_interval}
    \varUpsilon(\alpha_{j+1},t_j + \Delta t~;\alpha_j,t_j) =& \int\frac{d^2\eta_{j+1}}{\pi}{\rm exp}\Biggl[\eta_{j+1}^*(\alpha_{j+1} - \alpha_j) - \eta_{j+1}(\alpha_{j+1}^* - \alpha_j^*)\Biggr.\nonumber \\
    &+ \mleft.\frac{i\Delta t}{\hbar}\mleft\{\sum_{n=0,1}(-1)^nH_W\mleft(\alpha_j + (-1)^n\frac{\eta_{j+1}}{2}\mright) - i\hbar\mathfrak{L}\mleft(\alpha_j + \frac{\eta_{j+1}}{2},\alpha_j - \frac{\eta_{j+1}}{2}\mright)\mright\} + o(\Delta t)\mright].
\end{align}
Finally, substituting this propagator into the right-hand side of Eq.~\eqref{eq:Wigner_functional_Markovian_open_system_using_propagator} and ignoring the terms of the order $o(\Delta t)$, we obtain the path-integral representation of the GKSL equation:
\begin{gather}
    \label{eq:Wigner_functional_open_single_discrete}
    W(\alpha_{\rm f},t) = \lim_{\Delta t \to 0}\prod_{j=0}^{N_t-1}\int\frac{d^2\alpha_j d^2\eta_{j+1}}{\pi^2}e^{i\Delta t s_j/\hbar} W(\alpha_0,t_0), \\
    \label{eq:Wigner_functional_open_single_discrete_action}
    s_j = i\hbar\mleft\{\eta_{j+1}\Biggl(\frac{\alpha_{j+1}^* - \alpha_j^*}{\Delta t}\Biggr) - \eta_{j+1}^*\Biggl(\frac{\alpha_{j+1} - \alpha_j}{\Delta t}\Biggr)\mright\} + \sum_{n=0,1}(-1)^nH_W\mleft(\alpha_j + (-1)^n\frac{\eta_{j+1}}{2}\mright) - i\hbar\mathfrak{L}\mleft(\alpha_j + \frac{\eta_{j+1}}{2},\alpha_j - \frac{\eta_{j+1}}{2}\mright).
\end{gather}
When we choose $\gamma_k = 0$ for $\forall k$, this result reduces to the one for an isolated system \cite{Polkovnikov2010,Polkovnikov2009}.
From this correspondence, we can regard $\alpha_j$ and $\eta_{j+1}$ as the classical and quantum fields, respectively.
The classical field characterizes the classical motion of the system, whereas the quantum field characterizes the quantum fluctuations around the classical motion \cite{Polkovnikov2010,Polkovnikov2009,Polkovnikov2003,Marinov,Dittrich2006,Dittrich2010}.

In the continuous limit, we formally represent Eq.~\eqref{eq:Wigner_functional_open_single_discrete} as
\begin{gather}
    \label{eq:Wigner_functional_open_single_continuous}
    W(\alpha,t) = \int\mathscr{D}^2\alpha\mathscr{D}^2\eta e^{iS[\alpha,\eta]/\hbar}W(\alpha_0,t_0), \\
    \label{eq:Wigner_functional_open_single_action}
    S\mleft[\alpha,\eta\mright] = \int^t_{t_0}d\tau \mleft\{i\hbar\mleft(\eta\frac{\partial\alpha^*}{\partial\tau} - \eta^*\frac{\partial\alpha}{\partial\tau}\mright) + \sum_{n=0,1}(-1)^nH_W\mleft(\alpha + (-1)^n\frac{\eta}{2}\mright) - i\hbar\mathfrak{L}\mleft(\alpha + \frac{\eta}{2},\alpha - \frac{\eta}{2}\mright)\mright\},
\end{gather}
where $S[\alpha,\eta]$ is the action of the system.
At the boundaries, while the classical field takes $\alpha(t_0) = \alpha_0$ and $\alpha(t) = \alpha$, the quantum field is unconstrained.
\begin{figure}[t]
	\centering 
	\includegraphics[width = \linewidth]{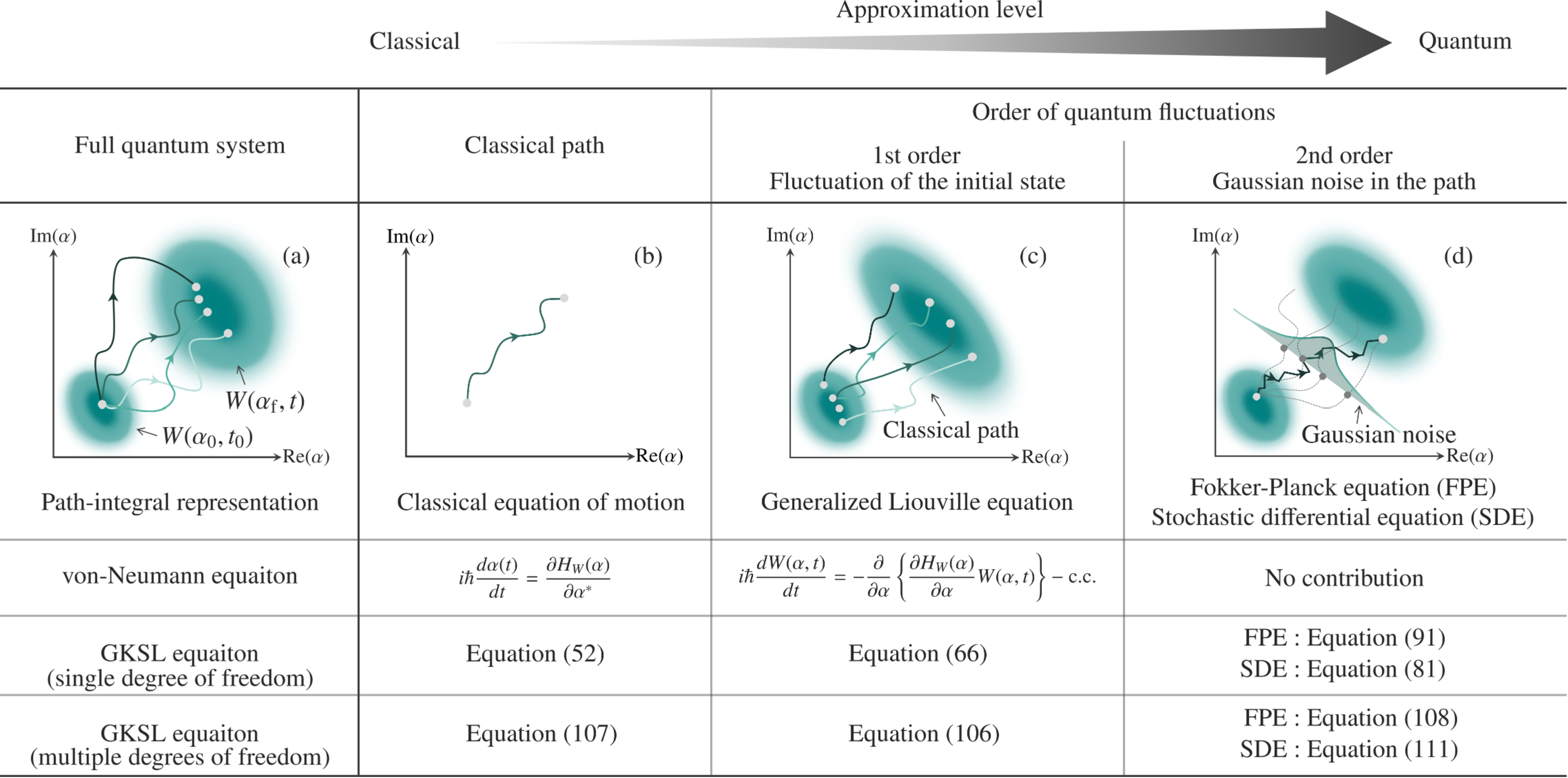}
	\caption{Schematic images of (a) the path-integral representation Eq.~\eqref{eq:Wigner_functional_open_single_continuous} and (b)-(d) the approximations of the GKSL equation in the phase space.
    (b) In the absence of quantum fluctuations, the system follows the classical path depicted as a single path in the phase space.
(c) Within the first order of quantum fluctuations, the GKSL equation is approximated into the generalized Liouville equation \cite{Gerlich,Steeb}, where each points distributed by the initial Wigner function follows the classical path.
(d) The effects of the second order of the quantum fluctuations are incorporated into the classical path as Gaussian noises,
where each points follows the stochastic differential equation.
Then, within the second order of the quantum fluctuations, the GKSL equation is approximated into the Fokker-Planck
equation, and this is the TWA.
In the last three rows of the table, we compare the equations of the von Neumann and GKSL equations.
In the absence of couplings with environments, there is no contribution of the second order quantum fluctuations, and the TWA for isolated systems approximates the von Neumann equation into the Liouville equation.
This means that the effect of the second order of the quantum fluctuations in the GKSL equation purely comes from the jump operators.
}
	\label{fig:Path_integral_short_summary}
\end{figure}

Figure~\ref{fig:Path_integral_short_summary}(a) displays a schematic illustration for the path-integral presentation.
When we choose a point in the phase space as an initial state, it follows infinite paths in time evolution.
Equation \eqref{eq:Wigner_functional_open_single_continuous} says that we need to sum up all of the paths with multiplying the appropriate phase factor $e^{iS[\alpha,\eta]/\hbar}$.
Then, we can obtain the Wigner function $W(\alpha_{\rm f},t)$ by applying the same procedure for all points $\alpha_0$ in the phase space and taking the ensemble average of the results weighted by the initial Wigner function $W(\alpha_0,t_0)$.


\subsection{\label{subsec:Equation of motion in the phase space} Equation of motion in the phase space}
We summarize the results of this section in Fig.~\ref{fig:Path_integral_short_summary}.
We first obtain the classical path [Fig.~\ref{fig:Path_integral_short_summary}(b), Sec.~\ref{subsec:Saddle-point path}], or the classical equation of motion of the system,  by neglecting the quantum field, i.e., by choosing $\eta=0$, on the saddle point of the action. On the other hand,
in the formulation of the TWA, we take into account the effects of quantum fluctuations around the classical path by expanding the action with respect to $\eta$ as a perturbation order by order.
In the first-order calculation, the integration with respect to $\eta$ in the propagator gives rise to the classical path starting from each point in the phase space as in the case of an isolated system.
The Winger function at time $t$ is obtained by summing up the contributions from all the paths weighted by the initial Winger function [Fig.~\ref{fig:Path_integral_short_summary}(c), Sec.~\ref{subsec:first order of quantum fluctuations}].
The TWA accounts for taking the effects of second-order of quantum fluctuations, where the GKSL equation is approximated into the Fokker-Planck equation [Fig.~\ref{fig:Path_integral_short_summary}(d), Sec.~\ref{subsec:second order of quantum fluctuations and truncated Wigner approximation}].
However, the Fokker-Planck equation is not always reduced to stochastic differential equations because the diffusion matrix has negative eigenvalues depending on details of the jump operators.
In Sec.~\ref{subsec:second order of quantum fluctuations and truncated Wigner approximation}, we derive the conditions against the jump operators for obtaining stochastic differential equations and analytically describe the resulting equations.


\subsubsection{\label{subsec:Saddle-point path} Classical path}
Before formulating the TWA, we derive the classical equation of motion (equation of motion of the classical path) from Eq.~\eqref{eq:Wigner_functional_open_single_continuous}.
As in the conventional path-integral formulation of quantum mechanics, a saddle-point path of the Wigner function \cite{Gozzi,Pagani} is defined as a path making the action stationary.
The classical equation of motion is obtained by setting the quantum field being zero in the equations of the saddle-point path.

In the saddle-point path, the functional derivatives of the action with respect to the classical and quantum fields vanish, i.e,
\begin{align}
    \label{eq:saddle-point_path_condition}
    \frac{\delta S}{\delta \alpha^*}=0,\quad\frac{\delta S}{\delta \eta^*}=0,\quad\frac{\delta S}{\delta \alpha}=0,\quad\frac{\delta S}{\delta \eta}=0.
\end{align}
Introducing new variables
\begin{align}
    \psi_{\pm} = \alpha \pm \frac{\eta}{2},
\end{align}
and rewriting the action \eqref{eq:Wigner_functional_open_single_action} as
\begin{align}
    S\mleft[\alpha,\eta\mright] = \int^t_{t_0}d\tau \mleft\{i\hbar\mleft(\eta\frac{\partial\alpha^*}{\partial\tau} - \eta^*\frac{\partial\alpha}{\partial\tau}\mright) + H_W(\psi_+) - H_W(\psi_-) - i\hbar\mathfrak{L}(\psi_+,\psi_-)\mright\},
\end{align}
we obtain the saddle-point equations for $\alpha$ and $\eta$ as
\begin{align}
    \label{eq:saddle-point_equation_alpha}
    i\hbar\frac{d\alpha}{dt} &= \frac{1}{2}\frac{\partial H_W(\psi_+)}{\partial\psi^*_+} + \frac{1}{2}\frac{\partial H_W(\psi_-)}{\partial\psi^*_-} -\frac{i\hbar}{2}\mleft\{\frac{\partial\mathfrak{L}(\psi_+,\psi_-)}{\partial\psi^*_+} - \frac{\partial\mathfrak{L}(\psi_+,\psi_-)}{\partial\psi^*_-}\mright\},\\
    \label{eq:saddle-point_equation_eta}
    i\hbar\frac{d\eta}{dt} &= \frac{\partial H_W(\psi_+)}{\partial\psi^*_+} - \frac{\partial H_W(\psi_-)}{\partial\psi^*_-} -i\hbar\mleft\{\frac{\partial\mathfrak{L}(\psi_+,\psi_-)}{\partial\psi^*_+} + \frac{\partial\mathfrak{L}(\psi_+,\psi_-)}{\partial\psi^*_-}\mright\},
\end{align}
where the partial derivatives of $\mathfrak{L}(\psi_+,\psi_-)$ are given by
\begin{align}
    \label{eq:partial_diff_of_mathfracL_1}
    &\frac{\partial\mathfrak{L}(\psi_+,\psi_-)}{\partial\psi^*_+} = \sum_k\gamma_k\mleft[\frac{L^{\ast}_{kW}(\psi_+)}{\partial\psi^*_+}\star^{e} L_{kW}(\psi_-) - \frac{1}{2}\frac{\partial}{\partial\psi^*_+}\Biggl\{L^{\ast}_{kW}(\psi_+)\star L_{kW}(\psi_+)\Biggr\}\mright], \\
    \label{eq:partial_diff_of_mathfracL_2}
    &\frac{\partial\mathfrak{L}(\psi_+,\psi_-)}{\partial\psi^*_-} = \sum_k\gamma_k\mleft[L^{\ast}_{kW}(\psi_+)\star^{e} \frac{L_{kW}(\psi_-)}{\partial\psi^*_-} - \frac{1}{2}\frac{\partial}{\partial\psi^*_-}\Biggl\{L^{\ast}_{kW}(\psi_-)\star L_{kW}(\psi_-)\Biggr\} \mright].
\end{align}

The classical equation of motion is obtained by setting $\eta = 0$, i.e., $\psi_+=\psi_- = \alpha$ in Eq.~\eqref{eq:saddle-point_equation_alpha}.
Here, $\eta = 0$ is the solution of Eq.~\eqref{eq:saddle-point_equation_eta}:
From Eqs.~\eqref{eq:partial_diff_of_mathfracL_1} and \eqref{eq:partial_diff_of_mathfracL_2}, we can show that at $\eta = 0$, the right-hand side of Eq.~\eqref{eq:saddle-point_equation_eta} equals to zero;
It follows that when the initial value of the quantum field is zero, the saddle-point solution of $\eta$ remains zero for the whole time interval from $t_0$ to $t$.
Thus, the classical path is characterized only by the classical field $\alpha$.

Substituting $\eta = 0$ in Eq.~\eqref{eq:saddle-point_equation_alpha}, we obtain the classical equation of motion;
\begin{gather}
    \label{eq:equation_of_motion_alpha_classical_open}
    i\hbar\frac{d\alpha}{dt} = \frac{\partial H_W(\alpha)}{\partial\alpha^*} + \frac{i\hbar}{2}\sum_k\gamma_k\mleft\{L^{\ast}_{kW}(\alpha)\star \frac{\partial L_{kW}(\alpha)}{\partial\alpha^*} - \frac{\partial L^{\ast}_{kW}(\alpha)}{\partial\alpha^*}\star L_{kW}(\alpha)\mright\}.
\end{gather}
In the absence of couplings with environments, i.e., $\gamma_k = 0$ for $\forall k$, this equation reduces to the Hamilton equation.
Here, we stress that the non-unitary term on the right-hand side of Eq.~\eqref{eq:equation_of_motion_alpha_classical_open} comes from the first terms on the right-hand sides of Eqs.~\eqref{eq:partial_diff_of_mathfracL_1} and \eqref{eq:partial_diff_of_mathfracL_2}, originally from the quantum-jump term of the GKSL equation.
This means that the quantum-diffusion terms do not affect the classical path of the GKSL equation.

We note that Eq.~\eqref{eq:saddle-point_equation_eta} has solutions $\eta \neq 0$ depending on initial conditions.
However, we leave out further analysis of the behavior of the saddle-point path with $\eta \neq 0$ as it beyond the scope of this paper.


\subsubsection{\label{subsec:first order of quantum fluctuations} First order of quantum fluctuations}
Although we have derived the classical path in the previous section by setting $\eta=0$ in the saddle-point equation, in this section, we shall obtain the same equation by taking the effect of the quantum fluctuations up to the first order in the path-integral representation of the GKSL equation. Moreover, by considering the time evolution of the Wigner function, we can include the effect of the quantum fluctuations as the initial distribution of the Wigner function.
The Wigner function evolves following the generalized Liouville equation, which we derive in this section.

Expanding $s_j$ in Eq.~\eqref{eq:Wigner_functional_open_single_discrete_action}
with respect to the quantum field $\eta_{j+1}$ up to the first order, we obtain
\begin{align}
    s_j &= s_j^{(1)} + o\mleft(\eta_{j+1}\mright), \\
    \label{eq:discrete_action_single_first_order}
    s_j^{(1)} &= - \eta^*_{j+1}\mleft\{i\hbar\mleft(\frac{\alpha_{j+1} - \alpha_j}{\Delta t}\mright) - \frac{\partial H_W(\alpha_j)}{\partial\alpha^*_j} + i\hbar\mleft.\frac{\partial}{\partial\eta^*_{j+1}}\mathfrak{L}\mleft(\alpha_j + \frac{\eta_{j+1}}{2},\alpha_j - \frac{\eta_{j+1}}{2}\mright)\mright|_{\eta_{j+1} = 0}\mright\} + {\rm c.c.},
\end{align}
where c.c. stands for the complex conjugate of the proceeding term, and $s_j^{(1)}$ denotes the first-order terms of the quantum filed in $s_j$.
The non-unitary term of Eq.~\eqref{eq:discrete_action_single_first_order} is calculated as
\begin{align}
    \label{eq:deviation_of_L_with_quantum_field_single}
    \mleft.\frac{\partial}{\partial\eta^*_{j+1}}\mathfrak{L}\mleft(\alpha_j + \frac{\eta_{j+1}}{2},\alpha_j - \frac{\eta_{j+1}}{2}\mright)\mright|_{\eta_{j+1} = 0} = -\frac{1}{2}\sum_k\gamma_k\mleft\{L^*_{kW}(\alpha_j)\star \frac{\partial L_{kW}(\alpha_j)}{\partial\alpha^*_j} - \frac{\partial L^*_{kW}(\alpha_j)}{\partial\alpha^*_j}\star L_{kW}(\alpha_j)\mright\}.
\end{align}
Approximating $s_j$ in Eq.~\eqref{eq:Wigner_functional_open_single_discrete_action} with $s_j^{(1)}$, we obtain
\begin{align}
    W(\alpha_{\rm f},t) &\approx \lim_{\Delta t \to 0}\prod_{j=0}^{N_t-1}\int\frac{d^2\alpha_j d^2\eta_{j+1}}{\pi^2}e^{i\Delta ts_j^{(1)}/\hbar}W(\alpha_0,t_0) \\
    \label{eq:Wigner_1st_order_open_single}
    &= \lim_{\Delta t \to 0}\prod_{j=0}^{N_t-1}\int\frac{d^2\alpha_j d^2\eta_{j+1}}{\pi^2}{\rm exp}\mleft\{\eta^*_{j+1}\mleft(\alpha_{j+1} - \alpha_j - \frac{\Delta t}{i\hbar}\frac{\partial H_W}{\partial\alpha^*_j} + \Delta t\mleft.\frac{\partial\mathfrak{L}}{\partial\eta^*_{j+1}}\mright|_{\eta_{j+1} = 0}\mright) - {\rm c.c.}\mright\} W(\alpha_0,t_0).
\end{align}
Using Eq.~\eqref{eq:Dirac_delta_function_phase_space}, we can perform the integration with respect to $\eta_{j+1}$ in Eq.~\eqref{eq:Wigner_1st_order_open_single} and then obtain
\begin{align}
    \label{eq:Wigner_functional_open_single_discrete_first_order}
    W(\alpha_{\rm f},t) = \lim_{\Delta t \to 0}\prod_{j=0}^{N_t-1}\int \frac{d^2\alpha_j}{\pi}\varUpsilon^{(1)}(\alpha_{j+1},t_j + \Delta t~;\alpha_j,t_j) W(\alpha_0,t_0),
\end{align}
where $\varUpsilon^{(1)}(\alpha_{j+1},t_j + \Delta t~;\alpha_j,t_j)$ is the first-order propagator given by
\begin{align}
    \label{eq:propagator_infinitesimal_time_interval_first_order}
    \varUpsilon^{(1)}(\alpha_{j+1},t_j + \Delta t~;\alpha_j,t_j) = \pi\delta^{(2)}\mleft(\alpha_{j+1} - \alpha_j  - \frac{\Delta t}{i\hbar}\frac{\partial H_W}{\partial\alpha^*_j} - \frac{\Delta t}{2}\sum_k\gamma_k\mleft(L^*_{kW}\star \frac{\partial L_{kW}}{\partial\alpha^*_j} - \frac{\partial L^*_{kW}}{\partial\alpha^*_j}\star L_{kW}\mright)\mright).
\end{align}
In Eq.~\eqref{eq:propagator_infinitesimal_time_interval_first_order}, we use the expression of Eq.~\eqref{eq:deviation_of_L_with_quantum_field_single}.

Equation~\eqref{eq:Wigner_functional_open_single_discrete_first_order} is the formal solution of the GKSL equation within the first-order approximation,
whose schematic image is illustrated in Fig.~\ref{fig:Path_integral_short_summary}(c).
An initial point of the phase space distributed by the initial Wigner function follows the path determined by the argument of the Dirac delta function in Eq.~\eqref{eq:Wigner_functional_open_single_discrete_first_order}:
\begin{align}
    \label{eq:equation_of_motion_alpha_classical_open_discrete}
    \alpha_{j+1} - \alpha_j = \frac{\Delta t}{i\hbar}\frac{\partial H_W(\alpha_j)}{\partial\alpha^*_j} + \frac{\Delta t}{2}\sum_k\gamma_k\mleft\{L^*_{kW}(\alpha_j)\star \frac{\partial L_{kW}(\alpha_j)}{\partial\alpha^*_j} - \frac{\partial L^*_{kW}(\alpha_j)}{\partial\alpha^*_j}\star L_{kW}(\alpha_j)\mright\}.
\end{align}
One can see that the continuous limit of Eq.~\eqref{eq:equation_of_motion_alpha_classical_open_discrete} is identical to Eq.~\eqref{eq:equation_of_motion_alpha_classical_open}, which means that each points distributed by the initial Wigner function follows the classical path.
Using Eqs.~\eqref{eq:expectation_value_phase_space} and \eqref{eq:Wigner_functional_open_single_discrete_first_order}, we obtain
\begin{align}
    \label{eq:physical_quantity_first_order_single}
    \Braket{\hat{A}(t)} = \int\frac{d^2\alpha_{\rm f}}{\pi}A_W(\alpha_{\rm f})\lim_{\Delta t \to 0}\prod_{j=0}^{N_t-1}\int \frac{d^2\alpha_j}{\pi}\varUpsilon^{(1)}(\alpha_{j+1},t_j + \Delta t~;\alpha_j,t_j) W(\alpha_0,t_0).
\end{align}
Thus, we can calculate the physical quantity $\braket{\hat{A}(t)}$ by iteratively solving the classical equations of motion starting from various points $\alpha_0$ distributed from the initial Wigner function, calculating the value $A_W(\alpha_{\rm f})$, and taking an ensemble average over the results.

We derive the equation of motion of the Wigner function within the first-order approximation.
Expanding $H_W$ and $\mathfrak{L}$ in Eq.~\eqref{eq:GKSL_Wigner_Weyl_representation_infinitesimal_time_interval_tochu} with respect to the quantum field $\eta_{j+1}$ up to first order, we obtain
\begin{align}
    &W(\alpha_{j+1},t_j + \Delta t) = W(\alpha_j,t_j) + \frac{i\Delta t}{\hbar}\mleft[\int\frac{d^2\alpha_jd^2\eta_{j+1}}{\pi^2}e^{\eta^*_{j+1}(\alpha_{j+1} - \alpha_j) - \eta_{j+1}(\alpha^*_{j+1} - \alpha^*_j)}\mright. \nonumber \\
    &\hphantom{W(\alpha_{j+1},t_j + \Delta t) = W(\alpha_j,t_j) + \frac{i\Delta t}{\hbar}\int\frac{d^2\alpha_jd^2\eta_{j+1}}{\pi^2}}\mleft.~\times\mleft\{\eta^*_{j+1}\mleft(\frac{\partial H_W}{\partial\alpha^*_j} - i\hbar\mleft.\frac{\partial\mathfrak{L}}{\partial\eta^*_{j+1}}\mright|_{\eta_{j+1} = 0}\mright) + {\rm c.c.}\mright\}W(\alpha_j,t_j)\mright] \\
    \label{eq:derivation_of_generalized Liouville_eq_tochu_1}
    &\hphantom{W(\alpha_{j+1},t_j + \Delta t)} = W(\alpha_j,t_j) - \frac{i\Delta t}{\hbar}\mleft[\int d^2\alpha_j \frac{\partial}{\partial\alpha_j}\mleft\{\int\frac{d^2\eta_{j+1}}{\pi^2}e^{\eta^*_{j+1}(\alpha_{j+1} - \alpha_j) - \eta_{j+1}(\alpha^*_{j+1} - \alpha^*_j)}\mright\}\mright. \nonumber \\
    &\hphantom{W(\alpha_{j+1},t_j + \Delta t) = W(\alpha_j,t_j) + \frac{i\Delta t}{\hbar}\frac{\partial}{\partial\alpha_{j+1}}\int\frac{d^2\alpha_jd^2\eta_{j+1}}{\pi^2}}\mleft.~\times\mleft(\frac{\partial H_W}{\partial\alpha^*_j} - i\hbar\mleft.\frac{\partial\mathfrak{L}}{\partial\eta^*_{j+1}}\mright|_{\eta_{j+1} = 0}\mright)W(\alpha_j,t_j) - {\rm c.c.} \mright].
\end{align}
Using Eq.~\eqref{eq:Dirac_delta_function_phase_space}, we can rewrite Eq.~\eqref{eq:derivation_of_generalized Liouville_eq_tochu_1} as
\begin{align}
    W(\alpha_{j+1},t_j + \Delta t) - W(\alpha_j,t_j) = -\frac{i\Delta t}{\hbar}\int d^2\alpha_j\mleft\{\frac{\partial}{\partial\alpha_j}\delta^{(2)}(\alpha_{j+1} - \alpha_j)\mright\}\mleft(\frac{\partial H_W}{\partial\alpha^*_j} - i\hbar\mleft.\frac{\partial\mathfrak{L}}{\partial\eta^*_{j+1}}\mright|_{\eta_{j+1} = 0}\mright)W(\alpha_j,t_j) + {\rm c.c.} 
\end{align}
Performing the integration by part, we obtain
\begin{align}
    \label{eq:derivation_of_generalized Liouville_eq_tochu_2}
    W(\alpha_{j+1},t_j + \Delta t) - W(\alpha_j,t_j) = \frac{i\Delta t}{\hbar}\frac{\partial}{\partial\alpha_{j+1}}\mleft\{\mleft(\frac{\partial H_W}{\partial\alpha^*_{j+1}} - i\hbar\mleft.\frac{\partial\mathfrak{L}}{\partial\eta^*_{j+1}}\mright|_{\eta_{j+1} = 0}\mright)W(\alpha_{j+1},t_j)\mright\} + {\rm c.c.}
\end{align}
Taking the continuous limit of Eq.~\eqref{eq:derivation_of_generalized Liouville_eq_tochu_2}, we obtain the following generalized Liouville equation:
\begin{align}
    \label{generalized Liouville_eq_single_open}
    i\hbar\frac{dW(\alpha,t)}{dt} =-\frac{\partial}{\partial\alpha}\mleft[\mleft\{\frac{\partial H_W}{\partial\alpha^*} + \frac{i\hbar}{2}\sum_k\gamma_k\mleft(L^{\ast}_{kW}\star \frac{\partial L_{kW}}{\partial\alpha^*} - \frac{\partial L^{\ast}_{kW}}{\partial\alpha^*}\star L_{kW}\mright)\mright\}W(\alpha,t)\mright] - {\rm c.c.},
\end{align}
where we use Eq.~\eqref{eq:deviation_of_L_with_quantum_field_single}.


\subsubsection{\label{subsec:second order of quantum fluctuations and truncated Wigner approximation} Second order of quantum fluctuations and truncated Wigner approximation}
In the second-order calculation, we perform the integration with respect to $\eta$ in the propagator by introducing a complex auxiliary field and derive the Fokker-Planck equation for the Wigner function and the corresponding stochastic differential equation.
In particular, we analytically obtain the general condition to ensure the positive-semidefiniteness of the diffusion matrix, which equals to the condition for obtaining the stochastic differential equation from the Fokker-Planck equation. 
This implies that the TWA does not necessarily lead to the usual Fokker-Planck equation,
which was pointed out in previous works \cite{Plimak,Huber} using specific models.

Expanding $s_j$ in Eq.~\eqref{eq:Wigner_functional_open_single_discrete_action} with respect to the quantum field up to second order, we obtain
\begin{align}
    s_j &= s_j^{(1)} + s_j^{(2)} + o\mleft(\eta^2_{j+1}\mright), \\ 
    \label{eq:discrete_action_single_second_order}
    s_j^{(2)} &= i\hbar\mleft\{\lambda(\alpha_j)\eta_{j+1}^{*2} + 2\Lambda(\alpha_j)\bigl|\eta_{j+1}\bigr|^2 + \lambda^*(\alpha_j)\eta_{j+1}^{2}\mright\},
\end{align}
where $s_j^{(2)}$ denotes the second-order term of $s_j$ with respect to the quantum field, the first order term $s_j^{(1)}$ is given by Eq.~\eqref{eq:discrete_action_single_first_order}, and $\lambda\in \mathbb{C}$ and $\Lambda\in \mathbb{R}$ are defined by
\begin{align}
    \label{eq:lambda_single_degree}
    \lambda(\alpha) &= \sum_k\frac{\gamma_k}{2}\frac{\partial L^*_{kW}(\alpha)}{\partial\alpha^*}\star \frac{\partial L_{kW}(\alpha)}{\partial\alpha^*}, \\
    \label{eq:Lambda_single_degree}
    \Lambda(\alpha) &= \sum_k\frac{\gamma_k}{4}\mleft\{\frac{\partial L^*_{kW}(\alpha)}{\partial\alpha^*}\star \frac{\partial L_{kW}(\alpha)}{\partial\alpha} + \frac{\partial L^*_{kW}(\alpha)}{\partial\alpha}\star \frac{\partial L_{kW}(\alpha)}{\partial\alpha^*}\mright\}.
\end{align}
Here, $\Lambda$ takes a non-negative real value.
In the derivation of Eqs.~\eqref{eq:lambda_single_degree} and \eqref{eq:Lambda_single_degree}, the quantum-jump and quantum-diffusion terms equally contribute to the second-order approximation, in contrast to the fact that only the quantum-jump term affects the classical path.
According to Eq.~\eqref{eq:discrete_action_single_second_order}, the path-integral representation of the GKSL equation \eqref{eq:Wigner_functional_open_single_discrete} within the second order of the quantum fluctuations becomes
\begin{align}
    W(\alpha_{\rm f},t) &\approx \lim_{\Delta t \to 0}\prod_{j=0}^{N_t-1}\int\frac{d^2\alpha_j d^2\eta_{j+1}}{\pi^2}e^{i\Delta t\mleft(s_j^{(1)} + s_j^{(2)}\mright)/\hbar}W(\alpha_0,t_0) \\
    \label{eq:Wigner_2nd_order_open_single}
    &= \lim_{\Delta t \to 0}\prod_{j=0}^{N_t-1}\int\frac{d^2\alpha_j d^2\eta_{j+1}}{\pi^2}e^{i\Delta ts_j^{(1)}/\hbar}{\rm exp}\Biggl[-\Delta t\biggl\{\lambda(\alpha_j)\eta_{j+1}^{*2} + 2\Lambda(\alpha_j)\bigl|\eta_{j+1}\bigr|^2 + \lambda^*(\alpha_j)\eta_{j+1}^{2}\biggr\}\Biggr] W(\alpha_0,t_0).
\end{align}

In order to perform the integration with respect to the quantum field $\eta_{j+1}$, we introduce the auxiliary field $\Delta \Xi_j$ as an integration variable of the Hubbard-Stratonovich transformation:
\begin{align}
    \label{eq:Hubbard-Stratonovich_transformation}
    e^{-2\Lambda\Delta t \mleft|\eta_{j+1}\mright|^2 - \lambda^*\Delta t\eta_{j+1}^2 - \lambda\Delta t\eta_{j+1}^{*2}} = \int d^2\Delta\Xi_j\frac{e^{-(\Delta\Xi^{\rm re}_j)^2/2\Delta t\mleft(\Lambda - |\lambda|\mright)}}{\sqrt{2\pi\Delta t\mleft(\Lambda - |\lambda|\mright)}}\frac{e^{-(\Delta\Xi^{\rm im}_j)^2/2\Delta t\mleft(\Lambda + |\lambda|\mright)}}{\sqrt{2\pi\Delta t\mleft(\Lambda + |\lambda|\mright)}}{\rm exp}\mleft(\eta^*_{j+1}e^{i\theta/2}\Delta\Xi_j - \eta_{j+1}e^{-i\theta/2}\Delta\Xi^*_j\mright),
\end{align}
where $\Delta\Xi_j = \Delta\Xi_j^{\rm re} + i\Delta\Xi_j^{\rm im}\in\mathbb{C}$ ($\Delta\Xi_j^{\rm re},\Delta\Xi_j^{\rm im}\in\mathbb{R}$), and $\theta = \theta(\alpha)$ is defined by
\begin{align}
    \theta(\alpha) = {\rm arg}\mleft(\lambda(\alpha)\mright).
\end{align}
This Hubbard-Stratonovich transformation is feasible only when
\begin{align}
    \label{eq:Hubbard_Storatonovich_transformation_condition}
    \Lambda(\alpha) \geq |\lambda(\alpha)|.
\end{align}
Especially when $\Lambda = |\lambda|$, the Hubbard-Stratonovich transformation becomes
\begin{align}
    \label{eq:Hubbard-Stratonovich_transformation_Lambda=lambda}
    e^{-2\Lambda\Delta t \mleft|\eta_{j+1}\mright|^2 - \lambda^*\Delta t\eta_{j+1}^2 - \lambda\Delta t\eta_{j+1}^{*2}} = \int d\Delta\Xi^{\rm im}_j\frac{e^{-(\Delta\Xi^{\rm im}_j)^2/2\Delta t\mleft(\Lambda + |\lambda|\mright)}}{\sqrt{2\pi\Delta t\mleft(\Lambda + |\lambda|\mright)}}{\rm exp}\mleft(i\eta^*_{j+1}e^{i\theta/2}\Delta\Xi^{\rm im}_j + i\eta_{j+1}e^{-i\theta/2}\Delta\Xi^{\rm im}_j\mright),
\end{align}
which corresponds to fixing the real part of the auxiliary field to be zero, $\Delta \Xi^{\rm re}_j = 0$, in Eq.~\eqref{eq:Hubbard-Stratonovich_transformation}.
We provide the derivations of these Hubbard-Stratonovich transformations in \ref{Hubbard-Stratonovich transformation}.

According to the Hubbard-Stratonovich transformation, Eq.~\eqref{eq:Wigner_2nd_order_open_single} is transformed into
\begin{align}
    W(\alpha_{\rm f},t) = \lim_{\Delta t \to 0}\prod_{j=0}^{N_t-1}\int&\frac{d^2\alpha_j d^2\eta_{j+1}d^2\Delta\Xi_j}{\pi^2}\frac{e^{-(\Delta\Xi^{\rm re}_j)^2/2\Delta t\mleft(\Lambda - |\lambda|\mright)}}{\sqrt{2\pi\Delta t\mleft(\Lambda - |\lambda|\mright)}}\frac{e^{-(\Delta\Xi^{\rm im}_j)^2/2\Delta t\mleft(\Lambda + |\lambda|\mright)}}{\sqrt{2\pi\Delta t\mleft(\Lambda + |\lambda|\mright)}} \nonumber \\
    &\times{\rm exp}\mleft[\eta^*_{j+1}\mleft\{\alpha_{j+1} - \alpha_j - \frac{\Delta t}{i\hbar}\frac{\partial H_W}{\partial\alpha^*_j} - \Delta t\mleft.\frac{\partial\mathfrak{L}}{\partial\eta^*_{j+1}}\mright|_{\eta_{j+1} = 0} + e^{i\theta/2}\Delta\Xi_j\mright\} - {\rm c.c.}\mright] W(\alpha_0,t_0).
\end{align}
Then, we can integrate out the quantum field $\eta_{j+1}$ using Eq.~\eqref{eq:Dirac_delta_function_phase_space}, obtaining
\begin{align}
    \label{eq:Wigner_functional_open_single_discrete_second_order}
    W(\alpha_{\rm f},t) = \lim_{\Delta t \to 0}\prod_{j=0}^{N_t-1}\int\frac{d^2\alpha_j}{\pi} \varUpsilon^{(2)}(\alpha_{j+1},t_j + \Delta t~;\alpha_j,t_j)W(\alpha_0,t_0),
\end{align}
where $\varUpsilon^{(2)}(\alpha_{j+1},t_j + \Delta t~;\alpha_j,t_j)$ is a second-order propagator given by
\begin{align}
    \label{eq:propagator_infinitesimal_time_interval_second_order}
    \varUpsilon^{(2)}(\alpha_{j+1},t_j + \Delta t~;\alpha_j,t_j) =&~ \pi\int d^2\Delta\Xi_j\frac{e^{-(\Delta\Xi^{\rm re}_j)^2/2\Delta t\mleft(\Lambda - |\lambda|\mright)}}{\sqrt{2\pi\Delta t\mleft(\Lambda - |\lambda|\mright)}}\frac{e^{-(\Delta\Xi^{\rm im}_j)^2/2\Delta t\mleft(\Lambda + |\lambda|\mright)}}{\sqrt{2\pi\Delta t\mleft(\Lambda + |\lambda|\mright)}} \nonumber \\
    &\times\delta^{(2)}\mleft(\alpha_{j+1} - \alpha_j - \frac{\Delta t}{i\hbar}\frac{\partial H_W}{\partial\alpha^*_j} - \frac{\Delta t}{2}\sum_k\gamma_k\mleft(L^*_{kW}\star \frac{\partial L_{kW}}{\partial\alpha^*_j} - \frac{\partial L^*_{kW}}{\partial\alpha^*_j}\star L_{kW}\mright) + e^{i\theta/2}\Delta\Xi_j\mright).
\end{align}
To derive Eq.~\eqref{eq:propagator_infinitesimal_time_interval_second_order}, we use the expression of Eq.~\eqref{eq:deviation_of_L_with_quantum_field_single}.

Equation~\eqref{eq:Wigner_functional_open_single_discrete_second_order} is the formal solution of the GKSL equation within the second order of the quantum fluctuations,
whose schematic image is illustrated in Fig.~\ref{fig:Path_integral_short_summary}(d).
Each points sampled from the initial Wigner function follows the path given by the argument of the Dirac delta function in Eq.~\eqref{eq:propagator_infinitesimal_time_interval_second_order}:
\begin{align}
    \label{eq:stochastic_differential_equation_discrete}
    \alpha_{j+1} - \alpha_j = \frac{\Delta t}{i\hbar}\frac{\partial H_W(\alpha_j)}{\partial\alpha^*_j} + \frac{\Delta t}{2}\sum_k\gamma_k\mleft\{L^*_{kW}(\alpha_j)\star \frac{\partial L_{kW}(\alpha_j)}{\partial\alpha^*_j} - \frac{\partial L^*_{kW}(\alpha_j)}{\partial\alpha^*_j}\star L_{kW}(\alpha_j)\mright\} + e^{i\theta(\alpha_j)/2}\Delta\Xi_j.
\end{align}
In the continuous limit, Eq.~\eqref{eq:stochastic_differential_equation_discrete} becomes
\begin{align}
    \label{eq:stochastic_differential_equation_continuous}
    i\hbar d\alpha(t) = \mleft[\frac{\partial H_W(\alpha)}{\partial\alpha^*} + \frac{i\hbar}{2}\sum_k\gamma_k\mleft\{L^*_{kW}(\alpha)\star \frac{\partial L_{kW}(\alpha)}{\partial\alpha^*} - \frac{\partial L^*_{kW}(\alpha)}{\partial\alpha^*}\star L_{kW}(\alpha)\mright\}\mright]dt + i\hbar e^{i\theta(\alpha)/2}\cdot d\Xi(\alpha),
\end{align}
where $\cdot$ is an Ito product \cite{Risken} and $\Xi(\alpha)$ is a complex stochastic process, whose real and imaginary parts are independent of each other and have Gaussian increments with a mean of $0$, i.e., the changes of $\Delta\Xi = \Delta\Xi^{\rm re} + i\Delta\Xi^{\rm im}\in\mathbb{C}$ ($\Delta\Xi^{\rm re},\Delta\Xi^{\rm im}\in\mathbb{R}$) in the time interval $\Delta t$
\begin{align}
    \Delta \Xi^{\rm re}(\alpha) &= \Xi^{\rm re}(\alpha(t + \Delta t)) - \Xi^{\rm re}(\alpha(t)), \\
    \Delta \Xi^{\rm im}(\alpha) &= \Xi^{\rm im}(\alpha(t + \Delta t)) - \Xi^{\rm im}(\alpha(t))
\end{align}
obey the following Gaussian distribution functions:
\begin{align}
    \label{eq:complex_stochastic_process_real}
    P\mleft[\Delta \Xi^{\rm re}(\alpha)\mright] &= \frac{1}{\sqrt{2\pi\Delta t\mleft(\Lambda(\alpha) - |\lambda(\alpha)|\mright)}}{\rm exp}\mleft[-\frac{\Bigl(\Delta \Xi^{\rm re}(\alpha)\Bigr)^2}{2\Delta t\mleft(\Lambda(\alpha) - |\lambda(\alpha)|\mright)}\mright], \\
    \label{eq:complex_stochastic_process_imag}
    P\mleft[\Delta \Xi^{\rm im}(\alpha)\mright] &= \frac{1}{\sqrt{2\pi\Delta t\mleft(\Lambda(\alpha) + |\lambda(\alpha)|\mright)}}{\rm exp}\mleft[-\frac{\Bigl(\Delta \Xi^{\rm im}(\alpha)\Bigr)^2}{2\Delta t\mleft(\Lambda(\alpha) + |\lambda(\alpha)|\mright)}\mright].
\end{align}
We note that the complex stochastic process $\Xi(\alpha)$ has the different variances in its real and imaginary parts, which are given by $(\Lambda(\alpha) - |\lambda(\alpha)|)\Delta t$ and $(\Lambda(\alpha) + |\lambda(\alpha)|)\Delta t$, respectively.
The difference between the classical equation of motion [Eq.~\eqref{eq:equation_of_motion_alpha_classical_open}] and the stochastic differential equation [Eq.~\eqref{eq:stochastic_differential_equation_continuous}] is only the existence of the complex stochastic process $\Xi(\alpha)$, which means that the effect of the second order of the quantum fluctuations is introduced as the complex Gaussian noises for the classical path.
Using Eqs.~\eqref{eq:expectation_value_phase_space} and \eqref{eq:Wigner_functional_open_single_discrete_second_order}, we obtain
\begin{align}
\label{eq:physical_quantity_second_order_single}
\Braket{\hat{A}(t)} = \int\frac{d^2\alpha_{\rm f}}{\pi}A_W(\alpha_{\rm f})\lim_{\Delta t \to 0}\prod_{j=0}^{N_t-1}\int \frac{d^2\alpha_j}{\pi}\varUpsilon^{(2)}(\alpha_{j+1},t_j + \Delta t~;\alpha_j,t_j) W(\alpha_0,t_0).
\end{align}
Thus, we can calculate the physical quantity $\braket{\hat{A}(t)}$ by iteratively solving the stochastic differential equation \eqref{eq:stochastic_differential_equation_continuous} with different initial conditions sampled from the initial Wigner function $W(\alpha_0,t_0)$, calculating the value $A_W(\alpha_{\rm f})$, and taking an ensemble average with respect to the results.

We derive the equation of motion of the Wigner function within the second-order approximation.
Expanding $H_W$ and $\mathfrak{L}$ in Eq.~\eqref{eq:GKSL_Wigner_Weyl_representation_infinitesimal_time_interval_tochu} with respect to the quantum field $\eta_{j+1}$ up to second order, we obtain
\begin{align}
    W(\alpha_{j+1},t_j + \Delta t) = ~& W(\alpha_j,t_j) + \frac{i\Delta t}{\hbar}\mleft[\frac{\partial}{\partial\alpha_{j+1}}\mleft\{\mleft(\frac{\partial H_W}{\partial\alpha^*_{j+1}} - i\hbar\mleft.\frac{\partial\mathfrak{L}}{\partial\eta^*_{j+1}}\mright|_{\eta_{j+1} = 0}\mright)W(\alpha_{j+1},t_j)\mright\} - {\rm c.c.}\mright] \nonumber \\
    &-\Delta t\int\frac{d^2\alpha_j d^2\eta_{j+1}}{\pi^2}e^{\eta^*_{j+1}(\alpha_{j+1} - \alpha_j) - \eta_{j+1}(\alpha^*_{j+1} - \alpha^*_j)}\Biggl\{\lambda(\alpha_j)\eta_{j+1}^{*2} + 2\Lambda(\alpha_j)\bigl|\eta_{j+1}\bigr|^2 + \lambda^*(\alpha_j)\eta_{j+1}^{2}\Biggr\}W(\alpha_j,t_j) \\
    \label{eq:derivation_of_FP_eq_tochu_1}
    = ~& W(\alpha_j,t_j) + \frac{i\Delta t}{\hbar}\mleft[\frac{\partial}{\partial\alpha_{j+1}}\mleft\{\mleft(\frac{\partial H_W}{\partial\alpha^*_{j+1}} - i\hbar\mleft.\frac{\partial\mathfrak{L}}{\partial\eta^*_{j+1}}\mright|_{\eta_{j+1} = 0}\mright)W(\alpha_{j+1},t_j)\mright\} - {\rm c.c.}\mright] \nonumber \\
    &-\Delta t\int d^2\alpha_j\Biggl\{\lambda(\alpha_j)\frac{\partial^2}{\partial\alpha^2_j} + 2\Lambda(\alpha_j)\frac{\partial^2}{\partial\alpha_j\alpha^*_j} + \lambda^*(\alpha_j)\frac{\partial^2}{\partial\alpha^{*2}_j}\Biggr\}\mleft\{\int\frac{ d^2\eta_{j+1}}{\pi^2}e^{\eta^*_{j+1}(\alpha_{j+1} - \alpha_j) - \eta_{j+1}(\alpha^*_{j+1} - \alpha^*_j)}\mright\}W(\alpha_j,t_j),
\end{align}
where we use Eq.~\eqref{eq:derivation_of_generalized Liouville_eq_tochu_2} to obtain the first line.
Using Eq.~\eqref{eq:Dirac_delta_function_phase_space}, we can rewrite Eq.~\eqref{eq:derivation_of_FP_eq_tochu_1} as
\begin{align}
    W(\alpha_{j+1},t_j + \Delta t) = ~& W(\alpha_j,t_j) + \frac{i\Delta t}{\hbar}\mleft[\frac{\partial}{\partial\alpha_{j+1}}\mleft\{\mleft(\frac{\partial H_W}{\partial\alpha^*_{j+1}} - i\hbar\mleft.\frac{\partial\mathfrak{L}}{\partial\eta^*_{j+1}}\mright|_{\eta_{j+1} = 0}\mright)W(\alpha_{j+1},t_j)\mright\} - {\rm c.c.}\mright] \nonumber \\
    &-\Delta t\int d^2\alpha_jW(\alpha_j,t_j)\Biggl\{\lambda(\alpha_j)\frac{\partial^2}{\partial\alpha^2_j} + 2\Lambda(\alpha_j)\frac{\partial^2}{\partial\alpha_j\alpha^*_j} + \lambda^*(\alpha_j)\frac{\partial^2}{\partial\alpha^{*2}_j}\Biggr\}\delta^{(2)}(\alpha_{j+1} - \alpha_j).
\end{align}
Performing the integration by part, we obtain
\begin{align}
    \label{eq:derivation_of_FP_eq_tochu_2}
    W(\alpha_{j+1},t_j + \Delta t) - W(\alpha_j,t_j) =~& \frac{i\Delta t}{\hbar}\frac{\partial}{\partial\alpha_{j+1}}\mleft[\mleft(\frac{\partial H_W}{\partial\alpha^*_{j+1}} - i\hbar\mleft.\frac{\partial\mathfrak{L}}{\partial\eta^*_{j+1}}\mright|_{\eta_{j+1} = 0}\mright)W(\alpha_{j+1},t_j)\mright]\nonumber \\
    & -\Delta t\frac{\partial^2}{\partial\alpha_{j+1}^2}\biggl[\lambda W(\alpha_{j+1},t_j)\biggr] + \Delta t\frac{\partial^2}{\partial\alpha_{j+1}\partial\alpha_{j+1}^*}\biggl[\Lambda W(\alpha_{j+1},t_j)\biggr] + {\rm c.c.}
\end{align}
Taking the continuous limit of Eq.~\eqref{eq:derivation_of_FP_eq_tochu_2}, we obtain the following Fokker-Planck equation:
\begin{align}
    \label{Fokker_Planck_eq_single_open}
    i\hbar\frac{dW(\alpha,t)}{dt} =& -\frac{\partial}{\partial\alpha}\mleft[\mleft\{\frac{\partial H_W}{\partial\alpha^*} + \frac{i\hbar}{2}\sum_k\gamma_k\mleft(L^{\ast}_{kW}\star \frac{\partial L_{kW}}{\partial\alpha^*} - \frac{\partial L^{\ast}_{kW}}{\partial\alpha^*}\star L_{kW}\mright)\mright\}W(\alpha,t)\mright] \nonumber \\
    &-i\hbar\frac{\partial^2}{\partial\alpha^2}\biggl[\lambda W(\alpha,t)\biggr] + i\hbar\frac{\partial^2}{\partial\alpha\partial\alpha^*}\biggl[\Lambda W(\alpha,t)\biggr]- {\rm c.c.},
\end{align}
where we have used Eq.~\eqref{eq:deviation_of_L_with_quantum_field_single}.

Before closing this section, we remark on the diffusion term of the Fokker-Planck equation \eqref{Fokker_Planck_eq_single_open}.
The feasible condition of the Hubbard-Stratonovich transformation, $\Lambda \geq |\lambda|$, corresponds to the positive-semidefiniteness condition on the diffusion matrix of the Fokker-Planck equation \eqref{Fokker_Planck_eq_single_open}: Only when $\Lambda \geq |\lambda|$, we can reduce Eq.~\eqref{Fokker_Planck_eq_single_open} into the stochastic differential equation \eqref{eq:stochastic_differential_equation_continuous}.


\subsection{\label{subsec:Extension to a multi-mode system} Extension to a multi-mode system}
In the previous sections, we formulate the path-integral representation of the GKSL equation and derive its analytical expressions within the first- and second-order quantum fluctuations for a system with a single degree of freedom.
In this section, we extend the results into a system with multiple degrees of freedom.
In the following, we consider a system with total $M$ degrees of freedom and denote each mode by subscript $m~(=1,2,\cdots, M)$.
Here, we remark that the second-order results hold only for local jump operators that do not couple different degrees of freedom. See setup $2$ in Sec.~\ref{sec:setup}.
We also note that the path-integral representation and the first-order results hold for arbitrary jump operators.

We first introduce the multi-mode displacement operator $\hat{D}(\vec{\alpha})$ as
\begin{align}
    \hat{D}(\vec{\alpha}) &= \bigotimes_m\hat{D}(\alpha_m), \\
    \hat{D}(\alpha_m) &= e^{\alpha_m\hat{a}^{\dagger}_m - \alpha_m^*\hat{a}_m},
\end{align}
where $\hat{a}^{\dagger}_m$ and $\hat{a}_m$ are the creation and annihilation operator of bosons in mode $m$ satisfying the commutation relation $[\hat{a}_m,\hat{a}^{\dagger}_{m'}]_- = \delta_{mm'}$.
The Weyl-Wigner transformation maps an arbitrary operator $\hat{A}$ consisting $\hat{a}^{\dagger}_m$ and $\hat{a}_m$ for $m=1,2,\cdots, M$ into a $c$-number function $A_W(\vec{\alpha})$ in the $2M$-dimensional phase space:
\begin{align}
    A_W(\vec{\alpha}) &= \int\frac{d^2\vec{\eta}}{\pi^M}\chi_A(\vec{\eta})e^{\vec{\alpha}^*\cdot\vec{\eta} - \vec{\alpha}\cdot\vec{\eta}^*}, \\
    \chi_A(\vec{\eta}) &= {\rm Tr}\mleft[\hat{A}\hat{D}^{\dagger}(\vec{\eta})\mright],
\end{align}
where $\vec{\alpha}=(\alpha_1, \alpha_2, \cdots, \alpha_M)$ and $\int d^2\vec{\eta} = \prod_{m}\int d^2\eta_m$.
The trace and the Weyl-Wigner representation of a product of two operators are given by
\begin{align}
    \label{eq:Tr_AB_Weyl_Wigner_multi}
    {\rm Tr}\mleft[\hat{A}\hat{B}\mright] &= \int\frac{d^2\vec{\alpha}}{\pi^M}A_W(\vec{\alpha})B_W(\vec{\alpha}), \\
    \label{eq:AB_Weyl_Wigner_representation_1_multi}
    \mleft[\hat{A}\hat{B}\mright]_W(\vec{\alpha}) &= \int\frac{d^2\vec{\alpha}_0d^2\vec{\eta}}{\pi^{2M}}A_W(\vec{\alpha}_0)B_W\mleft(\vec{\alpha}_0 + \frac{\vec{\eta}}{2}\mright)e^{(\vec{\alpha} - \vec{\alpha}_0)\cdot \vec{\eta}^* - (\vec{\alpha}^* - \vec{\alpha}^*_0)\cdot \vec{\eta}} \\
    \label{eq:AB_Weyl_Wigner_representation_2_multi}
    &= \int\frac{d^2\vec{\alpha}_0d^2\vec{\eta}}{\pi^{2M}}A_W\mleft(\vec{\alpha}_0 - \frac{\vec{\eta}}{2}\mright)B_W\mleft(\vec{\alpha}_0\mright)e^{(\vec{\alpha} - \vec{\alpha}_0)\cdot \vec{\eta}^* - (\vec{\alpha}^* - \vec{\alpha}^*_0)\cdot \vec{\eta}}.
\end{align}
Following the same procedure in Secs.~\ref{subsec:Markov condition in the phase space} and \ref{subsec:Path-integral representation} with Eqs.~\eqref{eq:Tr_AB_Weyl_Wigner_multi}, \eqref{eq:AB_Weyl_Wigner_representation_1_multi}, and \eqref{eq:AB_Weyl_Wigner_representation_2_multi}, we can straightforwardly formulate the path-integral representation of GKSL equation for a multi-mode system as
\begin{gather}
    \label{eq:Wigner_functional_open_multi_discrete}
    W(\vec{\alpha}_{\rm f},t) = \lim_{\Delta t \to 0}\prod_{j=0}^{N_t-1}\int\frac{d^2\vec{\alpha}_j d^2\vec{\eta}_{j+1}}{\pi^{2M}}e^{i\Delta t s_j/\hbar} W(\vec{\alpha}_0,t_0), \\
    \label{eq:Wigner_functional_open_multi_discrete_action}
    s_j = i\hbar\mleft\{\vec{\eta}_{j+1}\cdot\Biggl(\frac{\vec{\alpha}_{j+1}^* - \vec{\alpha}_j^*}{\Delta t}\Biggr) - \vec{\eta}_{j+1}^*\cdot\mleft(\frac{\vec{\alpha}_{j+1} - \vec{\alpha}_j}{\Delta t}\mright)\mright\} + \sum_{n=0,1}(-1)^nH_W\mleft(\vec{\alpha}_j + (-1)^n\frac{\vec{\eta}_{j+1}}{2}\mright) - i\hbar\mathfrak{L}\mleft(\vec{\alpha}_j+\frac{\vec{\eta}_{j+1}}{2},\vec{\alpha}_j-\frac{\vec{\eta}_{j+1}}{2}\mright),
\end{gather}
where $\cdot$ indicates the inner product, $\vec{\alpha}_j$ is the classical field vector, $\vec{\eta}_{j+1}$ is the quantum field vector, and the non-unitary term $\mathfrak{L}$ is given by
\begin{align}
    \mathfrak{L}(\vec{\alpha},\vec{\beta}) = \sum_k\gamma_k\mleft\{L^{\ast}_{kW}\mleft(\vec{\alpha}\mright)\star^{e} L_{kW}(\vec{\beta}) - \frac{1}{2}L^{\ast}_{kW}\mleft(\vec{\alpha}\mright)\star L_{kW}\mleft(\vec{\alpha}\mright) - \frac{1}{2}L^{\ast}_{kW}(\vec{\beta})\star L_{kW}(\vec{\beta})\mright\}.
\end{align}
Here, the Moyal product $\star$ and the extended Moyal product $\star^{e}$ for a multi-mode system are defined by
\begin{align}
    \label{eq:Moyal_product_multi_system}
    A_W(\vec{\alpha})\star B_W(\vec{\alpha}) &= A_W(\vec{\alpha}){\rm exp}\mleft\{\frac{1}{2}\sum_m\mleft(\frac{\overleftarrow{\partial}}{\partial \alpha_m}\frac{\overrightarrow{\partial}}{\partial \alpha_m^*} - \frac{\overleftarrow{\partial}}{\partial \alpha_m^*}\frac{\overrightarrow{\partial}}{\partial \alpha_m}\mright)\mright\}B_W(\vec{\alpha}), \\
    \label{eq:extended_Moyal_product_multi_system}
    A_W(\vec{\alpha})\star^{e}B_W(\vec{\beta}) &= A_W(\vec{\alpha}){\rm exp}\mleft\{\frac{1}{2}\sum_m\mleft(\frac{\overleftarrow{\partial}}{\partial \alpha_m}\frac{\overrightarrow{\partial}}{\partial \beta_m^*} - \frac{\overleftarrow{\partial}}{\partial \alpha_m^*}\frac{\overrightarrow{\partial}}{\partial \beta_m}\mright)\mright\}B_W(\vec{\beta}).
\end{align}
In the continuous limit, Eq.~\eqref{eq:Wigner_functional_open_multi_discrete} can be formally expressed  as
\begin{gather}
    \label{eq:Wigner_functional_open_multi_continuous}
    W(\vec{\alpha},t) = \int\mathscr{D}^2\vec{\alpha}\mathscr{D}^2\vec{\eta} e^{iS[\vec{\alpha},\vec{\eta}]/\hbar}W(\vec{\alpha}_0,t_0), \\
    \label{eq:Wigner_functional_open_multi_action}
    S\mleft[\vec{\alpha},\vec{\eta}\mright] = \int^t_{t_0}d\tau \mleft\{i\hbar\mleft(\vec{\eta}\cdot \frac{\partial\vec{\alpha}^*}{\partial\tau} - \vec{\eta}^*\cdot \frac{\partial\vec{\alpha}}{\partial\tau}\mright) + \sum_{n=0,1}(-1)^nH_W\mleft(\vec{\alpha} + (-1)^n\frac{\vec{\eta}}{2}\mright) - i\hbar\mathfrak{L}\mleft(\vec{\alpha}+\frac{\vec{\eta}}{2},\vec{\alpha}-\frac{\vec{\eta}}{2}\mright)\mright\},
\end{gather}
where $S[\vec{\alpha},\vec{\eta}]$ is the action.

Within the first order of the quantum fluctuations, the GKSL equation is approximated into the following generalized Liouville equation for the Wigner function:
\begin{align}
    \label{generalized Liouville_eq_multi_open}
    i\hbar\frac{dW(\vec{\alpha},t)}{dt} =-\sum_m\mleft(\frac{\partial}{\partial\alpha_m}\mleft[\mleft\{\frac{\partial H_W}{\partial\alpha_m^*} + \frac{i\hbar}{2}\sum_k\gamma_k\mleft(L^{\ast}_{kW}\star \frac{\partial L_{kW}}{\partial\alpha_m^*} - \frac{\partial L^{\ast}_{kW}}{\partial\alpha_m^*}\star L_{kW}\mright)\mright\}W(\vec{\alpha},t)\mright]\mright) + {\rm c.c.},
\end{align}
which can be derived by the same procedures in Sec.~\ref{subsec:first order of quantum fluctuations}.
In order to solve this generalized Liouville equation, we need to calculate the classical equation of motion, which is given by
\begin{gather}
    \label{eq:equation_of_motion_alpha_classical_open_multi}
    i\hbar\frac{d\alpha_m}{dt} = \frac{\partial H_W(\vec{\alpha})}{\partial\alpha_m^*} + \frac{i\hbar}{2}\sum_k\gamma_k\mleft\{L^{\ast}_{kW}(\vec{\alpha})\star \frac{\partial L_{kW}(\vec{\alpha})}{\partial\alpha_m^*} - \frac{\partial L^{\ast}_{kW}(\vec{\alpha})}{\partial\alpha_m^*}\star L_{kW}(\vec{\alpha})\mright\}.
\end{gather}
We can derive the same equation of motion by following the same procedures in Sec.~\ref{subsec:Saddle-point path} by using the action in Eq.~\eqref{eq:Wigner_functional_open_multi_action}.
Then, according to the discussion below Eq.~\eqref{eq:physical_quantity_first_order_single}, we can calculate the physical quantity by iteratively solving the classical equations of motion \eqref{eq:equation_of_motion_alpha_classical_open_multi} with respect to $\alpha_m$ for $\forall m$.

The extension of the second-order calculation is not straightforward because it is intractable to perform the Hubbard-Stratonovich transformation analytically for a general multi-mode system. However, the transformation becomes tractable when we assume that the jump operators are local, i.e., each $L_{kW}(\vec{\alpha})$ involves only one component of $\vec{\alpha}$ and the jump operators do not induce transitions of bosons between different modes.
Under this assumption, we rewrite the subscript $k$ of $L_{kW}(\vec{\alpha})$ with $k,m$ to indicate that $L_{k,mW}(\vec{\alpha})$ is a function of only $\alpha_m$.
Then, we can straightforwardly extends the results in Sec.~\ref{subsec:second order of quantum fluctuations and truncated Wigner approximation} to the system with multiple degrees of freedom and obtain the following Fokker-Planck equation:
\begin{align}
    \label{Fokker_Planck_eq_multi_open}
    i\hbar\frac{dW\mleft(\vec{\alpha},t\mright)}{dt} =& -\sum_m\frac{\partial}{\partial\alpha_m}\mleft[\mleft\{\frac{\partial H_W}{\partial\alpha_m^*} + \frac{i\hbar}{2}\sum_{k}\gamma_{k,m}\mleft(L^{\ast}_{k,mW}\star \frac{\partial L_{k,mW}}{\partial\alpha_m^*} - \frac{\partial L^{\ast}_{k,mW}}{\partial\alpha_m^*}\star L_{k,mW}\mright)\mright\}W\mleft(\vec{\alpha},t\mright)\mright] + {\rm c.c.}\nonumber \\
    & -i\hbar\sum_m\mleft(\frac{\partial^2}{\partial\alpha_m^2}\mleft[\lambda_m W\mleft(\vec{\alpha},t\mright)\mright] - \frac{\partial^2}{\partial\alpha_m\partial\alpha_m^*}\mleft[\Lambda_m W\mleft(\vec{\alpha},t\mright)\mright]\mright)- {\rm c.c.},
\end{align}
where $\lambda_m = \lambda_m(\alpha_m)\in \mathbb{C}$ and $\Lambda_m = \Lambda_m(\alpha_m)\in \mathbb{R}$ are given by
\begin{align}
    \label{lambda_multi_degreee}
    \lambda_m(\alpha_m) &= \sum_{k}\frac{\gamma_{k,m}}{2}\frac{\partial L^*_{k,mW}(\alpha_m)}{\partial\alpha^*_m}\star \frac{\partial L_{k,mW}(\alpha_m)}{\partial\alpha^*_m}, \\
    \label{Lambda_Multi_degreee}
    \Lambda_m(\alpha_m) &= \sum_{k}\frac{\gamma_{k,m}}{4}\mleft\{\frac{\partial L^*_{k,mW}(\alpha_m)}{\partial\alpha_m^*}\star \frac{\partial L_{k,mW}(\alpha_m)}{\partial\alpha_m} + \frac{\partial L^*_{k,mW}(\alpha_m)}{\partial\alpha_m}\star \frac{\partial L_{k,mW}(\alpha_m)}{\partial\alpha_m^*}\mright\}.
\end{align}
The corresponding stochastic differential equation is given by
\begin{align}
    \label{eq:stochastic_differential_equation_continuous_multi}
    i\hbar d\alpha_m(t) = \mleft[\frac{\partial H_W(\vec{\alpha})}{\partial\alpha^*_m} + \frac{i\hbar}{2}\sum_{k}\gamma_{k,m}\mleft\{L^*_{k,mW}(\alpha_m)\star \frac{\partial L_{k,mW}(\alpha_m)}{\partial\alpha^*_m} - \frac{\partial L^*_{k,mW}(\alpha_m)}{\partial\alpha^*_m}\star L_{k,mW}(\alpha_m)\mright\}\mright]dt + i\hbar e^{i\theta_m(\alpha_m)/2}\cdot d\Xi_m(\alpha_m),
\end{align}
where $\theta_m$ is defined by $\theta_m = {\rm arg}(\lambda_m )$ and the complex stochastic process $\Xi_m(\alpha_m)$ has Gaussian increments with mean 0: 
\begin{align}
    \label{eq:complex_stochastic_process_real_multi}
    P\mleft[\Delta \Xi_m^{\rm re}(\alpha_m)\mright] &= \frac{1}{\sqrt{2\pi\Delta t\mleft(\Lambda_m(\alpha_m) - |\lambda_m(\alpha_m)|\mright)}}{\rm exp}\mleft[-\frac{\Bigl(\Delta \Xi_m^{\rm re}(\alpha_m)\Bigr)^2}{2\Delta t\mleft(\Lambda_m(\alpha_m) - |\lambda_m(\alpha_m)|\mright)}\mright], \\
    \label{eq:complex_stochastic_process_imag_multi}
    P\mleft[\Delta \Xi_m^{\rm im}(\alpha_m)\mright] &= \frac{1}{\sqrt{2\pi\Delta t\mleft(\Lambda_m(\alpha_m) + |\lambda_m(\alpha_m)|\mright)}}{\rm exp}\mleft[-\frac{\Bigl(\Delta \Xi_m^{\rm im}(\alpha_m)\Bigr)^2}{2\Delta t\mleft(\Lambda_m(\alpha_m) + |\lambda_m(\alpha_m)|\mright)}\mright].
\end{align}
Then, following the procedure below Eq.~\eqref{eq:physical_quantity_second_order_single}, we can calculate the physical quantity by iteratively solving the stochastic differential equation \eqref{eq:stochastic_differential_equation_continuous_multi} with respect to $\alpha_m$ for $\forall m$.

In the second-order calculation, we assume that the jump operators are local because otherwise, we cannot analytically perform the Hubbard-Stratonovich transformation.
However, when we adopt to numerically diagonalize the quadratic terms, we can extend our theory to the system with the general jump operators, which will be published elsewhere.


\subsection{\label{subsec:Non-equal time correlation functions} Non-equal time correlation functions}
The detailed expressions of the propagator of the Wigner function enables us to calculate the non-equal time correlation functions within the first- and second-order quantum fluctuations.
Here, we first consider a system with a single degree of freedom, and subsequently extend the results to a multi-mode system.
The non-equal two-time correlation is defined by \cite{Breuer}
\begin{align}
    \label{eq:def_of_two_time_correlation}
    \Braket{\hat{A}(t)\hat{B}(t_0)} = {\rm Tr}\mleft[\hat{A}\hat{\mathcal{V}}(t,t_0)\hat{B}\hat{\rho}(t_0)\mright]\quad t_0 \leq t,
\end{align}
where $\hat{\mathcal{V}}(t,t_0)$ is defined by Eq.~\eqref{eq:def_of_dynamical_map}.
In the phase space, Eq.~\eqref{eq:def_of_two_time_correlation} becomes
\begin{align}
    \label{eq:first_line_of_two_time_correlation_phase_space}
    \Braket{\hat{A}(t)\hat{B}(t_0)} &= \int\frac{d^2\alpha_{\rm f}}{\pi}A_W(\alpha_{\rm f})\mleft[\hat{\mathcal{V}}(t,t_0)\hat{B}\hat{\rho}(t_0)\mright]_W(\alpha_{\rm f}) \\
    \label{eq:two_time_correlation_phase_space}
    &= \int\frac{d^2\alpha_{\rm f}d^2\alpha_0}{\pi^2}A_W(\alpha_{\rm f})\varUpsilon\mleft(\alpha_{\rm f},t~;\alpha_0,t_0\mright)\Bigl[B_W(\alpha_0)\star W(\alpha_0,t_0)\Bigr],
\end{align}
where we use Eq.~\eqref{eq:Tr_AB_Weyl_Wigner} to transform Eq.~\eqref{eq:def_of_two_time_correlation} into Eq.~\eqref{eq:first_line_of_two_time_correlation_phase_space}.
In order to obtain Eq.~\eqref{eq:two_time_correlation_phase_space}, we follow the same procedure as to obtain Eq.~\eqref{eq:Wigner_function_and_propagator}.
Equation~\eqref{eq:two_time_correlation_phase_space} is a general phase space representation of the non-equal two-time correlation function.
In the same way, the non-equal three-time correlation is defined by
\begin{align}
    \Braket{\hat{A}(t)\hat{B}(t_j)\hat{C}(t_0)} = {\rm Tr}\mleft[\hat{A}\hat{\mathcal{V}}(t,t_j)\hat{B}\hat{\mathcal{V}}(t_j,t_0)\hat{C}\hat{\rho}(t_0)\mright]\quad t_0\leq t_j \leq t,
\end{align}
and its phase space representation is given by
\begin{align}
    \label{eq:three_time_correlation_phase_space}
    \Braket{\hat{A}(t)\hat{B}(t_j)\hat{C}(t_0)} = \int\frac{d^2\alpha_{\rm f}d^2\alpha_j d^2\alpha_0}{\pi^4}A_W(\alpha_{\rm f})\varUpsilon(\alpha_{\rm f},t~;\alpha_j,t_j)B_W(\alpha_j)\star\mleft[\varUpsilon(\alpha_j,t_j~;\alpha_0,t_0)\biggl\{C_W(\alpha_0)\star W(\alpha_0,t_0)\biggr\}\mright].
\end{align}
In a similar manner, we can obtain the phase space representation of a non-equal multi-time correlation function for a product of an arbitrary number of operators.

Now we focus on the non-equal two-time correlation function for the Markoian open quantum system.
By applying the Markov condition Eq.~\eqref{eq:Markov_condition_phase_space} to Eq.~\eqref{eq:two_time_correlation_phase_space}, we obtain
\begin{align}
    \label{eq:two_time_correlation_Markov}
    \Braket{\hat{A}(t)\hat{B}(t_0)} = \int\frac{d^2\alpha_{\rm f}}{\pi}A_W(\alpha_{\rm f})\lim_{\Delta t\to 0}\prod_{j=0}^{N_t - 1}\int\frac{d^2\alpha_j}{\pi}\varUpsilon(\alpha_{j+1},t_j + \Delta t~;\alpha_j,t_j)\Bigl[B_W(\alpha_0)\star W(\alpha_0,t_0)\Bigr].
\end{align}
This is a general expression of the non-equal two-time correlation under the Markov condition.
As a specific case, we choose $\hat{B} = \hat{a}$ and assume the initial state is given by a pure coherent state $\hat{\rho}(t_0) = \ket{\alpha_{\rm I}}\bra{\alpha_{\rm I}}$, whose corresponding Wigner function is a two-dimensional Gaussian function $W(\alpha,t_0) = 2e^{-2|\alpha - \alpha_{\rm I}|^2}$.
Then, Eq.~\eqref{eq:two_time_correlation_Markov} becomes
\begin{align}
    \label{eq:two_time_correlation_Markov_specific_model}
    \Braket{\hat{A}(t)\hat{a}(t_0)} = \int\frac{d^2\alpha_{\rm f}}{\pi}\alpha_{\rm I}A_W(\alpha_{\rm f})\lim_{\Delta t\to 0}\prod_{j=0}^{N_t - 1}\int\frac{d^2\alpha_j}{\pi}\varUpsilon(\alpha_{j+1},t_j + \Delta t~;\alpha_j,t_j)W(\alpha_0,t_0).
\end{align}
Using the analytical expressions of the propagator in Eqs.~\eqref{eq:propagator_infinitesimal_time_interval_first_order} and \eqref{eq:propagator_infinitesimal_time_interval_second_order}, we can approximately calculate the non-equal two-time correlation function Eq.~\eqref{eq:two_time_correlation_Markov_specific_model}.
For example, within the first-order [second-order] quantum fluctuations, we first solve the equation of motion of $\alpha$, Eq.~\eqref{eq:equation_of_motion_alpha_classical_open} [Eq.~\eqref{eq:stochastic_differential_equation_continuous}] iteratively with different initial conditions sampled from the initial Wigner function $W(\alpha_0,t_0)$ and calculate the value $\alpha_{\rm I}A_W(\alpha_{\rm f})$.
Then, we can calculate Eq.~\eqref{eq:two_time_correlation_Markov_specific_model} by taking an ensemble average of the results.
In particular, when we choose $\hat{A}=\hat{a}^\dagger$, Eq.~\eqref{eq:two_time_correlation_Markov_specific_model} gives the time-normally ordered correlation function.
Here, we note that when we choose $\gamma_k = 0$ for $\forall k$,
the TWA of Eq.~\eqref{eq:two_time_correlation_Markov_specific_model} reduces to the one for an isolated system \cite{Polkovnikov2009}.

The extension to the system with the multiple degrees of freedom is straightforward.
Here, we consider the two-time correlation function.
As in the case of Eq.~\eqref{eq:two_time_correlation_Markov_specific_model}, we choose $\hat{B}=\hat{a}_m$ ($m=1,2,\cdots,M$) and assume the initial density matrix $\hat{\rho}(t_0)=\bigotimes_m|\alpha_{{\rm I}m}\rangle\langle \alpha_{{\rm I}m}|$ where $\hat{a}_m|\alpha_{{\rm I}m}\rangle=\alpha_{{\rm I}m}|\alpha_{{\rm I}m}\rangle$.
The corresponding initial Wigner function is a Gaussian function $\prod_m 2e^{-2|\alpha_m-\alpha_{{\rm I}m}|^2}$.
The resulting two-time correlation function $\braket{\hat{A}(t)\hat{a}_m(t_0)}$ is given by the right-hand side of Eq.~\eqref{eq:two_time_correlation_Markov_specific_model} with replacing $\alpha_{\rm I}$ to $\alpha_{{\rm I}m}$ and the other $\alpha$'s to $\vec{\alpha}$, which is numerically evaluated by solving the equation of motion of $\alpha_m$ for $\forall m$ according to Eq.~\eqref{eq:equation_of_motion_alpha_classical_open_multi} (first order) or Eq.~\eqref{eq:stochastic_differential_equation_continuous_multi} (second order) and taking an ensemble average of the results weighted by the initial Wigner function. This procedure enables us to calculate, e.g., the one-body non-equal time correlation $\braket{\hat{a}_m^\dagger(t)\hat{a}_n(t_0)}$.


\section{\label{sec:Benchmark calculations} Benchmark calculations}
We shall numerically study the validity of the equations of motion derived in the previous sections by focusing on the relaxation dynamics of the GKSL equation for two models whose exact solutions are numerically obtainable.
The quantities used in our numerical calculations are particle numbers, spins, and non-equal time correlation functions, and we calculate them using the classical equation of motion, the generalized Liouville equation, and the Fokker-Planck equation, which are respectively abbreviated as ``CEOM”, ``GLE” and ``FPE”, and compare them with the numerically exact results. Below, we consider the systems with a single and multiple degrees of freedom in Secs.~\ref{subsec:Single degree of freedom} and \ref{subsec:Multiple degrees of freedom}, respectively.


\subsection{\label{subsec:Single degree of freedom} Model 1: Single degree of freedom}
We first consider the following GKSL equation:
\begin{align}
    \label{eq:GKSL_single_degree}
    \frac{d\hat{\rho}(t)}{dt} &= -\frac{i}{\hbar}\mleft[\hat{H},\hat{\rho}(t)\mright]_- + \gamma\mleft(\hat{L}\hat{\rho}(t)\hat{L}^{\dagger} - \frac{1}{2}\mleft[\hat{L}^{\dagger}\hat{L},\hat{\rho}(t)\mright]_+\mright), \\
    \hat{H} &= -\mu \hat{a}^{\dagger}\hat{a} + \frac{U}{2}\hat{a}^{\dagger}\hat{a}^{\dagger}\hat{a}\hat{a},\\
    \hat{L} &= \hat{a}\hat{a}.
\end{align}
This Hamiltonian describes a system of Bose-Einstein condensed atoms trapped in a strongly confined potential with chemical potential $\mu$ and atom-atom interaction energy $U$.
The jump operator describes two-body losses that can be introduced by using a photoassociation technique in cold atomic systems and $\gamma$ represents the loss strength \cite{Menegatti}.
We prepare the system initially in a pure coherent state $\rho(0) = \ket{\alpha_{\rm I}}\bra{\alpha_{\rm I}}$ where $\alpha_{\rm I} = \sqrt{N_{\rm I}}e^{i\pi/4}$ with its mean atomic number $N_{\rm I} = 50$.
The corresponding Wigner function is a two-dimensional Gaussian function, $W(\alpha) = 2e^{-2|\alpha - \alpha_{\rm I}|^2}$.
We then numerically calculate the time evolutions of the remaining fraction of atoms and the non-equal time correlation function, which are respectively defined by
\begin{align}
    \label{eq:quantities_scalar}
    n(t) = \frac{\Braket{\hat{a}^{\dagger}(t)\hat{a}(t)}}{N_{\rm I}},\quad G(t,0) = \frac{\Braket{\hat{a}^{\dagger}(t)\hat{a}(0)}}{N_{\rm I}}
\end{align}
under parameter settings of $\mu/(\hbar N_{\rm I}\gamma)=U/(\hbar\gamma)=1$.
Here, the non-equal time correlation function takes a complex value $G(t,0) = G^{\rm re}(t,0) + iG^{\rm im}(t,0) \in \mathbb{C}$ ($G^{\rm re}(t,0), G^{\rm im}(t,0) \in \mathbb{R}$).

\begin{figure}[t]
	\centering 
	\includegraphics[width = 9cm]{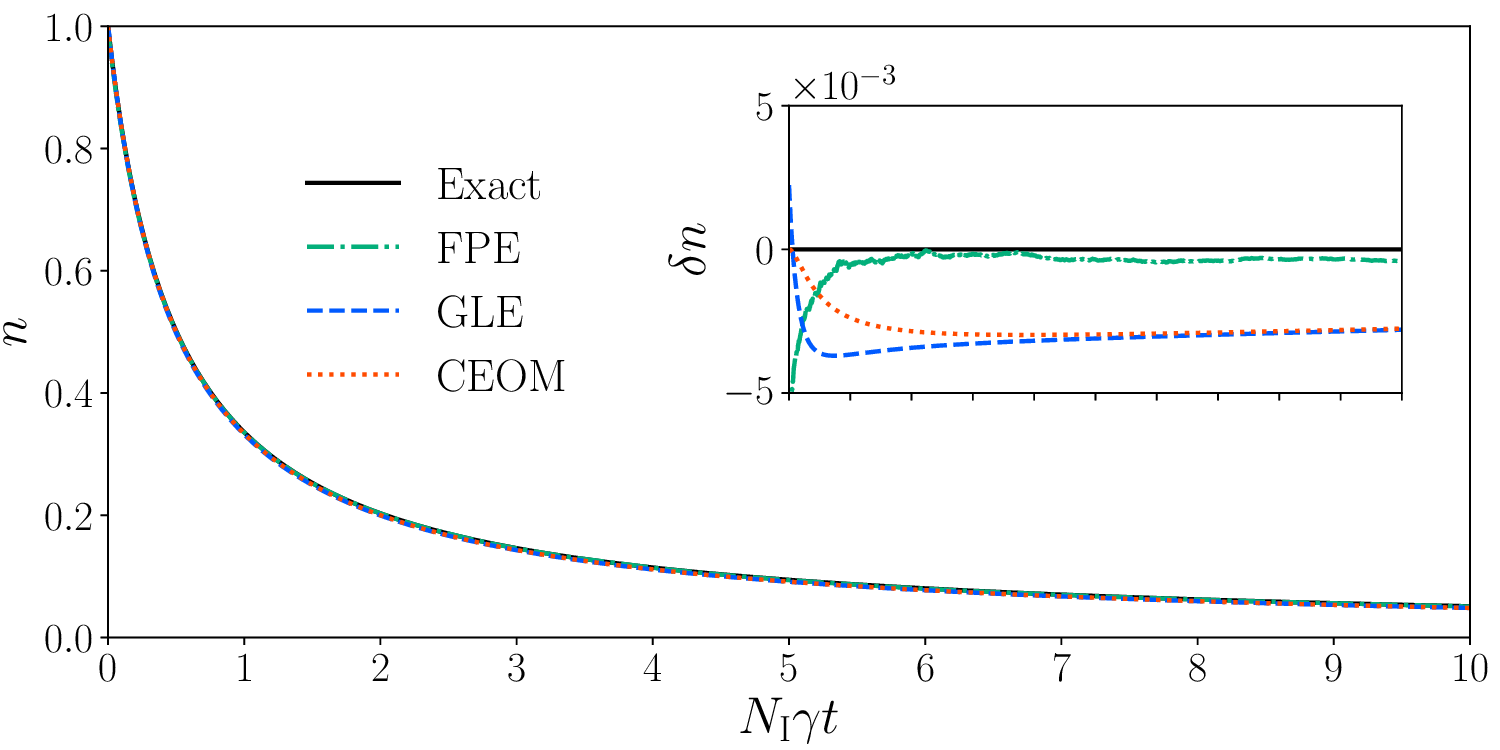}
	\caption{Time evolution of the remaining fraction of particles in the system obeying the GKSL equation \eqref{eq:GKSL_single_degree} starting from a pure coherent state $\rho(0)=\ket{\alpha_{\rm I}}\bra{\alpha_{\rm I}}$ where $\alpha_{\rm I}=\sqrt{N_{\rm I}}e^{i\pi/4}$ with $N_{\rm I}=50$. We compare the numerically exact results (Exact) with the approximated ones which are obtained by numerically solving the classical equation of motion (CEOM), the generalized Liouville equation (GLE), and the Fokker-Plank equation (FPE) with $\mu/(\hbar N_{\rm I}\gamma)=U/(\hbar\gamma)=1$, where we take 1000 samples for the initial value of $\alpha$ (GLE and FPE) and 100 samples for the stochastic processes (FPE). 
The inset depicts the difference between the result of each approximation and the exact one.}
	\label{fig:Single_Calculation}
\end{figure}
Using the Weyl-Wigner representation $L_W(\alpha) = [\hat{a}\hat{a}]_W(\alpha) =\alpha^2$, we obtain the classical equation of motion [Eq.~\eqref{eq:equation_of_motion_alpha_classical_open}] and the stochastic differential equation [Eq.~\eqref{eq:stochastic_differential_equation_continuous}] as
\begin{align}
    \label{eq:saddle_point_equation_scalar_model}
    i\hbar\frac{d\alpha}{dt} &= -\mu\alpha + U\alpha\mleft(|\alpha|^2 - 1\mright) - i\hbar\gamma \alpha\mleft(|\alpha|^2 - 1\mright), \\
    \label{eq:stochastic_differential_equation_scalar_benchmark}
    i\hbar d\alpha &= \biggl\{-\mu\alpha + U\alpha\mleft(|\alpha|^2 - 1\mright) - i\hbar\gamma \alpha\mleft(|\alpha|^2 - 1\mright)\biggr\}dt + i\hbar\cdot d\Xi,
\end{align}
respectively. Here, because Eqs.~\eqref{eq:lambda_single_degree} and \eqref{eq:Lambda_single_degree} reduce to $\lambda(\alpha)=0$ and $\Lambda(\alpha)=\gamma(|\alpha|^2-1/2)$, respectively, we can choose $\theta=0$ in Eq.~\eqref{eq:stochastic_differential_equation_continuous},
and the complex stochastic process $\Delta \Xi(\alpha)$ follows the Gaussian distributions:
\begin{align}
    P\mleft[\Delta \Xi^{\rm re}(\alpha)\mright] &= \frac{1}{\sqrt{2\pi\Delta t\gamma\mleft(|\alpha|^2 - \frac{1}{2}\mright)}}{\rm exp}\mleft[-\frac{\Bigl(\Delta \Xi^{\rm re}(\alpha)\Bigr)^2}{2\Delta t\gamma\mleft(|\alpha|^2 - \frac{1}{2}\mright)}\mright], \\
    P\mleft[\Delta \Xi^{\rm im}(\alpha)\mright] &= \frac{1}{\sqrt{2\pi\Delta t\gamma\mleft(|\alpha|^2 - \frac{1}{2}\mright)}}{\rm exp}\mleft[-\frac{\Bigl(\Delta \Xi^{\rm im}(\alpha)\Bigr)^2}{2\Delta t\gamma\mleft(|\alpha|^2 - \frac{1}{2}\mright)}\mright].
\end{align}
In the calculation of the classical equation of motion, we solve Eq.~\eqref{eq:saddle_point_equation_scalar_model} starting from the initial state $\alpha=\alpha_{\rm I}$.
In the first [second] order of quantum fluctuations, we solve Eq.~\eqref{eq:saddle_point_equation_scalar_model} [Eq.~\eqref{eq:stochastic_differential_equation_scalar_benchmark}] with various initial values of $\alpha$ sampled from the initial Wigner function and taking an ensemble average over the results.
In particular, in the second order calculation, we also need to take an ensemble average with respect to the stochastic processes to solve Eq.~\eqref{eq:stochastic_differential_equation_scalar_benchmark}.
In this benchmark calculation, we take 1000 samples for the initial states and 100 samples for the stochastic processes.
We use the 4th order Runge–Kutta method to solve Eq.~\eqref{eq:GKSL_single_degree} (exact calculation) and Eq.~\eqref{eq:saddle_point_equation_scalar_model}, and use the Platen method to solve Eq.~\eqref{eq:stochastic_differential_equation_scalar_benchmark}.

Figure~\ref{fig:Single_Calculation} shows the time evolution of $n(t)$, and the inset depicts the deviation between the result of each approximation and the exact one, $\delta n = n_{\rm Exact} - n_{\rm Approx}$ with ``Approx" $=$ ``CEOM", ``GLE", and ``FPE".
Although all the approximations in the main panel exhibit good agreement with the exact one, the inset shows the calculation of the second order of quantum fluctuations agrees most, which is consistent with the perturbation theory in Secs.~\ref{subsec:first order of quantum fluctuations} and \ref{subsec:second order of quantum fluctuations and truncated Wigner approximation}.
We note that the relatively large deviation between the results of the second order of quantum fluctuations and exact calculations at the early times becomes smaller by increasing the number of samples for the initial states.
In addition, the curve becomes smoother by increasing the number of samples for the stochastic processes.

\begin{figure}[t]
	\centering 
	\includegraphics[width = \linewidth]{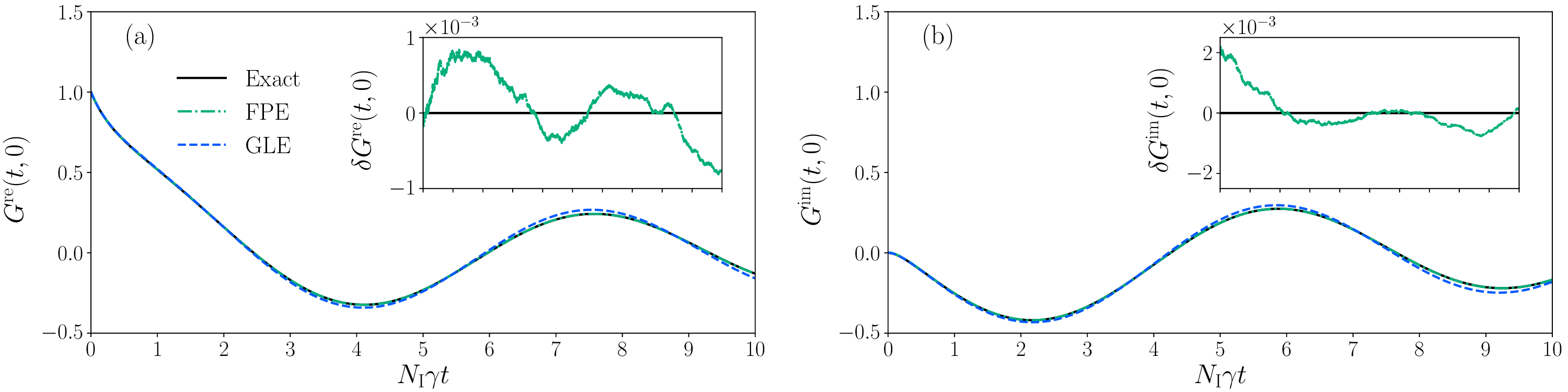}
	\caption{Real (a) and imaginary part (b) of the non-equal time correlation defined by Eq.~\eqref{eq:quantities_scalar} in the relaxation dynamics described by the GKSL equation \eqref{eq:GKSL_single_degree}, where the meaning of each curve and the parameters used are the same as those in Fig.~\ref{fig:Single_Calculation}. The inset in each panel shows the deviation of the results of the second order of quantum fluctuations (FPE) from the exact ones (Exact), indicating the calculations of the second order of quantum fluctuations well reproduces the exact results.}
	\label{fig:Single_multitime_Calculation}
\end{figure}

The difference in each approximation becomes more prominent when we compare the results of non-equal time correlation as shown in Fig.~\ref{fig:Single_multitime_Calculation}, where one can identify the difference between the first- and second-order calculation in the main panel and the latter agrees more with the exact one.
The inset shows the difference between the results of second order of quantum fluctuations and exact calculations, $\delta G^{\rm re}(t,0) = G^{\rm re}(t,0)_{\rm Exact} - G^{\rm re}(t,0)_{\rm FPE}$ and $\delta G^{\rm im}(t,0) = G^{\rm im}(t,0)_{\rm Exact} - G^{\rm im}(t,0)_{\rm FPE}$, supporting the validity of the TWA.


\subsection{\label{subsec:Multiple degrees of freedom} Model 2: Multiple degrees of freedom}
As a case of a system with multiple degrees of freedom, we consider spin-1 Bose atoms obeying the following GKSL equation:
\begin{align}
    \label{eq:GKSL_Multi_degree}
    \frac{d\hat{\rho}(t)}{dt} &= -\frac{i}{\hbar}\mleft[\hat{H},\hat{\rho}(t)\mright]_- + \gamma\sum_{k=1}^4\mleft(\hat{L}_k\hat{\rho}(t)\hat{L}_k^{\dagger} - \frac{1}{2}\mleft[\hat{L}_k^{\dagger}\hat{L}_k,\hat{\rho}(t)\mright]_+\mright), \\
    \label{eq:multi_degrees_Hamiltonian}
    \hat{H} &= \mu\hat{N} - \frac{c_0}{2}\hat{N}\mleft(\hat{N} - 1\mright) + \frac{c_1}{2}\mleft(\hat{\bm{F}}^2 - 2\hat{N}\mright), \\
    \label{eq:multi_degrees_jump_operators}
    \hat{L}_1 &= \hat{a}_1^{\dagger}\hat{a}_1,~\hat{L}_2 = \hat{a}_0^{\dagger}\hat{a}_0,~\hat{L}_3 = \hat{a}_0\hat{a}_0,~\hat{L}_4 = \hat{a}_{-1}^{\dagger}\hat{a}_{-1}.
\end{align}
Here, we assume that atoms are strongly confined in a trapping potential such that the spatial degrees of freedom are frozen and the atoms have only the three degrees of freedom corresponding to the magnetic sublevels $m = 1,0,-1$.
In the Hamiltonian \eqref{eq:multi_degrees_Hamiltonian}, $\mu$ is the chemical potential, $c_0$ and $c_1$ are the strengths of the spin-independent and spin-dependent inter-atomic interactions, respectively, and the total atomic number operator $\hat{N}$ and the spin operator $\hat{F}_{\nu={\rm x},{\rm y},{\rm z}}$ are defined by  
\begin{align}
    \hat{N} &= \sum_{m = 0,\pm1}\hat{a}_m^{\dagger}\hat{a}_m, \\
    \hat{F}_{\nu} &= \sum_{m,m'=0,\pm1}\hat{a}^{\dagger}_m(f_{\nu})_{mm'}\hat{a}_{m'},
\end{align}
where $\hat{a}_m$ is the annihilation operator of an atom in a magnetic sublevel $m$, and $f_{\nu={\rm x},{\rm y},{\rm z}}$ is the spin-1 matrix given by
\begin{align}
    &f_{\rm x} = \frac{1}{\sqrt{2}}
    \begin{bmatrix}
        0&1&0 \\
        1&0&1 \\
        0&1&0
    \end{bmatrix},~
    f_{\rm y} = \frac{1}{\sqrt{2}}
    \begin{bmatrix}
        0&-i&0 \\
        i&0&-i \\
        0&i&0
    \end{bmatrix},~
    f_{\rm z} = 
    \begin{bmatrix}
        1&0&0 \\
        0&0&0 \\
        0&0&-1
    \end{bmatrix}.
\end{align}
As for the jump operators, we consider the four operators given by Eq.~\eqref{eq:multi_degrees_jump_operators} that do not mix different internal states. The jumps cause the dephasing of atoms in each sublevel and the two-body loss of atoms in the $m=0$ state. For simplicity, we choose the same strength $\gamma$ for all jumps.

We prepare the initial state in a pure coherent state of $m=0$ atoms, i.e., $\rho(0) = \ket{0,\alpha_{\rm I},0}\bra{0,\alpha_{\rm I},0}$, where $\alpha_{\rm I} = \sqrt{N_{\rm I}}e^{i\pi/4}$ and $\hat{a}_m\ket{0,\alpha_{\rm I},0} = \delta_{m0}\alpha_{\rm I}\ket{0,\alpha_{\rm I},0}$ with $N_{\rm I} = 10$.
The corresponding Wigner function is given by $W(\vec{\alpha}) = 8e^{-2|\alpha_1|^2 -2|\alpha_{0} - \alpha_{\rm I}|^2 -2|\alpha_{-1}|^2}$ with $\vec{\alpha} = (\alpha_1,\alpha_0,\alpha_{-1})$.
We then study the time evolution of the remaining fraction of atoms in the $m=0$ state, $n_0(t)$, the transverse magnetization fluctuation, $f(t)$, and the non-equal time correlation function of atoms in the $m=0$ state, $G_0(t,0)$, which are respectively defined by
\begin{align}
    \label{eq:quantities_multi}
    n_0(t) = \frac{\Braket{\hat{a}_0^{\dagger}(t)\hat{a}_0(t)}}{N_{\rm I}},\quad f(t) = \frac{\Braket{\hat{F}_{\rm x}(t)^2} + \Braket{\hat{F}_{\rm y}(t)^2}}{N_{\rm I}},\quad G_0(t,0) = \frac{\Braket{\hat{a}_0^{\dagger}(t)\hat{a}_0(0)}}{N_{\rm I}}.
\end{align}
In the numerical calculation, we set $\mu/(\hbar N_{\rm I} \gamma) = c_0 /(\hbar\gamma)=c_1 /(\hbar\gamma)=1$.
Here, the non-equal time correlation function takes a complex value $G_0(t,0) = G_0^{\rm re}(t,0) + iG_0^{\rm im}(t,0) \in \mathbb{C}$ ($G_0^{\rm re}(t,0), G_0^{\rm im}(t,0) \in \mathbb{R}$).

Applying the results in Sec.~\ref{subsec:Extension to a multi-mode system} to the present system, we obtain the classical equations of motion [Eq.~\eqref{eq:equation_of_motion_alpha_classical_open_multi}]:

\begin{subequations}
    \label{eq:saddle_point_equation_benchmark_multi}
    \begin{align}
        \label{eq:saddle_point_equation_benchmark_multi_1}
        i\hbar\frac{d\alpha_1}{dt} &= \mu\alpha_1 - c_0\alpha_1(N - 2) + c_1\mleft\{\alpha_1\mleft(N - 2|\alpha_{-1}|^2 + \frac{1}{2}\mright) + \alpha^2_0\alpha_{-1}^*\mright\} - \frac{i\hbar}{2}\gamma\alpha_1, \\
        \label{eq:saddle_point_equation_benchmark_multi_0}
        i\hbar\frac{d\alpha_0}{dt} &= \mu\alpha_0 - c_0\alpha_0(N - 2) + c_1\mleft\{\alpha_0\mleft(N - |\alpha_0|^2 + \frac{1}{2}\mright) + 2\alpha^*_0\alpha_1\alpha_{-1}\mright\} -i\hbar\gamma\alpha_0\mleft(|\alpha_0|^2 - \frac{1}{2}\mright), \\
        \label{eq:saddle_point_equation_benchmark_multi_-1}
        i\hbar\frac{d\alpha_{-1}}{dt} &= \mu\alpha_{-1} - c_0\alpha_{-1}(N - 2) + c_1\mleft\{\alpha_{-1}\mleft(N - 2|\alpha_{1}|^2 + \frac{1}{2}\mright) + \alpha^2_0\alpha_{1}^*\mright\} - \frac{i\hbar}{2}\gamma\alpha_{-1},
    \end{align}
\end{subequations}
and the stochastic differential equations [Eq.~\eqref{eq:stochastic_differential_equation_continuous_multi}]:
\begin{subequations}
    \label{eq:stochastic_differential_equation_benchmark_multi}
    \begin{align}
    \label{eq:stochastic_differential_equation_benchmark_multi_1}
    i\hbar d\alpha_1 &= \mleft[\mu\alpha_1 - c_0\alpha_1(N - 2) + c_1\mleft\{\alpha_1\mleft(N - 2|\alpha_{-1}|^2 + \frac{1}{2}\mright) + \alpha^2_0\alpha_{-1}^*\mright\} - \frac{i\hbar\gamma}{2}\alpha_1\mright]dt + i\hbar e^{i\theta_1/2}\cdot d\Xi_1, \\
    \label{eq:stochastic_differential_equation_benchmark_multi_0}
    i\hbar d\alpha_0 &= \mleft[\mu\alpha_0 - c_0\alpha_0(N - 2) + c_1\mleft\{\alpha_0\mleft(N - |\alpha_0|^2 + \frac{1}{2}\mright) + 2\alpha^*_0\alpha_1\alpha_{-1}\mright\} -i\hbar\gamma\alpha_0\mleft(|\alpha_0|^2 - \frac{1}{2}\mright)\mright]dt + i\hbar e^{i\theta_0/2}\cdot d\Xi_0, \\
    \label{eq:stochastic_differential_equation_benchmark_multi_-1}
    i\hbar d\alpha_{-1} &= \mleft[\mu\alpha_{-1} - c_0\alpha_{-1}(N - 2) + c_1\mleft\{\alpha_{-1}\mleft(N - 2|\alpha_{1}|^2 + \frac{1}{2}\mright) + \alpha^2_0\alpha_{1}^*\mright\} - \frac{i\hbar\gamma}{2}\alpha_{-1}\mright]dt + i\hbar e^{i\theta_{-1}/2}\cdot d\Xi_{-1},
    \end{align}
\end{subequations}
where $N = N(\vec{\alpha})$ in the both equations is the Weyl-Wigner representation of the total number of atoms given by
\begin{align}
    N(\vec{\alpha}) = [\hat{N}]_W(\vec{\alpha}) = \sum_{m=0,\pm 1}|\alpha_m|^2 - \frac{3}{2}.
\end{align}
As for the stochastic processes in Eqs.~\eqref{eq:stochastic_differential_equation_benchmark_multi_1}-\eqref{eq:stochastic_differential_equation_benchmark_multi_-1}, Eqs.~\eqref{lambda_multi_degreee} and \eqref{Lambda_Multi_degreee} lead to 
\begin{align}
\lambda_m(\alpha_m)&=\frac{\gamma}{2}\alpha_m^2\ \ (m=0,\pm 1),\\
\Lambda_{\pm 1}(\alpha_{\pm 1})&=\frac{\gamma}{2}|\alpha_{\pm 1}|^2,\\
\Lambda_{0}(\alpha_0)&=\gamma\mleft(\frac{3}{2}|\alpha_0|^2-\frac{1}{2}\mright),
\end{align}
from which we obtain $\theta={\rm arg}(\alpha_m^2)$, $\Xi^{\rm re}_{m=\pm 1}=0$ (because of $|\lambda_{\pm 1}|=\Lambda_{\pm 1}$), and the distributions of the stochastic processes are given by 
\begin{align}
    \label{eq:stochastic_process_increments_adaggera}
    P\mleft[\Delta \Xi_{m=\pm 1}^{\rm im}(\alpha_{m=\pm 1})\mright] &= \frac{1}{\sqrt{2\pi\Delta t\gamma|\alpha_{m=\pm 1}|^2}}{\rm exp}\mleft[-\frac{\Bigl(\Delta \Xi_{m=\pm 1}^{\rm im}(\alpha_{m=\pm 1})\Bigr)^2}{2\Delta t\gamma|\alpha_{m=\pm 1}|^2}\mright], \\
    P\mleft[\Delta \Xi_0^{\rm re}(\alpha_0)\mright] &= \frac{1}{\sqrt{2\pi\Delta t\gamma\mleft(|\alpha_0|^2 - \frac{1}{2}\mright)}}{\rm exp}\mleft[-\frac{\Bigl(\Delta \Xi^{\rm re}(\alpha_0)\Bigr)^2}{2\Delta t\gamma\mleft(|\alpha_0|^2 - \frac{1}{2}\mright)}\mright], \\
    P\mleft[\Delta \Xi_0^{\rm im}(\alpha_0)\mright] &= \frac{1}{\sqrt{2\pi\Delta t\gamma\mleft(2|\alpha_0|^2 - \frac{1}{2}\mright)}}{\rm exp}\mleft[-\frac{\Bigl(\Delta \Xi^{\rm im}(\alpha_0)\Bigr)^2}{2\Delta t\gamma\mleft(2|\alpha_0|^2 - \frac{1}{2}\mright)}\mright].
\end{align}
As in the case of Sec.~\ref{subsec:Single degree of freedom}, we numerically study the dynamics according to Eqs.~\eqref{eq:saddle_point_equation_benchmark_multi} and \eqref{eq:stochastic_differential_equation_benchmark_multi} using the 4th-order Runge-Kutta method and the Platen method, respectively, and take an ensemble average of 1000 samples for the initial states and 100 samples for the stochastic processes.

\begin{figure}[t]
	\centering 
	\includegraphics[width = \linewidth]{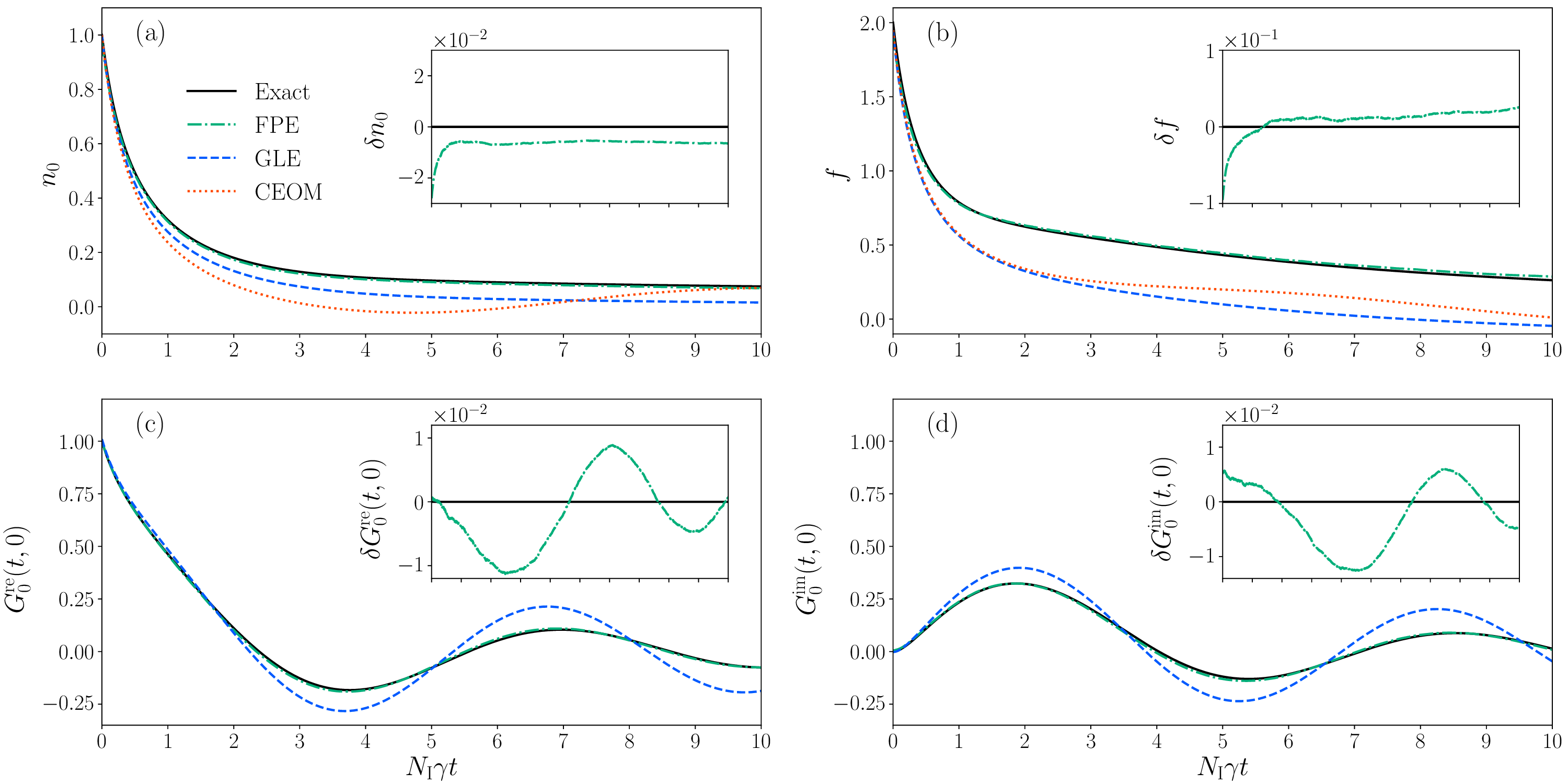}
	\caption{Relaxation dynamics of spin-1  Bose atoms obeying the GKSL equation \eqref{eq:GKSL_Multi_degree} starting from a pure coherent state $\rho(0)=\ket{0,\alpha_{\rm I},0}\bra{0,\alpha_{\rm I},0}$ where $\alpha_{\rm I}=\sqrt{N_{\rm I}}e^{i\pi/4}$ with $N_{\rm I}=10$.
Shown are (a) remaining fraction of atoms in the $m=0$ state, $n_0$, (b) transverse magnetization fluctuation, $f$, and real (c) and imaginary (d) parts of the non-equal time correlation function of the $m=0$ atoms, $G_0(t,0)$, which are defined in Eq.~\eqref{eq:quantities_multi}.
In each panel, we compare the results for the numerically exact calculation (Exact) and the approximated ones which
are obtained by numerically solving the classical equation of motion (CEOM) [shown only in (a) and (b)], the generalized Liouville equation (GLE), and the Fokker-Plank equation (FPE) with $\mu/(\hbar N_{\rm I}\gamma)=c_0/(\hbar \gamma)=c_1/(\hbar \gamma)=1$, where we take 1000 samples for the initial value of $\alpha$ (GLE and FPE) and 100 samples for the stochastic processes (FPE). 
The insets depict the difference between the results of the second order of quantum fluctuations (FPE) and the exact ones (Exact).
}
	\label{fig:Multi_Calculation}
\end{figure}
Figure~\ref{fig:Multi_Calculation} shows the relaxation dynamics of the remaining fraction $n_0$ of atoms in the $m=0$ state [Fig.~\ref{fig:Multi_Calculation}(a)], the transverse magnetization fluctuation $f$ [Fig.~\ref{fig:Multi_Calculation}(b)], and the non-equal time correlation function $G_0(t,0)$ of atoms in the $m=0$ state [Figs.~\ref{fig:Multi_Calculation}(c) and \ref{fig:Multi_Calculation}(d)] defined by Eq.~\eqref{eq:quantities_multi}.
In all the panels, we can see the significant deviations of results of the classical equations of motion and first order of quantum fluctuations from the exact one, whereas the calculations of the second order of quantum fluctuations reproduce the numerically exact dynamics well (See also the inset of each panel where we show the difference between the 2nd-order result and the exact one defined by $\delta A = A_{\rm Exact} - A_{\rm FPE}$ with $A = n_0,f,G^{\rm re}_0(t,0),$ and $G^{\rm im}_0(t,0)$).


\section{\label{sec:Discussions} Discussions}
In the benchmark calculation, all the approximations reasonably agree with the exact one in Sec.~\ref{subsec:Single degree of freedom}, whereas only the calculation with the Fokker-Plank equation well reproduces the exact result in Sec.~\ref{subsec:Multiple degrees of freedom}.
The difference comes from the Hermiticity of the jump operators. In this section, we show that the effects of the second-order quantum fluctuations become indispensable in determining relaxation dynamics when the system includes a Hermitian jump operator $\hat{L} = \hat{L}^{\dagger}$, i.e., $L_{W}(\alpha) = L^*_{W}(\alpha)$.
For simplicity, we consider a system with a single degree of freedom and only one jump operator.

Let us suppose that the initial state is set to be a pure coherent state with the mean number of particles $N_{\rm I}$.
In Secs.~\ref{subsec:Single degree of freedom} and \ref{subsec:Multiple degrees of freedom}, the strengths of many-body interactions and jumps are properly scaled so that they do not vanish or diverge in the thermodynamic limit $N_{\rm I}\to\infty$:
Since $|\alpha|^2 \sim O(N_{\rm I})$, the order of the classical field $\alpha$ and that of the left-hand side of Eq.~\eqref{eq:stochastic_differential_equation_continuous} are $O(\sqrt{N_{\rm I}})$;
It follows that the right-hand side of Eq.~\eqref{eq:stochastic_differential_equation_continuous} should be $O(\sqrt{N_{\rm I}})$;
Suppose the jump operator is a $n$-body operator with $2n \in \mathbb{Z}_{>0}$, i.e., $\hat{L}$ is constructed with $2n$ products of $\hat{a}^{\dagger}$ and $\hat{a}$, $\gamma$ should be scaled such that $\gamma N_{\rm I}^{2n-1}$ is a constant in the limit of $N_{\rm I}\to\infty$, where we impose no restriction on $\hat{L}$ and consider a general $\hat{L}$.
Taking the constant as $\Gamma$, 
we can rewrite Eq.~\eqref{eq:stochastic_differential_equation_continuous} as
\begin{align}
    \label{eq:saddle-point_path_discussion}
    i\hbar d\alpha = \mleft[\frac{\partial H_W(\alpha)}{\partial\alpha^*} + \frac{i\hbar\Gamma}{N^{2n-1}_{\rm I}}\mleft\{L^{\ast}_W(\alpha)\star \frac{\partial L_W(\alpha)}{\partial\alpha^*} - \frac{\partial L^{\ast}_W(\alpha)}{\partial\alpha^*}\star L_W(\alpha)\mright\}\mright]dt  + i\hbar e^{i\theta(\alpha)/2} \cdot d\Xi(\alpha).
\end{align}
where we scale the strengths of the many-body interaction in $\hat{H}$ such that the order of $H_W(\alpha)$ is $O(N_{\rm I})$, and $\lambda(\alpha)$ and $\Lambda(\alpha)$ in the complex stochastic term are given by Eqs.~\eqref{eq:lambda_single_degree} and \eqref{eq:Lambda_single_degree} with $\gamma=\Gamma/N_{\rm I}^{2n-1}$.
The benchmark calculations in Secs.~\ref{subsec:Single degree of freedom} and \ref{subsec:Multiple degrees of freedom} correspond to the case of $n=1$ in Eq.~\eqref{eq:saddle-point_path_discussion}.
When the jump operator is non-Hermitian, while the order of the left-hand side and the classical-path term in the right-hand of Eq.~\eqref{eq:saddle-point_path_discussion} is $O(\sqrt{N_{\rm I}})$, the order of the stochastic term is $O(d\Xi) = O(\sqrt{\lambda\pm\Lambda})  = O(1)$.
Hence, the classical path dominates the relaxation dynamics resulting that the classical equation of motion well reproduces the exact dynamics as in Sec.~\ref{subsec:Single degree of freedom}.
On the other hand, when the jump operator is Hermitian, Eq.~\eqref{eq:saddle-point_path_discussion} becomes
\begin{align}
    \label{eq:saddle-point_path_Hermitian_jump_discussion}
    i\hbar d\alpha = \mleft[\frac{\partial H_W(\alpha)}{\partial\alpha^*} + \frac{i\hbar\Gamma}{2N^{2n-1}_{\rm I}}L_W(\alpha){\rm sinh}\frac{\hat{\phi}}{2}\frac{\partial L_W(\alpha)}{\partial\alpha^*}\mright]dt + i\hbar e^{i\theta(\alpha)/2} \cdot d\Xi(\alpha),
\end{align}
where the differential operator $\hat{\phi}$ is defined by Eq.~\eqref{eq:def_of_hat_phi}.
Because the leading order of ${\rm sinh}\hat{\phi}/2$ is $\hat{\phi}\sim O(N_{\rm I}^{-1})$, the non-unitary term of the classical path is in the order $O(N_{\rm I}^{-1/2})$, which is smaller than that of the stochastic term $O(1)$.
This means that the non-unitary dynamics of the GKSL equation is strongly affected by the stochastic term, and hence, the classical equation of motion fails to reproduce the exact dynamics as shown in Sec.~\ref{subsec:Multiple degrees of freedom}.

When we consider the thermodynamic limit in Eq.~\eqref{eq:saddle-point_path_Hermitian_jump_discussion}, since the non-unitary term of the classical-path term and the stochastic term become negligible, Eq.~\eqref{eq:saddle-point_path_Hermitian_jump_discussion} is reduced to the Hamilton equation and fails to describe the non-unitary dynamics in the thermodynamic limit.
In order to capture the non-unitary dynamics of the GKSL equation with a Hermitian jump operator in the thermodynamic limit, we need to rescale $\gamma$ such that $\tilde{\Gamma} = \gamma N_{\rm I}^{2n-2}$ becomes a constant, in which Eq.~\eqref{eq:saddle-point_path_Hermitian_jump_discussion} becomes
\begin{align}
    \label{eq:saddle-point_path_Hermitian_jump_discussion_rescale}
    i\hbar d\alpha = \mleft[\frac{\partial H_W(\alpha)}{\partial\alpha^*} + \frac{i\hbar\tilde{\Gamma}}{2N^{2n-2}_{\rm I}}L_W(\alpha){\rm sinh}\frac{\hat{\phi}}{2}\frac{\partial L_W(\alpha)}{\partial\alpha^*}\mright]dt + i\hbar e^{i\theta(\alpha)/2} \cdot d\Xi(\alpha),
\end{align}
where all the terms on the right-hand side, including the stochastic term, have the same order $O(\sqrt{N_{\rm I}})$.
We note that the above argument is valid only for $n\geq 1$.
In the case of $n = 1/2$, where the jump operator is linear with respect to $\hat{a}$ and $\hat{a}^{\dagger}$, the non-unitary term in the classical path becomes completely zero and thus the non-unitary effect comes only from the second order of quantum fluctuations.
To summarize, the Hermitian jump operator induces non-negligible effects of the second order of quantum fluctuations on the relaxation dynamics.
This indicates that neglecting quantum fluctuations (mean-field approximation) is not appropriate for simulating the dynamics of the GKSL equation with Hermitian jump operators.
Below, we introduce two examples that clearly show the necessity of the stochastic term.

The first example is the jump operator $\hat{L} = \hat{a}^{\dagger}\hat{a}$,
which is the case of $n=1$ in Eq.~\eqref{eq:saddle-point_path_Hermitian_jump_discussion_rescale}.
Under the assumption $[\hat{H},\hat{a}^{\dagger}\hat{a}]_- = 0$, the total number of particles $N = \braket{\hat{a}^{\dagger}\hat{a}}$ is preserved throughout the time evolution.
However, the classical path for this situation is analytically solved as
\begin{align}
    i\hbar\frac{d\alpha(t)}{dt} = \frac{\partial H_W(\alpha)}{\partial\alpha^*} - \frac{i\hbar\tilde{\Gamma}}{2} \alpha(t)\quad \longrightarrow\quad N(t) = |\alpha(t_0)|^2e^{-\tilde{\Gamma} (t-t_0)} - \frac{1}{2},
\end{align}
and the total number of particles is not a conserved quantity but exponentially decreases with decay time $1/\tilde{\Gamma}$.
To deal with the dynamics while preserving the total number of particles, we need to consider the second-order quantum fluctuations.
We can analytically prove the conservation law from the stochastic differential equation:
\begin{align}
    i\hbar d\alpha(t) = \mleft[\frac{\partial H_W(\alpha)}{\partial\alpha^*} - \frac{i\hbar\tilde{\Gamma}}{2} \alpha(t)\mright]dt + i\hbar e^{i\theta(\alpha)/2}\cdot d\Xi(\alpha)\quad \longrightarrow\quad N(t) = N(t_0),
\end{align}
where $\theta(\alpha)$ is given by $\theta(\alpha) = {\rm arg}(\alpha^2)$, and the complex stochastic process is purely imaginary with the increments following the Gaussian distribution with mean 0 and variance $\Delta t\tilde{\Gamma}|\alpha|^2$:
\begin{align}
    P\mleft[\Delta \Xi^{\rm im}(\alpha)\mright] = \frac{1}{\sqrt{2\pi\Delta t\tilde{\Gamma}|\alpha|^2}}{\rm exp}\mleft[-\frac{\Bigl(\Delta \Xi^{\rm im}(\alpha)\Bigr)^2}{2\Delta t\tilde{\Gamma}|\alpha|^2}\mright].
\end{align}
This result indicates that the quantum fluctuations play an important rule to determine the relaxation dynamics of the system with the jump operator $\hat{L}= \hat{a}^{\dagger}\hat{a}$.

The second example is the jump operator $\hat{L} = c_{x}\hat{x} + c_{p}\hat{p}$ with $c_{x},c_{p}\in\mathbb{R}$, $\hat{x} = (\hat{a} + \hat{a}^{\dagger})/\sqrt{2}$ and $\hat{p} = i(\hat{a}^{\dagger} - \hat{a})/\sqrt{2}$.
This is the case of $n=1/2$ in Eq.~\eqref{eq:saddle-point_path_Hermitian_jump_discussion_rescale}, where the classical path has zero non-unitary term.
The corresponding stochastic differential equation is given by
\begin{align}
    i\hbar d\alpha(t) = \frac{\partial H_W(\alpha)}{\partial\alpha^*}dt + i\hbar \cdot d\Xi(\alpha),
\end{align}
where the complex stochastic process is purely imaginary with the increments following the Gaussian distribution with mean 0 and variance $N_{\rm I}\tilde{\Gamma}(c_x^2 + c_y^2)/2$:
\begin{align}
    P\mleft[\Delta \Xi^{\rm im}(\alpha)\mright] = \frac{1}{\sqrt{\pi\Delta t\tilde{\Gamma}(c_x^2+c_y^2)}}{\rm exp}\mleft[-\frac{\Bigl(\Delta \Xi^{\rm im}(\alpha)\Bigr)^2}{\Delta t\tilde{\Gamma}(c_x^2+c_y^2)}\mright].
\end{align}
Since the jump operator is linear with respect to $\hat{a}$ and $\hat{a}^{\dagger}$, there is no non-unitary terms which are higher than third order of quantum fluctuations.
Hence, the jump operator $\hat{L} = c_{x}\hat{x} + c_{p}\hat{p}$ affect the dynamics only as the second order of quantum fluctuations, which act on the second-order moment of the Wigner function in the phase space.

We comment that the results in this section have been already obtained in the literature of the semi-classical approximation for the GKSL equation \cite{Strunz}.
In the semi-classical approximation, the GKSL equation is approximated into the Fokker-Planck equation based on the perturbative expansion with respect to the Planck constant $\hbar$.
Hence, the detailed forms of the Fokker-Planck and stochastic equations differ from those of TWA based on the expansion concerning quantum fields in the path-integral.
However, they become identical in the thermodynamic limit.
This fact ensures the validity of analytical expressions of the stochastic differential equations \eqref{eq:stochastic_differential_equation_continuous} and \eqref{eq:stochastic_differential_equation_continuous_multi}.


\section{\label{sec:Summary and conclusions} Summary and conclusions}
The TWA is an efficient tool to study the quantum dynamics of bosonic many-body systems.
In the TWA, the GKSL equation is approximated into the Fokker-Planck equation for the Wigner function, which we calculate by solving the corresponding stochastic differential equations.
The main result of this work is the derivation of the analytical expressions of the Fokker-Planck equation and stochastic differential equations.
For systems with arbitrary Hamiltonian and the jump operators that do not couple different degrees of freedom, we derived the condition [Eq.~\eqref{eq:Hubbard_Storatonovich_transformation_condition}] for the jump operators so that the Fokker-Planck equation [Eqs.~\eqref{Fokker_Planck_eq_single_open} and \eqref{Fokker_Planck_eq_multi_open}] can be reduced to the stochastic differential equations [Eqs.~\eqref{eq:stochastic_differential_equation_continuous} and \eqref{eq:stochastic_differential_equation_continuous_multi}].
This result enables us to shortcut the complicated mappings from operators to $c$-numbers in the conventional procedure of the TWA.
We hope that these expressions will become a powerful tool in the practical calculations.

In the course of the derivations, we formulated the path-integral representation of the GKSL equation based on the Weyl-Wigner transformation [Eqs.~\eqref{eq:Wigner_functional_open_single_discrete} and \eqref{eq:Wigner_functional_open_multi_discrete}], which takes a form of a natural extension from the path-integral representation for isolated systems \cite{Polkovnikov2010,Polkovnikov2009}.
The path-integral representation is characterized by the classical and quantum fields describing the classical motion and quantum fluctuations of systems, respectively.
To obtain the general expressions of the Fokker-Planck equation and stochastic differential equations, we assume small quantum fluctuations and expand the action with respect to the quantum field order by order.
Within the first-order quantum fluctuations, the GKSL equation is approximated into the generalized Liouville equation [Eqs.~\eqref{generalized Liouville_eq_single_open} and \eqref{generalized Liouville_eq_multi_open}], where each point in the phase space on which the initial Wigner function distributes follows the classical equation of motion [Eqs.~\eqref{eq:equation_of_motion_alpha_classical_open} and \eqref{eq:equation_of_motion_alpha_classical_open_multi}].
The second-order fluctuations introduce stochastic processes into the classical motion, resulting in the Fokker-Planck equation and stochastic differential equations.
Following Ref.~\cite{Polkovnikov2009}, we also provided the formula for calculating the non-equal time correlation functions described by the GKSL equation by using the TWA.
In the benchmark calculations, we numerically investigated the relaxation dynamics of some observables and non-equal time correlation functions for two models. 
Although there is a clear discrepancy between the first-order calculation and the exact one, the second-order calculation agrees well with the exact one.
These results confirm the validity of our formulation of the TWA for the GKSL equation.

In the last part of this paper, we discussed the relaxation dynamics of systems with Hermitian jump operators.
In systems with non-Hermitian jump operators, the effect of the second-order of quantum fluctuations is small and being negligible in the thermodynamic limit, and thus, the dynamics is well reproduced by the classical equation of motion.
On the other hand in systems with Hermitian jump operators, we saw that the second-order quantum fluctuations have non-negligible contributions to the dynamics even in the thermodynamic limit.
In particular, when jump operators include only the linear terms of the creation and annihilation operators, they do not modify the classical path, i.e., within the first-order approximation we obtain the same dynamics as the corresponding isolated system. The effect of the jump operators appears first in the second-order approximation as the stochastic process.
Our result indicates that neglecting quantum fluctuations (mean-field approximation) is not appropriate for simulating the quantum dynamics with Hermitian jump operators.

There are several remaining issues.
One is to remove the restriction on the jump operators for the case with multiple degrees of freedom. As discussed in Sec.~\ref{subsec:Extension to a multi-mode system}, we introduced the restriction that jump operators do not mix the different degrees of freedom to analytically diagonalize the diffusion matrix. If we numerically diagonalize the diffusion matrix while solving the stochastic differential equation, we expect it will become possible to deal with more general jump operators.
Another issue is to clarify the correspondence between our approach and the Schwinger-Keldysh path-integral approach.
The Schwinger-Keldysh path-integral rerepsentation for the GKSL equation has been obtained in literatures of the Keldysh field theory for the Markovian open quantum systems \cite{Diehl,Kamenev}.
However, their action and the distribution function are different from ours.
This is due to the difference in the phase-space representation used in the path integrals:
The Keldysh field theory is for systems written in normal-ordered operators, while our formulation is for systems written in Weyl-ordered operators.
Since these differently ordered systems are transferred with each other by commutation relations, their path integral representations are physically equivalent.
This point is also noted in Ref. \cite{Clerk}.
We are working on these issues on-going.


\section*{Acknowledgements}
This work was supported by JSPS KAKENHI (Grant Numbers JP23K13029, JP23K20817, and JP24K00557) and JST SPRING (Grant Number JPMJSP2125).

\appendix


\section{\label{Kraus representation in the phase space} Kraus representation in the phase space}
\subsection{\label{The propagator of the Wigner function} The propagator of the Wigner function{\rm :} Derivation of Eq.~\texorpdfstring{\eqref{propagator_kraus_representation}}{TEXT}}
According to the definition of the Weyl-Wigner representation Eq.~\eqref{eq:def_of_Weyl_Wigner_ transformation}, the Weyl-Wigner transformation of the left- and the right-hand sides of Eq.~\eqref{eq:Kraus_representation} becomes
\begin{align}
    \label{eq:Weyl_Wigner_rep_Kraus_Appendix_0}
    W\mleft(\alpha_{\rm f},t\mright) &=  \mleft[\sum_k\hat{M}_k(t,t_0)\hat{\rho}(t_0)\hat{M}^{\dagger}_k(t,t_0)\mright]_W(\alpha_{\rm f}) \\
    &= \int\frac{d^2\xi }{\pi}\sum_k{\rm Tr}\mleft[\hat{D}^{\dagger}\mleft(\xi \mright)\hat{M}_k(t,t_0)\hat{\rho}(t_0)\hat{M}^{\dagger}_k(t,t_0)\mright]e^{-\alpha_{\rm f}\xi^* + \alpha_{\rm f}^*\xi } \\
    \label{eq:Weyl_Wigner_rep_Kraus_Appendix_1}
    &= \int\frac{d^2\alpha_0d^2\xi }{\pi^2}\sum_k\mleft[\hat{M}^{\dagger}_k(t,t_0)\hat{D}^{\dagger}\mleft(\xi \mright)\hat{M}_k(t,t_0)\mright]_W(\alpha_0)e^{-\alpha_{\rm f}\xi^* + \alpha_{\rm f}^*\xi }W\mleft(\alpha_0,t_0\mright),
\end{align}
where we use the detailed expression of the characteristic function Eq.~\eqref{eq:charactratic_function_appendix} for the Kraus representation for the second equality and obtain the last line by using the cyclic property of the trace and Eq.~\eqref{eq:Tr_AB_Weyl_Wigner}.
Here, we express $[\hat{M}^{\dagger}_k(t,t_0)\hat{D}^{\dagger}\mleft(\xi \mright)\hat{M}_k(t,t_0)]_W(\alpha_0)$ on the right-hand side of Eq.~\eqref{eq:Weyl_Wigner_rep_Kraus_Appendix_1} in the integral form by using Eqs.~\eqref{eq:def_of_Weyl_Wigner_ transformation} and \eqref{eq:charactratic_function_appendix} as follow:
\begin{align}
    \label{eq:Weyl_Wigner_rep_appendix_1}
    \mleft[\hat{M}^{\dagger}_k(t,t_0)\hat{D}^{\dagger}\mleft(\xi \mright)\hat{M}_k(t,t_0)\mright]_W(\alpha_0) = \int\frac{d^2\eta }{\pi}{\rm Tr}\mleft[\hat{D}^{\dagger}\mleft(\eta \mright)\hat{M}^{\dagger}_k(t,t_0)\hat{D}^{\dagger}\mleft(\xi \mright)\hat{M}_k(t,t_0)\mright]e^{-\alpha_0\eta ^* + \alpha^*_0\eta}.
\end{align}
Substituting this expression into Eq.~\eqref{eq:Weyl_Wigner_rep_Kraus_Appendix_1}, we obtain
\begin{align}
    W\mleft(\alpha_{\rm f},t\mright) &= \int\frac{d^2\alpha_0}{\pi}\varUpsilon\mleft(\alpha_{\rm f},t~;\alpha_0,t_0\mright)W\mleft(\alpha_0,t_0\mright), \\
    \label{eq:Weyl_propagator_Appendix}
    \varUpsilon\mleft(\alpha_{\rm f},t~;\alpha_0,t_0\mright) &= \int\frac{d^2\xi d^2\eta }{\pi^2}\sum_k{\rm Tr}\mleft[\hat{D}^{\dagger}(\xi )\hat{M}_k\mleft(t,t_0\mright)\hat{D}^{\dagger}\mleft(\eta \mright)\hat{M}^{\dagger}_k\mleft(t,t_0\mright)\mright]e^{-\alpha_{\rm f}\xi^* + \alpha_{\rm f}^*\xi }e^{-\alpha_0\eta^* + \alpha^*_0\eta },
\end{align}
where $\varUpsilon\mleft(\alpha_{\rm f},t~;\alpha_0,t_0\mright)$ is the propagator of the Wigner function.


\subsection{\label{Markov condition for the propagator} Markov condition for the propagator{\rm :} Derivation of Eq.~\texorpdfstring{\eqref{eq:Markov_condition_phase_space}}{TEXT}}
Using the dynamical map, we can write the Markov dynamics of the system as $\hat{\rho}(t) = \hat{\mathcal{V}}(t,t_0)[\hat{\rho}(t_0)] = \hat{\mathcal{V}}(t,t_j)[\hat{\mathcal{V}}(t_j,t_0)[\hat{\rho}(t_0)]]$ for $t \geq t_j \geq t_0$, which is equivalent to
\begin{align}
    \label{eq:Markov_conditoin_kraus_representation}
    \sum_k\hat{M}_k(t,t_0)\hat{\rho}(t_0)\hat{M}^{\dagger}_k(t,t_0) = \sum_{k,k'}\hat{M}_{k'}(t,t_j)\hat{M}_{k}(t_j,t_0)\hat{\rho}(t_0)\hat{M}^{\dagger}_{k}(t_j,t_0)\hat{M}^{\dagger}_{k'}(t,t_j)
\end{align}
in the Kraus representation.
Applying the same procedure from Eq.~\eqref{eq:Weyl_Wigner_rep_Kraus_Appendix_0} to Eq.~\eqref{eq:Weyl_propagator_Appendix} to the right-hand side of \eqref{eq:Markov_conditoin_kraus_representation}, we obtain
\begin{align}
    &\varUpsilon\mleft(\alpha_{\rm f},t~;\alpha_0,t_0\mright) = \int\frac{d^2\xi' d^2\eta }{\pi^2}\sum_{k,k'}{\rm Tr}\mleft[\hat{D}^{\dagger}(\xi' )\hat{M}_{k'}(t,t_j)\hat{M}_k(t_j,t_0)\hat{D}^{\dagger}\mleft(\eta \mright)\hat{M}^{\dagger}_k(t_j,t_0)\hat{M}^{\dagger}_{k'}(t,t_j)\mright]e^{-\alpha_{\rm f}\xi'^* + \alpha_{\rm f}^*\xi' }e^{-\alpha_0\eta^* + \alpha^*_0\eta } \\
    &\hphantom{\varUpsilon\mleft(\alpha_{\rm f},t~;\alpha_0,t_0\mright)} = \int\frac{d^2\alpha_j}{\pi}\int\frac{d^2\xi'}{\pi}\sum_{k'}\mleft[\hat{M}^{\dagger}_{k'}(t,t_j)\hat{D}^{\dagger}(\xi' )\hat{M}_{k'}(t,t_j)\mright]_W(\alpha_j)e^{-\alpha_{\rm f}\xi'^* + \alpha_{\rm f}^*\xi' }\nonumber \\
    \label{eq:Markov_conditoin_phase_space_appendix_1}
    &\hphantom{\hphantom{\varUpsilon\mleft(\alpha_{\rm f},t~;\alpha_0,t_0\mright)} = \int\frac{d^2\alpha_j}{\pi}}\!\!\!\!\!\times\int\frac{d^2\eta}{\pi}\sum_k\mleft[\hat{M}_k(t_j,t_0)\hat{D}^{\dagger}\mleft(\eta \mright)\hat{M}^{\dagger}_k(t_j,t_0)\mright]_W(\alpha_j)e^{-\alpha_0\eta^* + \alpha^*_0\eta },
\end{align}
where we have used the cyclic property of the trace and Eq.~\eqref{eq:Tr_AB_Weyl_Wigner} for the second equal sign.
Similarly to Eq.~\eqref{eq:Weyl_Wigner_rep_appendix_1}, the integral forms of the Weyl-Wigner transformation in the integrands are given by
\begin{align}
    \mleft[\hat{M}^{\dagger}_{k'}(t,t_j)\hat{D}^{\dagger}(\xi' )\hat{M}_{k'}(t,t_j)\mright]_W(\alpha_j) &= \int\frac{d^2\eta'}{\pi}{\rm Tr}\mleft[\hat{D}^{\dagger}(\xi' )\hat{M}_{k'}(t,t_j)\hat{D}^{\dagger}(\eta')\hat{M}^{\dagger}_{k'}(t,t_j)\mright]e^{-\alpha_j\eta'^* + \alpha^*_j\eta'},\\
    \mleft[\hat{M}_k(t_j,t_0)\hat{D}^{\dagger}\mleft(\eta \mright)\hat{M}^{\dagger}_k(t_j,t_0)\mright]_W(\alpha_j) &= \int\frac{d^2\xi}{\pi}{\rm Tr}\mleft[\hat{D}^{\dagger}(\xi)\hat{M}_k(t_j,t_0)\hat{D}^{\dagger}\mleft(\eta \mright)\hat{M}^{\dagger}_k(t_j,t_0)\mright]e^{-\alpha_j\xi^* + \alpha^*_j\xi}.
\end{align}
Substituting these expressions into Eq.~\eqref{eq:Markov_conditoin_phase_space_appendix_1}, we obtain
\begin{align}
    &\varUpsilon\mleft(\alpha_{\rm f},t~;\alpha_0,t_0\mright) = \int\frac{d^2\alpha_j}{\pi}\int\frac{d^2\xi'd^2\eta'}{\pi^2}\sum_{k'}{\rm Tr}\mleft[\hat{D}^{\dagger}(\xi' )\hat{M}_{k'}(t,t_j)\hat{D}^{\dagger}(\eta')\hat{M}^{\dagger}_{k'}(t,t_j)\mright]e^{-\alpha_{\rm f}\xi'^* + \alpha_{\rm f}^*\xi' }e^{-\alpha_j\eta'^* + \alpha^*_j\eta'}\nonumber \\
    &\hphantom{\hphantom{\varUpsilon\mleft(\alpha_{\rm f},t~;\alpha_0,t_0\mright)} = \int\frac{d^2\alpha_j}{\pi}}\!\!\!\!\!\times\int\frac{d^2\xi d^2\eta}{\pi^2}\sum_k{\rm Tr}\mleft[\hat{D}^{\dagger}(\xi)\hat{M}_k(t_j,t_0)\hat{D}^{\dagger}\mleft(\eta \mright)\hat{M}^{\dagger}_k(t_j,t_0)\mright]e^{-\alpha_j\xi^* + \alpha^*_j\xi}e^{-\alpha_0\eta^* + \alpha^*_0\eta }.
\end{align}
Finally, using the expressions
\begin{align}
    \varUpsilon(\alpha_{\rm f},t~;\alpha_j,t_j) &= \int\frac{d^2\xi'd^2\eta'}{\pi^2}\sum_{k'}{\rm Tr}\mleft[\hat{D}^{\dagger}(\xi' )\hat{M}_{k'}(t,t_j)\hat{D}^{\dagger}(\eta')\hat{M}^{\dagger}_{k'}(t,t_j)\mright]e^{-\alpha_{\rm f}\xi'^* + \alpha_{\rm f}^*\xi' }e^{-\alpha_j\eta'^* + \alpha^*_j\eta'},\\
    \varUpsilon(\alpha_j,t_j~;\alpha_0,t_0) &= \int\frac{d^2\xi d^2\eta}{\pi^2}\sum_k{\rm Tr}\mleft[\hat{D}^{\dagger}(\xi)\hat{M}_k(t_j,t_0)\hat{D}^{\dagger}\mleft(\eta \mright)\hat{M}^{\dagger}_k(t_j,t_0)\mright]e^{-\alpha_j\xi^* + \alpha^*_j\xi}e^{-\alpha_0\eta^* + \alpha^*_0\eta },
\end{align}
we obtain
\begin{align}
    \varUpsilon\mleft(\alpha_{\rm f},t~;\alpha_0,t_0\mright) = \int\frac{d^2\alpha_j}{\pi}\varUpsilon(\alpha_{\rm f},t~;\alpha_j,t_j)\varUpsilon(\alpha_j,t_j~;\alpha_0,t_0).
\end{align}
This is the Markov condition for the propagator of the Wigner function.


\section{\label{Weyl-Wigner representation of a product of three operators} Weyl-Wigner representation of a product of three operators: Derivation of Eq.~\texorpdfstring{\eqref{eq:ABC_Weyl_Wigner_representation}}{TEXT}}
First, we regard the product of three operators as a product of two operators $\hat{A}$ and $\hat{B}\hat{C}$ in order to use Eq.~\eqref{eq:AB_Weyl_Wigner_representation_1} and apply Eq.~\eqref{eq:AB_Weyl_Wigner_representation_2} for $[\hat{B}\hat{C}]_W(\alpha)$:
\begin{align}
    \mleft[\hat{A}\hat{B}\hat{C}\mright]_W(\alpha) &= \int\frac{d^2\zeta d^2\eta}{\pi^2}e^{\eta^*(\alpha - \zeta) - \eta(\alpha^* - \zeta^*)}A_W\mleft(\zeta - \frac{\eta}{2}\mright)\mleft[\hat{B}\hat{C}\mright]_W(\zeta) \\
    &= \int\frac{d^2\alpha_0 d^2\xi d^2\zeta d^2\eta}{\pi^4}e^{\eta^*(\alpha - \zeta) - \eta(\alpha^* - \zeta^*)}e^{\xi^*(\zeta - \alpha_0) - \xi(\zeta^* - \alpha^*_0)}C_W\mleft(\alpha_0 + \frac{\xi}{2}\mright)A_W\mleft(\zeta - \frac{\eta}{2}\mright)B_W(\alpha_0) \\
    \label{eq:ABC_Weyl_Wigner_representation_appendix_tochu1}
    &= \int\frac{d^2\alpha_0 d^2\xi d^2\delta\zeta d^2\eta}{\pi^4}e^{\eta^*(\alpha - \alpha_0) - \eta(\alpha^* - \alpha^*_0)}e^{\delta\zeta^*(\eta - \xi) - \delta\zeta(\eta^* - \xi^*)}C_W\mleft(\alpha_0 + \frac{\xi}{2}\mright)A_W\mleft(\alpha_0 - \frac{\eta}{2} + \delta\zeta\mright)B_W(\alpha_0),
\end{align}
where the last line is obtained by variable transformation $\zeta\to\alpha_0 + \delta\zeta$.
We expand the function $A_W(\alpha_0 - \eta/2 + \delta\zeta)$ with respect to $\delta\zeta$ and $\delta\zeta^*$ as
\begin{align}
    A_W\mleft(\alpha_0 - \frac{\eta}{2} + \delta\zeta\mright) &= \mleft.\sum_{n=0}^{\infty}\frac{1}{n!}\mleft(\delta\zeta\frac{\partial}{\partial\sigma_a} + \delta\zeta^*\frac{\partial}{\partial\sigma_a^*}\mright)^nA_W(\sigma_a)\mright|_{\sigma_a = \alpha_0 - \eta/2} = \mleft.\sum_{p,q = 0}^{\infty}\frac{\delta\zeta^p\delta\zeta^{*q}}{p!q!}\frac{\partial^{p+q}A_W(\sigma_a)}{\partial\sigma_a^p\partial\sigma_a^{*q}}\mright|_{\sigma_a = \alpha_0 - \eta/2}.
\end{align}
Substituting this expression into Eq.~\eqref{eq:ABC_Weyl_Wigner_representation_appendix_tochu1}, we obtain
\begin{align}
    \mleft[\hat{A}\hat{B}\hat{C}\mright]_W(\alpha) = \sum_{p,q = 0}^{\infty}\frac{1}{p!q!}&\mleft.\int\frac{d^2\alpha_0 d^2\xi d^2\eta}{\pi^2}e^{\eta^*(\alpha - \alpha_0) - \eta(\alpha^* - \alpha^*_0)}C_W\mleft(\alpha_0 + \frac{\xi}{2}\mright)\frac{\partial^{p+q}A_W(\sigma_a)}{\partial\sigma_a^p\partial\sigma_a^{*q}}\mright|_{\sigma_a = \alpha_0 - \eta/2}B_W(\alpha_0) \nonumber\\
    &\times \int\frac{d^2\delta\zeta}{\pi^2}\delta\zeta^p\delta\zeta^{*q}e^{\delta\zeta^*(\eta - \xi) - \delta\zeta(\eta^* - \xi^*)} \\
    \hphantom{\mleft[\hat{A}\hat{B}\hat{C}\mright]_W(\alpha)} = \sum_{p,q = 0}^{\infty}\frac{(-1)^q}{p!q!}&\mleft.\int\frac{d^2\alpha_0 d^2\xi d^2\eta}{\pi^2}e^{\eta^*(\alpha - \alpha_0) - \eta(\alpha^* - \alpha^*_0)}C_W\mleft(\alpha_0 + \frac{\xi}{2}\mright)\frac{\partial^{p+q}A_W(\sigma_a)}{\partial\sigma_a^p\partial\sigma_a^{*q}}\mright|_{\sigma_a = \alpha_0 - \eta/2}B_W(\alpha_0) \nonumber\\
    &\times \frac{\partial^{p+q}}{\partial\xi^{*p}\partial\xi^q}\int\frac{d^2\delta\zeta}{\pi^2}e^{\delta\zeta^*(\eta - \xi) - \delta\zeta(\eta^* - \xi^*)} \\
    \hphantom{\mleft[\hat{A}\hat{B}\hat{C}\mright]_W(\alpha)} = \sum_{p,q = 0}^{\infty}\frac{(-1)^q}{p!q!}&\mleft.\int\frac{d^2\alpha_0 d^2\xi d^2\eta}{\pi^2}e^{\eta^*(\alpha - \alpha_0) - \eta(\alpha^* - \alpha^*_0)}C_W\mleft(\alpha_0 + \frac{\xi}{2}\mright)\frac{\partial^{p+q}A_W(\sigma_a)}{\partial\sigma_a^p\partial\sigma_a^{*q}}\mright|_{\sigma_a = \alpha_0 - \eta/2}B_W(\alpha_0) \nonumber \\
    &\times\frac{\partial^{p+q}}{\partial\xi^{*p}\partial\xi^q}\delta^{(2)}\mleft(\eta - \xi\mright),
\end{align}
where we have used Eq.~\eqref{eq:Dirac_delta_function_phase_space} in the last line.
Finally, performing integration by parts $p+q$ times with respect to $\xi$, we obtain
\begin{align}
    &\mleft[\hat{A}\hat{B}\hat{C}\mright]_W(\alpha) = \int\frac{d^2\alpha_0 d^2\xi d^2\eta}{\pi^2}e^{\eta^*(\alpha - \alpha_0) - \eta(\alpha^* - \alpha^*_0)}\sum_{p,q = 0}^{\infty}\frac{(-1)^p}{2^{p+q}p!q!}\mleft\{\mleft.\frac{\partial^{p+q}C_W\mleft(\sigma_c\mright)}{\partial\sigma_c^{*p}\partial\sigma_c^q}\mright|_{\sigma_c = \alpha_0 + \eta/2 }\mleft.\frac{\partial^{p+q}A_W(\sigma_a)}{\partial\sigma_a^p\partial\sigma_a^{*q}}\mright|_{\sigma_a = \alpha_0 - \eta/2}\mright\}B_W(\alpha_0) \\
    &\hphantom{\mleft[\hat{A}\hat{B}\hat{C}\mright]_W(\alpha)} = \int\frac{d^2\alpha_0 d^2\eta}{\pi^2}e^{\eta^*(\alpha - \alpha_0) - \eta(\alpha^* - \alpha_0^*)}C_W\mleft(\alpha_0 + \frac{\eta}{2}\mright)\star^{e}A_W\mleft(\alpha_0 - \frac{\eta}{2}\mright)B_W(\alpha_0),
\end{align}
where $\star^{e}$ is the extended Moyal product defined by Eq.~\eqref{eq:extended_Moyal_product}.


\section{\label{Hubbard-Stratonovich transformation} Hubbard-Stratonovich transformation: Derivation of Eqs.~\texorpdfstring{\eqref{eq:Hubbard-Stratonovich_transformation}}{TEXT} and \texorpdfstring{\eqref{eq:Hubbard-Stratonovich_transformation_Lambda=lambda}}{TEXT}}
\subsection{\label{Phase-space Gaussian integral} Phase-space Gaussian integral}
We first derive the following Gaussian integral in the phase space necessary for deriving Eqs.~\eqref{eq:Hubbard-Stratonovich_transformation} and \eqref{eq:Hubbard-Stratonovich_transformation_Lambda=lambda}:
\begin{align}
    \label{eq:Gaussian_integral_phase_space}
    {\rm exp}\mleft(-\mleft[\eta^*, \eta\mright]
    \bm{A}
    \begin{bmatrix}
        \eta \\
        \eta^*
    \end{bmatrix}
    \mright)
    =
    \frac{1}{\sqrt{\text{det}\bm{A}}}\int\frac{d^2\xi}{\pi}{\rm exp}\mleft(-\frac{1}{2}\mleft[\xi^*, \xi\mright]\bm{A}^{-1}
    \begin{bmatrix}
        \xi \\
        \xi^*
    \end{bmatrix}
    + \sqrt{2}i\mleft(\eta^*\xi + \eta\xi^*\mright)
    \mright),
\end{align}
where $\bm{A}$ is a $2\times 2$ positive-definite matrix and $\eta$ is a complex value.
In the following, we calculate the right-hand side and show it agrees with the left-hand side.
For this purpose, we introduce the transformation matrix $\bm{V}$ as
\begin{align}
    \bm{V} =
    \begin{bmatrix}
        1 & i \\
        1 & -i
    \end{bmatrix},\quad
    \bm{V}^{-1} = \frac{1}{2}
    \begin{bmatrix}
        1 & 1 \\
        -i & i
    \end{bmatrix},
\end{align}
which acts on the vector $[\xi,\xi^*]^{\rm T}$ as
\begin{align}
    \bm{V}^{-1}
    \begin{bmatrix}
        \xi \\
        \xi^*
    \end{bmatrix} =
    \begin{bmatrix}
        \xi^{\rm re} \\
        \xi^{\rm im}
    \end{bmatrix},\quad
    \begin{bmatrix}
        \xi^*,\xi
    \end{bmatrix}\bm{V} = 2
    \begin{bmatrix}
        \xi^{\rm re}, \xi^{\rm im}
    \end{bmatrix}.
\end{align}
Substituting the $2\times 2$ identity matrix $\bm{V}\bm{V}^{-1} = \bm{1}$ into both side of $\bm{A}^{-1}$ on the right-hand side of Eq.~\eqref{eq:Gaussian_integral_phase_space} and using the fact that $\eta^*\xi + \eta\xi^* = [\eta^{\rm re},\eta^{\rm im}][\xi^{\rm re},\xi^{\rm im}]^{\rm T}$, we obtain
\begin{align}
    \label{eq:phase_space_Gaussian_integral_tochu_appendix}
    \text{RHS of Eq.~\eqref{eq:Gaussian_integral_phase_space}}
    &=
    \frac{2}{\sqrt{\text{det}\bm{A}}}\int\frac{d^2\xi}{2\pi}{\rm exp}\mleft(-\frac{1}{2}\mleft[\xi^{\rm re}, \xi^{\rm im}\mright]2\bm{V}^{-1}\bm{A}^{-1}\bm{V}
    \begin{bmatrix}
        \xi^{\rm re} \\
        \xi^{\rm im}
    \end{bmatrix}
    + 2\sqrt{2}i
    \begin{bmatrix}
        \eta^{\rm re},\eta^{\rm im}
    \end{bmatrix}
    \begin{bmatrix}
        \xi^{\rm re} \\
        \xi^{\rm im}
    \end{bmatrix}
    \mright).
\end{align}
Here, we can perform the integral by using the multiple-variables Gaussian integral formula:
\begin{align}
    \int \frac{d^2\xi}{2\pi}~{\rm exp}\mleft(-\frac{1}{2}\mleft[\xi^{\rm re},\xi^{\rm im}\mright]\bm{X}
    \begin{bmatrix}
        \xi^{\rm re} \\
        \xi^{\rm im}
    \end{bmatrix}
    + \mleft[u,v\mright]
    \begin{bmatrix}
        \xi^{\rm re} \\
        \xi^{\rm im}
    \end{bmatrix}
    \mright)
    =
    \frac{1}{\sqrt{\text{det}\bm{X}}}
    {\rm exp}\mleft(-\frac{1}{2}\mleft[u, v\mright]
    \bm{X}^{-1}
    \begin{bmatrix}
        u \\
        v
    \end{bmatrix}
    \mright),
\end{align}
where $\bm{X}$ is a $2\times 2$ positive-definite matrix and $u$ and $v$ are complex variables.
Substituting Eq.~\eqref{eq:phase_space_Gaussian_integral_tochu_appendix} with $\bm{X} = 2\bm{V}^{-1}\bm{A}^{-1}\bm{V}$ and $[u,v] = [2\sqrt{2}i\eta^{\rm re}, 2\sqrt{2}i\eta^{\rm im}]$ into the right-hand side of Eq.~\eqref{eq:phase_space_Gaussian_integral_tochu_appendix}, we obtain the left-hand side of Eq.~\eqref{eq:Gaussian_integral_phase_space}.


\subsection{\label{sub:Hubbard-Stratonovich transformation} Hubbard-Stratonovich transformation for \texorpdfstring{$\Lambda > |\lambda|$}{TEXT}{\rm :} Derivation of Eq.~\texorpdfstring{\eqref{eq:Hubbard-Stratonovich_transformation}}{TEXT}}
In the case of $\Lambda>|\lambda|$, the left-hand side of Eq.~\eqref{eq:Hubbard-Stratonovich_transformation} can be written in the following form:
\begin{align}
    \label{eq:Hubbard_Stratonovich_appendix_1}
    e^{-2\Lambda\Delta t |\eta|^2 - \lambda^*\Delta t\eta^2 - \lambda\Delta t \eta^{*2}}
    =
    {\rm exp}\mleft(-\mleft[\eta^*  \eta\mright]
    \bm{A}
    \begin{bmatrix}
        \eta \\
        \eta^*
    \end{bmatrix}
    \mright), 
\end{align}
where we have replaced $\eta_{j+1}$ with $\eta$ for simplicity and defined
\begin{align}
    \bm{A} = \Delta t
    \begin{bmatrix}
        \Lambda & \lambda \\
        \lambda^* & \Lambda
    \end{bmatrix}.
\end{align}
According to Eq.~\eqref{eq:Gaussian_integral_phase_space}, when the matrix $\bm{A}$ is positive-definite, i.e., $\Lambda > |\lambda|$, we can transform the right-hand side of Eq.~\eqref{eq:Hubbard_Stratonovich_appendix_1} into an integral form as
\begin{align}
    \label{eq:Hubbard_Stratonovich_appendix_2}
    {\rm exp}\mleft(-\mleft[\eta^*  \eta\mright]
    \bm{A}
    \begin{bmatrix}
        \eta \\
        \eta^*
    \end{bmatrix}
    \mright)
    =
    \frac{1}{\sqrt{\text{det}\bm{A}}}\int\frac{d^2\xi}{\pi}{\rm exp}\mleft(-\frac{1}{2}\mleft[\xi^*, \xi\mright]\bm{A}^{-1}
    \begin{bmatrix}
        \xi \\
        \xi^*
    \end{bmatrix}
    + \sqrt{2}i\mleft(\eta^*\xi + \eta\xi^*\mright)
    \mright).
\end{align}
The matrix $\bm{A}$ is diagonalized as
\begin{align}
    \bm{U}^{-1}\bm{A}\bm{U} =
    \begin{bmatrix}
        \Lambda - |\lambda| & 0 \\
        0 & \Lambda + |\lambda|
    \end{bmatrix},
\end{align}
where the unitary matrix $\bm{U}$ is given by
\begin{align}
    \bm{U} =
    \begin{bmatrix}
        -ie^{i\theta/2} & e^{i\theta/2} \\
        ie^{-i\theta/2} & e^{-i\theta/2}
    \end{bmatrix}
\end{align}
with $\theta = {\rm arg}(\lambda)$.
Inserting the identity matrix $\bm{U}\bm{U}^{-1} = \bm{1}$ into the both side of the matrix $\bm{A}^{-1}$ in the right-hand side of Eq.~\eqref{eq:Hubbard_Stratonovich_appendix_2}, we obtain
\begin{align}
    \label{eq:Hubbard_Stratonovich_appendix_3}
    {\rm exp}\mleft(-\mleft[\eta^*  \eta\mright]
    \bm{A}
    \begin{bmatrix}
        \eta \\
        \eta^*
    \end{bmatrix}
    \mright)
    &=
    \frac{1}{\sqrt{\Delta t^2\mleft(\Lambda^2 - |\lambda|^2\mright)}}\int\frac{d^2\xi}{\pi}{\rm exp}\mleft(-\frac{1}{2\Delta t}\mleft[\xi^*, \xi\mright]\bm{U}
    \begin{bmatrix}
        \Lambda - |\lambda| & 0 \\
        0 & \Lambda + |\lambda|
    \end{bmatrix}^{-1}
    \bm{U}^{-1}
    \begin{bmatrix}
        \xi \\
        \xi^*
    \end{bmatrix}
    + \sqrt{2}i\mleft(\eta^*\xi + \eta\xi^*\mright)
    \mright),
\end{align}
where we use $\text{det}\bm{A} = \Delta t^2(\Lambda^2 - |\lambda|^2)$.
Here, we transform the integral variable $\xi$ to $\Delta \Xi$ such that
\begin{align}
    \sqrt{2}\bm{U}^{-1}
    \begin{bmatrix}
        \xi \\
        \xi^*
    \end{bmatrix}
    =
    \begin{bmatrix}
        \Delta \Xi^{\rm re} \\
        \Delta \Xi^{\rm im}
    \end{bmatrix},
\end{align}
and $\Delta\Xi = \Delta\Xi^{\rm re} + i\Delta\Xi^{\rm im}$, where $\Delta\Xi^{\rm re}$ and $\Delta\Xi^{\rm im}$ are always real, and obtaining
\begin{align}
    \label{eq:Hubbard_Stratonovich_appendix_4}
    {\rm exp}\mleft(-\Delta t\mleft[\eta^*  \eta\mright]
    \bm{A}
    \begin{bmatrix}
        \eta \\
        \eta^*
    \end{bmatrix}
    \mright)
    &=
    \int d^2\Delta\Xi\frac{e^{-(\Delta\Xi^{\rm re})^2/2\Delta t\mleft(\Lambda - |\lambda|\mright)}}{\sqrt{2\pi\Delta t\mleft(\Lambda - |\lambda|\mright)}}\frac{e^{-(\Delta\Xi^{\rm im})^2/2\Delta t\mleft(\Lambda + |\lambda|\mright)}}{\sqrt{2\pi\Delta t\mleft(\Lambda + |\lambda|\mright)}}{\rm exp}\mleft(\eta^*e^{i\theta/2}\Delta\Xi - \eta e^{-i\theta/2}\Delta\Xi^*\mright).
\end{align}
Finally, by comparing Eqs.~\eqref{eq:Hubbard_Stratonovich_appendix_1} and \eqref{eq:Hubbard_Stratonovich_appendix_4}, we obtain
\begin{align}
    \label{eq:derivation_Hubbard-Stratonovich_transformation_appendix1}
    e^{-2\Lambda\Delta t|\eta|^2 - \lambda\Delta t\eta^2 - \lambda^*\Delta t\eta^{*2}} = \int d^2\Delta\Xi\frac{e^{-(\Delta\Xi^{\rm re})^2/2\Delta t\mleft(\Lambda - |\lambda|\mright)}}{\sqrt{2\pi\Delta t\mleft(\Lambda - |\lambda|\mright)}}\frac{e^{-(\Delta\Xi^{\rm im})^2/2\Delta t\mleft(\Lambda + |\lambda|\mright)}}{\sqrt{2\pi\Delta t\mleft(\Lambda + |\lambda|\mright)}}{\rm exp}\mleft(\eta^*e^{i\theta/2}\Delta\Xi - \eta e^{-i\theta/2}\Delta\Xi^*\mright).
\end{align}
This completes the derivation of Eq.~\eqref{eq:Hubbard-Stratonovich_transformation}.


\subsection{\label{sub:Hubbard-Stratonovich transformation Lambda = lambda} Hubbard-Stratonovich transformation for \texorpdfstring{$\Lambda = |\lambda|$}{TEXT}{\rm :} Derivation of Eq.~\texorpdfstring{\eqref{eq:Hubbard-Stratonovich_transformation_Lambda=lambda}}{TEXT}}
Our starting point is again the left-hand side of Eq.~\eqref{eq:Hubbard_Stratonovich_appendix_1}. 
For the case of $\Lambda=|\lambda|$, we can not use Eq.~\eqref{eq:Gaussian_integral_phase_space} because the matrix $\bm{A}$ has a zero eigenvalue and breaks the positive-definiteness.
Instead, by noting that the argument of the exponential function on the left-hand side of Eq.~\eqref{eq:Hubbard_Stratonovich_appendix_1} is always real, given by
$-2\Lambda\Delta t|\eta|^2 - \lambda^*\Delta t\eta^2 - \lambda\Delta t\eta^{*2} = -\Lambda\Delta t(\eta^*e^{i\theta/2} + \eta e^{-i\theta/2})^2 = -\Lambda\Delta t \tilde{\eta}^2$, $\tilde{\eta}\in \mathbb{R}$,
we can use the single-variable Gaussian integral formula
\begin{align*}
    e^{-\Lambda \Delta t\tilde{\eta}^2} = \frac{1}{\sqrt{4\pi\Lambda\Delta t}}\int_{-\infty}^\infty d\tilde{\xi} \exp\mleft(-\frac{\tilde{\xi}^2}{4\Lambda \Delta t}+i\tilde{\eta}\tilde{\xi}\mright)
\end{align*}
to rewrite the left-hand side of Eq.~\eqref{eq:Hubbard_Stratonovich_appendix_1}.
In particular, denoting $\tilde{\xi}$ as $\Delta\Xi^{\rm im}$ and recovering the detailed form of $\tilde{\eta}$, we derive Eq.~\eqref{eq:Hubbard-Stratonovich_transformation_Lambda=lambda} as
\begin{align}
    \label{eq:derivation_Hubbard-Stratonovich_transformation_appendix2}
    e^{-2\Lambda\Delta t|\eta|^2 - \lambda\Delta t\eta^2 - \lambda^*\Delta t\eta^{*2}} = \int d\Delta\Xi^{\rm im}\frac{e^{-(\Delta\Xi^{\rm re})^2/4\Delta t\Lambda}}{\sqrt{4\pi\Delta t\Lambda}}{\rm exp}\mleft(i\eta^*e^{i\theta/2}\Delta\Xi^{\rm im} + i\eta e^{-i\theta/2}\Delta\Xi^{\rm im}\mright).
\end{align}

We can also derive Eq.~\eqref{eq:derivation_Hubbard-Stratonovich_transformation_appendix2} as the $|\lambda|\to\Lambda$ limit of Eq.~\eqref{eq:derivation_Hubbard-Stratonovich_transformation_appendix1}.
In this limit, one of the Gaussian functions in the integrand of Eq.~\eqref{eq:derivation_Hubbard-Stratonovich_transformation_appendix1} behaves as the Dirac delta function,
\begin{align}
    \lim_{|\lambda|\to\Lambda}\frac{e^{-(\Delta\Xi^{\rm re})^2/2\Delta t\mleft(\Lambda - |\lambda|\mright)}}{\sqrt{2\pi\Delta t\mleft(\Lambda - |\lambda|\mright)}} = \delta\mleft(\Delta \Xi^{\rm re}\mright).
\end{align}
We then obtain the $|\lambda|\to\Lambda$ limit of Eq.~\eqref{eq:derivation_Hubbard-Stratonovich_transformation_appendix1} as
\begin{align}
    \lim_{|\lambda|\to\Lambda} e^{-2\Lambda\Delta t|\eta|^2 - \lambda\Delta t\eta^2 - \lambda^*\Delta t\eta^{*2}} &= \int d\Delta\Xi^{\rm im}\frac{e^{-(\Delta\Xi^{\rm re})^2/4\Delta t\Lambda}}{\sqrt{4\pi\Delta t\Lambda}}{\rm exp}\mleft(i\eta^*e^{i\theta/2}\Delta\Xi^{\rm im} + i\eta e^{-i\theta/2}\Delta\Xi^{\rm im}\mright),
\end{align}
which is identical to Eq.~\eqref{eq:derivation_Hubbard-Stratonovich_transformation_appendix1}.
According to this result, we can conclude that the Hubbard-Storatonovich transformation Eq.~\eqref{eq:Hubbard-Stratonovich_transformation} is feasible for $\Lambda \geq |\lambda|$ as in Eq.~\eqref{eq:Hubbard_Storatonovich_transformation_condition}.















\bibliographystyle{elsarticle-num}
\bibliography{elsarticle-num}




\end{document}